\def\mnhbodyspacing{1}
\def\mnhincludeappendix{1}
\PassOptionsToPackage{unicode}{hyperref}
\PassOptionsToPackage{hyphens}{url}
\PassOptionsToPackage{dvipsnames,svgnames,x11names}{xcolor}
\documentclass[12pt]{article}
\usepackage[pdftex]{graphicx}       % to include graphs
\usepackage[]{natbib}
\usepackage{amssymb,amsmath} % for maths symbols and equation numbering
\usepackage{xr-hyper}
\usepackage{hyperref}
\usepackage{enumitem}
\usepackage{xcolor}
\usepackage{colortbl}
\usepackage{mathtools}
\usepackage{makeidx}
\usepackage[T1]{fontenc}
\usepackage{lmodern}
\IfFileExists{microtype.sty}{%
  \usepackage[]{microtype}
  \UseMicrotypeSet[protrusion]{basicmath}
}{}
\usepackage{subcaption}
\usepackage{algorithm}
\usepackage{algpseudocode}
\usepackage{tikz}
\usepackage[section]{placeins}
\usepackage{lscape}
\usepackage{multirow}
\usepackage{setspace}
\usepackage{afterpage}
\usepackage{titlesec}

\definecolor{dark-red}{rgb}{0.6,0.15,0.15}
\definecolor{dark-blue}{rgb}{0.02,0.1,0.5}
\definecolor{medium-blue}{rgb}{0,0,0.5}
\definecolor{darkgreen}{rgb}{0.0, 0.5, 0.0} % This defines a darker shade of green.
\providecommand{\mnhpdfauthor}{Mante Zelvyte; Jim E. Griffin}
\providecommand{\mnhbodyspacing}{1.8}
\providecommand{\mnhincludeappendix}{0}
\hypersetup{
    pdftitle={A Multiplex Network Hawkes Model for Systemic Risk Measurement}, 
    pdfauthor={\mnhpdfauthor},
    pdfkeywords={Credit Default Swaps; Contagion Channels; Regression Modelling; Network Measures},
    colorlinks=true,
    linkcolor={dark-blue},
    citecolor={dark-red},
    urlcolor={dark-blue}
}
\usepackage{bm}

\newcommand{\E}{\mathbb{E}}

\newcommand{\diff}{\mbox{d}}

\makeatletter
\let\mnhsavedbibcite\bibcite
\def\bibcite#1#2{}
\if1\mnhincludeappendix
\else
\fi
\let\bibcite\mnhsavedbibcite
\makeatother

\newcommand*{\B}[1]{\ifmmode\bm{#1}\else\textbf{#1}\fi}
\titlespacing*{\paragraph}{0pt}{0.3ex plus 0.1ex minus 0.05ex}{0.4em}

\let\leq=\leqslant   % for nice-looking inequality signs
\let\geq=\geqslant

\numberwithin{equation}{section}  % equation numbers like (section#.equation#)

\providecommand{\anon}{1}

\makeindex            % create index
\DeclareUnicodeCharacter{2217}{*}

\begin{document}

\def\spacingset#1{\renewcommand{\baselinestretch}%
{#1}\small\normalsize} \spacingset{1}

%%%%%%%%%%%%%%%%%%%%%%%%%%%%%%%%%%%%%%%%%%%%%%%%%%%%%%%%%%%%%%%%%%%%%%%%%%%%%%

\if1\anon
{
  \title{\bf A Multiplex Network Hawkes Model for Systemic Risk Measurement}
  \author{Mante Zelvyte\\
    Department of Statistical Science, University College London\\
    and \\
    Jim E. Griffin \\
    Department of Statistical Science, University College London}
  \maketitle
} \fi

\if0\anon
{
  \bigskip
  \bigskip
  \bigskip
  \begin{center}
    {\LARGE\bf A Multiplex Network Hawkes Model for Systemic Risk Measurement} 
  \end{center}
  \medskip
} \fi

\bigskip
\begin{abstract}
We introduce the Multiplex Network Hawkes model, which extends the network Hawkes framework of \cite{linderman2014discovering} by allowing multiple excitation layers whose weights depend on observed edge and node covariates. We use the model to investigate how contagion in financial networks is affected by different transmission channels. The multiplex structure separates channel-specific contributions within a single inferred transmission network, allowing candidate propagation mechanisms to be compared directly rather than being absorbed into one homogeneous excitation layer. Covariate-dependent excitation allows us to investigate sources of transmission. We make posterior inference about the inferred directed network and its excitation dynamics using an MCMC sampler. The application uses a broad cross-industry credit default swap (CDS) dataset of 99 North American and European firms, including banks, insurers and non-financial firms over 2004--2022. We evaluate three candidate contagion channels associated with asset similarity, solvency and profitability. The results indicate sparse contagion pathways, with systemic-risk transmission concentrated in outward flows from a small number of influential institutions rather than in mutual feedback between institutions. The channel results show that industry similarity is the most consistently supported asset-similarity effect, while aggregate layer contributions indicate that asset-similarity, solvency and profitability channels all contribute to inferred excitation.
\end{abstract}
\noindent%

\vfill

\newpage
\spacingset{\mnhbodyspacing} % DON'T change the JBES spacing!

\section{Introduction} 
\label{sec:introduction}

The concept of systemic risk %\footnote{This is not to be confused with systematic risk, which refers to non-diversifiable risk driven by common factors. Systemic risk involves idiosyncratic shocks—such as a large bankruptcy—spreading through the financial system, often overwhelming diversification strategies \cite{barigozzi2021time}.}
has drawn sustained interest from central banks, regulators, and economists for over five decades. Research consistently shows that distress in a few key institutions---via liquidity or valuation shocks---can trigger widespread losses, a reality underscored by the 2008-2009 financial crisis and, more recently, the 2023 banking distress wave (e.g., Credit Suisse, SVB). Understanding and quantifying structural interconnectedness is therefore crucial for systemic risk monitoring and minimizing costly government interventions.

Systemically important institutions are typically identified via balance sheet data, such as the BIS's indicator-based G-SIB scores \citep{gsib}. However, because precise exposure data is siloed by regulators, and public balance-sheet metrics suffer from low reporting frequencies and quarter-end window-dressing \citep{behn2022behind}, their utility for dynamic, forward-looking risk assessment is severely limited.

To address data limitations, research has explored alternative systemic risk measures based on market data. Contagion modelling approaches generally fall into five categories. \textbf{Structural balance sheet models} simulate loss propagation through explicit balance sheet linkages \citep{eisenberg2001systemic}, but require full network specification.
\textbf{Co-dependency models} rely on market correlations, (G)ARCH, Granger causality, and related methods \citep{barigozzi2017network,Billio2012,Diebold2014}, but often assume linearity and struggle with crisis-induced clustering.
\textbf{Tail risk models} (e.g., CoVaR \citep{tobias2016covar}, MES \citep{Acharya2017}) link firm-specific and market-wide losses but may miss firm-level systemic relevance and depend heavily on historical data.
\textbf{Graphical models} infer network structures from data \citep{Bianchi2019}, capturing aggregate contagion effects but not individual transmission channels.
\textbf{Discrete process models} (e.g., Hawkes, Poisson) model shock timing and intensity \citep{agosto2022default}, avoiding Gaussian assumptions but facing scalability and interpretability challenges.

While each captures different facets of systemic risk, many struggle to represent complex endogenous network structures---such as small-world structure, clustering, preferential attachment, asymmetric influence, and homophily \citep{acemoglu2015systemic}---or aggregate all contagion channels (e.g., liquidity spirals, asset fire sales, information asymmetries \citep{allen2000financial}) into a single homogeneous network, obscuring the precise drivers of vulnerability \citep{bargigli2015multiplex}.

To address these challenges, we extend the Network Hawkes model of \citet{linderman2014discovering} to a multiplex setting and couple it with Bayesian inference driven by market-observable event data, enabling the recovery of latent, structured contagion networks. Motivated by the multiplex-network framework \citep{kivela2014multilayer}, we model contagion using additive excitation layers, with covariates that modulate layer weights to facilitate a transparent decomposition of spillovers into economically interpretable transmission mechanisms and to identify and quantify distinct risk propagation drivers. The additive layer representation is a modelling simplification that allows network excitation to be attributed to mechanism-specific contributions; it is not a claim that economic transmission channels are orthogonal in practice. Finally, we apply the model to a large multi-institution, multi-market dataset, and use Bayesian network summaries to characterize the inferred topology together with uncertainty, and to study how these structural features shape systemic risk transmission. %By leveraging this expansive dataset, we can conduct a more comprehensive assessment of the shock transmission mechanisms and identify critical nodes within the network.

The rest of the paper is structured as follows: Section \ref{sec:empiricaldataset} describes the empirical data and the covariates used to represent potential contagion channels; Section \ref{sec:multiplexnetworkhawkes} presents our model; Section \ref{sec:bayesianinference} details inference and network metrics; Section \ref{sec:simulationstudy} evaluates performance on synthetic data; Section \ref{sec:empirical} applies the model to real-world data; and Section \ref{sec:conclusion} concludes with findings and future directions.

\section{Empirical Data Description} \label{sec:empiricaldataset}

We combine market-observable, categorical, and balance sheet data to construct our model \citep{Bianchi2019,hautsch2015financial}. We derive event definitions from 5-year credit default swap (CDS) log-returns---the most liquid segment of the CDS market---rather than equity or bond returns. CDS spreads are a preferred measure of credit risk, incorporating risk-free rates, non-linear risk premia, and liquidity compensation into a comprehensive view of default probabilities \citep{blanco2005empirical,huang2012systemic}. An upward CDS log-return corresponds to a jump in the quoted credit spread and hence a deterioration in perceived credit risk. While CDS markets can be noisy due to speculative trading or short-term pricing staleness, our framework naturally mitigates these issues. First, we define an \emph{extreme credit-spread jump} as a CDS log-return exceeding the 99th percentile for its firm. Second, the model's multi-day influence horizon smoothly accommodates minor pricing delays without disrupting inferred causal links. Our sample comprises $K=99$ companies with sufficient CDS data from major indices (S\&P 500, DAX, Euro STOXX 50, and FTSE 100; see Appendix Table \ref{tab:scenarios_nodes_rd}) between January 2004 and January 2022, yielding $N=1905$ extreme credit-spread jumps over $T=4696$ business days.

Potential drivers of risk transmission between firms are represented as contagion channels, whose excitation weights are driven by firm-specific covariates (Equation \ref{eq:mnhmodel}). Guided by the systemic-risk literature on candidate propagation drivers \citep{altman1968financial,barboza2017machine}, we initially considered covariates capturing asset similarity, liquidity, solvency, and profitability. Following the variable selection procedures detailed in Appendix \ref{app:data}, we define three active channels for our final empirical analysis: an asset similarity channel driven by sector, industry, and primary revenue region; a solvency channel driven by the asset coverage ratio; and a profitability channel driven by book value per share. Finally, to mitigate the impact of extreme outliers, the asset coverage ratio and book value per share were transformed into quantile-based variables.

Our final dataset comprises 58 North American and 41 European firms spanning 11 diverse sectors and 18 industry groups. This broad cross-industry and cross-regional scope enables a more comprehensive analysis of multidimensional contagion than is typical in the systemic risk literature.

\begin{figure}[htbp!]
    %\centering
%\resizebox{}{}{}
\begin{subfigure}{0.46\textwidth}
\centering
\includegraphics[width=1\textwidth, angle=0, trim= 48 16 50 35,clip]{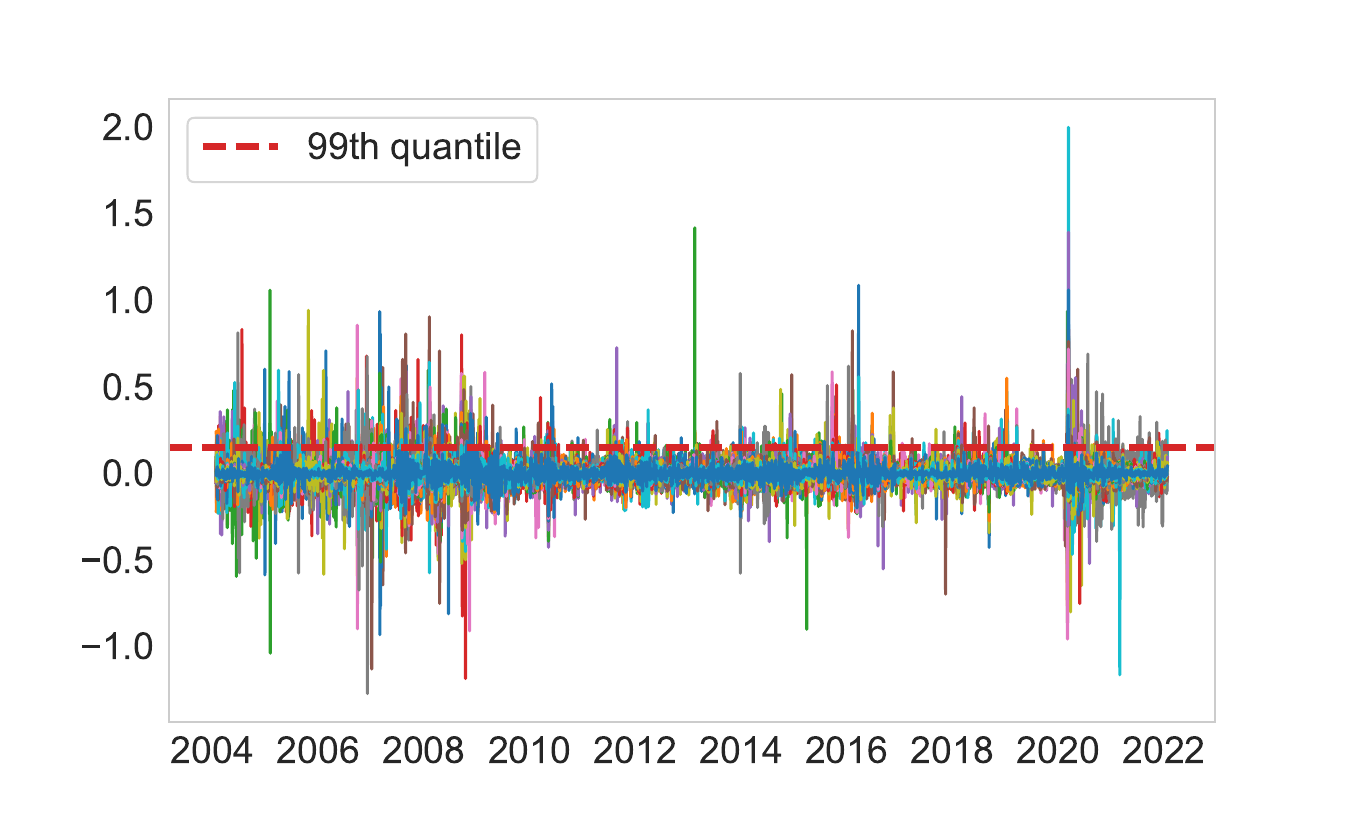}
% corresponding to CDS time series.
%\label{fig:process_rd3}
\end{subfigure}%
\begin{subfigure}{0.54\textwidth}
\centering
\includegraphics[width=1\textwidth]{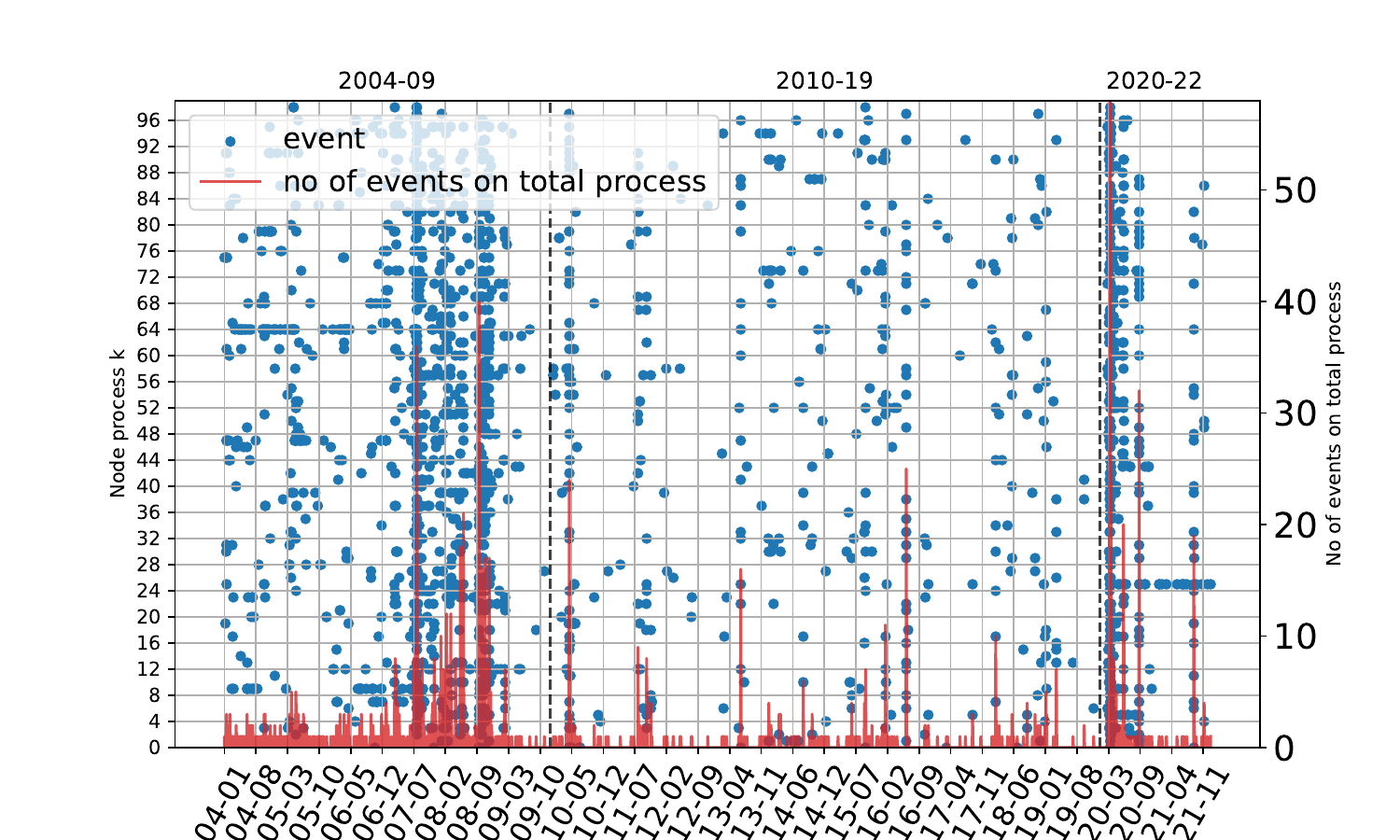}
\end{subfigure}
\caption{Extreme credit-spread jumps for 99 firms, 2004--2022. Left: CDS log returns, where values exceeding the firm's 99th percentile are extreme credit-spread jumps. Right: empirical event process with $N=1905$ events, $T=4696$ business days and $K=99$ nodes (firms). Dashed vertical lines mark the time-variation regimes: pre-GFC/GFC (2004--2009), euro crisis/calm (2010--2019), and COVID-19 (2020--January 2022).}
\label{fig:process_rd2}
\end{figure}
\FloatBarrier

The resulting event process, visualised in Figure \ref{fig:process_rd2}, averages 19 events per firm, with a few highly active nodes (the $99^{th}$ percentile reaches 55 events). In the pooled event process across firms, inter-event gaps are typically short after excluding same-day events (median $\le 2$ days), with significant temporal clustering around the Global Financial Crisis (GFC) and the onset of the COVID-19 pandemic.

\section{Multiplex Network Hawkes Process}
\label{sec:multiplexnetworkhawkes}
Hawkes processes are point processes whose conditional intensities depend on previous events. Suppose there are $K$ event processes, where $s_{k,1},s_{k,2},\ldots$ denote the event times for process $k$, $\mathcal{H}_t=\{s_{k,n}:s_{k,n}<t,\, k=1,\ldots,K\}$ is the history of event times across all processes before time $t$, and let $N_k(t)$ be the counting process for event process $k$. \citet{linderman2014discovering} define a Network Hawkes process to be a $K$-dimensional mutually-exciting point process whose conditional intensity function for process $k$ is
\[
\lambda_k(t\mid\mathcal{H}_t)=\lambda^{(0)}_{k} + \sum_{k'=1}^K\sum_{\{n: s_{k', n} < t\}} h_{k'\rightarrow k}(t - s_{k',n}),
\]
where $\lambda^{(0)}_k$ is a background intensity and $h_{k'\rightarrow k}(x)$ is the increase in the intensity of process $k$ at lag $x$ following an event in process $k'$. This allows an event in one process to increase the future intensity of another process through the impulse response function $h_{k'\rightarrow k}$, which naturally captures temporal clustering and propagation relevant to contagion modelling. The impulse response is decomposed as
\[
h_{k'\rightarrow k}(x)=G_{k'\rightarrow k}g_{\theta_{k'\rightarrow k}}(x)
=A_{k'\rightarrow k}W_{k'\rightarrow k}g_{\theta_{k'\rightarrow k}}(x), \qquad x>0,
\]
where $A_{k'\rightarrow k}\in\{0,1\}$ is the $k'\rightarrow k$ element of the adjacency matrix $A$ and determines whether an interaction is present, $W_{k'\rightarrow k}\in\mathbb{R}_{+}$ is the $k'\rightarrow k$ element of the weight matrix $W$ and controls the strength of the effect, and $g_{\theta_{k'\rightarrow k}}(x)$ is an interaction kernel describing its timing. We take this kernel to be a probability density with bounded support on $[0,\Delta t_{max}]$; the value used for $\Delta t_{max}$ is specified and assessed in the empirical and simulation analyses. As $g_{\theta_{k'\rightarrow k}}$ integrates to one, $G_{k'\rightarrow k}=A_{k'\rightarrow k}W_{k'\rightarrow k}$ is the expected number of offspring events on process $k$ induced by one event on process $k'$.
The multiplex formulation extends the Network Hawkes specification by expressing the excitation matrix as a sum of $L$ layer-specific matrices,
\[
G_{k'\rightarrow k}=\sum_{l=1}^{L}G_{k'\rightarrow k}^{(l)},
\qquad
G_{k'\rightarrow k}^{(l)}=A_{k'\rightarrow k}W_{k'\rightarrow k}^{(l)}.
\]
where $W_{k'\rightarrow k}^{(l)}$ is the $k'\rightarrow k$ element of the layer-specific weight matrix $W^{(l)}$. Equivalently, $G$ is defined on an $L$-layer multiplex graph, i.e. a multilayer network with at least some shared nodes \citep{kivela2014multilayer}. This decomposition lets each layer's weights depend on observable edge and node covariates through layer-specific regressions, so excitation strengths can vary systematically with measured characteristics such as industry similarity or solvency. In the systemic-risk application, the layer contributions are interpreted as candidate contagion channels.

Using this layer structure, we express the conditional intensity of the Multiplex Network Hawkes process as a sum of layers, for each $k\in\{1,\ldots,K\}$,
\begin{equation}\label{eq:mnhmodel}
\begin{aligned}
\lambda_k(t\mid\mathcal{H}_t)
&=\lambda^{(0)}_{k} + \sum_{k'=1}^{K}\sum_{\{n: s_{k',n}<t\}}\sum_{l=1}^{L} h_{k'\rightarrow k}^{(l)}(t-s_{k',n}),\\
h_{k'\rightarrow k}^{(l)}(t-s_{k',n})
&=G_{k'\rightarrow k}^{(l)}g_{\theta_{k'\rightarrow k}}(t-s_{k',n}).
\end{aligned}
\end{equation}
Here $G^{(l)}_{k'\rightarrow k}$ is the expected number of events induced through layer $l$, and $G_{k'\rightarrow k}$ is the total expected number induced across all layers. We use the terms ``network'' and ``excitation network'' interchangeably to refer to the graph $G$, and ``network layer'' or ``excitation layer'' to refer to a sub-graph $G^{(l)}$.

The assumption of a common, static $A$ across model layers is appropriate in our model because effective interactions are determined through the combination $A\odot W^{(l)}$: the main role of $A$ is to control network sparsity, with heterogeneity introduced through $W^{(l)}$. If $W^{(l)}_{k'\rightarrow k}$, which depends on the covariates assigned to layer $l$, is zero or near-zero, the corresponding interaction is effectively removed from $A\odot W^{(l)}$. This assumption can also contribute to stability within the sampler because event data across all model layers inform $A$.

The Poisson superposition theorem allows us to decompose the processes into $L$ independent layer-specific subprocesses, together with the background process, so the Multiplex Network Hawkes intensity is a superposition of Network Hawkes excitation components. For each observed event time $s_{k,n}$, we augment the data with a latent parent indicator $z_{k,n}$. We set $z_{k,n}=(0,0,0)$ when the event is allocated to the background-intensity component, and $z_{k,n}=(k',n',\ell)$ when it is allocated to excitation from a preceding event $s_{k',n'}$ through layer $\ell$. Conditional on the history, the event-allocation probabilities at $t=s_{k,n}$ are given by normalized intensity contributions, \[
\Pr\!\bigl(z_{k,n}=(0,0,0)\mid \mathcal H_{s_{k,n}}\bigr)
=\frac{\lambda^{(0)}_{k}}{\lambda_{k}(s_{k,n}\mid \mathcal H_{s_{k,n}})},~
\Pr\!\bigl(z_{k,n}=(k',n',\ell)\mid \mathcal H_{s_{k,n}}\bigr)
=\frac{h^{(\ell)}_{k'\to k}(s_{k,n}-s_{k',n'})}{\lambda_{k}(s_{k,n}\mid \mathcal H_{s_{k,n}})}.
\]
These allocations provide the marked-event representation used in inference and indicate which background or layer-specific subprocess explains each event. Under the representation, an event allocated to layer $\ell$ contributes through the layer-$\ell$ excitation component; overlap between economic mechanisms that is not captured by a layer-specific contribution is absorbed into the remaining components, including the background intensity. This modelling choice is motivated by our aim of identifying key transmitting institutions and separating candidate risk-propagation mechanisms discussed in the systemic-risk literature \citep[e.g.][]{Longstaff2010,brunnermeieroehmke2013,Kiyotaki2002,allen2000financial}.
%Under the layered decomposition,
%\[
%\lambda_{k}(t\mid\mathcal H_t)
%=\lambda^{(0)}_{k}+\sum_{k'=1}^K\sum_{\{n':\, s_{k',n'}<t\}}\sum_{\ell=1}^L h^{(\ell)}_{k'\to k}(t-s_{k',n'}).
%\]
\begin{figure}[htbp!]
\centering
\begin{subfigure}[b]{0.25\textwidth}
\includegraphics[width=1\textwidth, angle=0, trim= 300 105 295 110,clip]{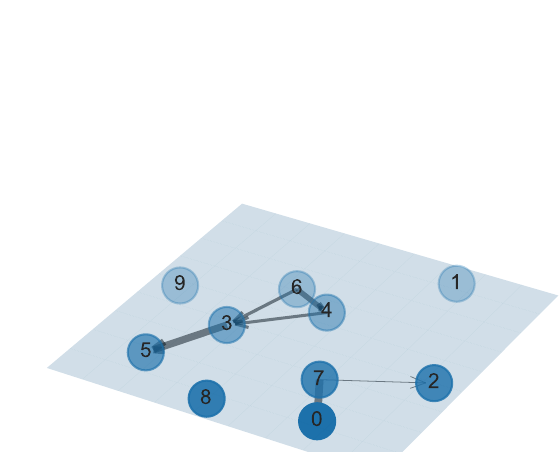}
\end{subfigure}%
\begin{subfigure}[b]{0.25\textwidth}
\includegraphics[width=1\textwidth, angle=0, trim= 300 105 293 110,clip]{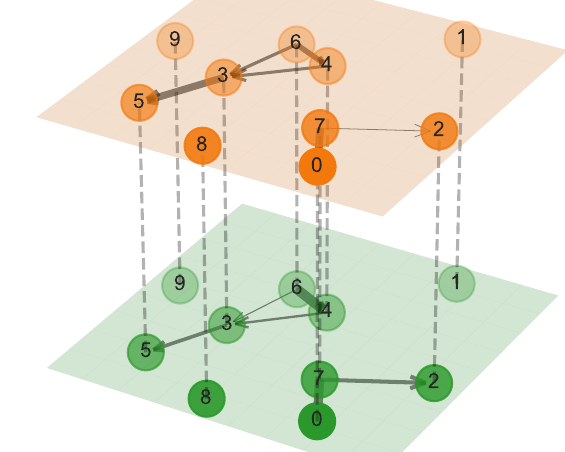}
\end{subfigure}
\caption{Aggregate excitation network $\mathbf{AW}=\sum_{l=1}^{2}\mathbf{AW}^{(l)}$ and its two layer-specific contributions. Darker directed edges indicate larger expected excitation strengths.}
\label{fig:network_gen}
\end{figure}

Figure \ref{fig:network_gen} illustrates the aggregate excitation network and its decomposition as a superposition of two layer-specific networks. Edge darkness indicates excitation strength and directed edges indicate the direction of an interaction.

Link formation in the adjacency matrix $\mathbf{A}$ is modelled through the simple Erd\H{o}s--R\'enyi prior
$A_{k'\rightarrow k}\sim\mbox{Bern}(\rho)$ for all $k'$ and $k$,
which imposes homogeneous link probabilities while controlling overall sparsity through $\rho$. For a directed graph with $n$ nodes and $M$ edges, the graph probability is $\rho^M(1-\rho)^{n(n-1)-M}$.

The layer-specific weight matrices $\mathbf{W}^{(l)}$ are modelled by gamma regressions. We replace the i.i.d.\ gamma weights of \citet{linderman2014discovering} with weights whose means depend on covariates for the ordered pair $k'\rightarrow k$, including pair-level covariates and sender or recipient node covariates:
\begin{align}
W^{(l)}_{k'\rightarrow k} &\sim  \mbox{Gamma}\left(\frac{1}{\kappa^{(l)}},\frac{1}{\kappa^{(l)} \mu_{k'\rightarrow k}^{(l)}}\right),\notag\\
\log \mu_{k'\rightarrow k}^{(l)}&=\eta_{k'\rightarrow k}^{(l)}=\bm{x_{k'\rightarrow k}}^{(l)}\bm{\beta}^{(l)},
\end{align}
where $\bm{\beta}^{(l)}=(\beta_0^{(l)},\ldots,\beta_p^{(l)})'$ are layer-$l$ regression coefficients, and $\bm{x_{k'\rightarrow k}}^{(l)}=(1,x_1,\ldots,x_p)$ is the covariate vector for pair $k'\rightarrow k$ (equivalently, $\mu_{k'\rightarrow k}^{(l)}=\exp\{\bm{x_{k'\rightarrow k}}^{(l)}\bm{\beta}^{(l)}\}$). Note the inverse-scale parameterisation of the gamma distribution is used throughout\footnote{$f(x)=\frac{\beta^\alpha}{\Gamma(\alpha)}x^{\alpha-1}e^{-\beta x}$}.

\iffalse
with the following interpretation of the components:
\begin{itemize}
    \item\textbf{Background intensity $\lambda_k^{(0)}$} - assumed constant, reflects process intensity that is not due past events.
    \item \textbf{Interaction kernel $g_{\theta_{k'\rightarrow k}}\left(\Delta t\right)$} - captures the temporal aspect of the interaction and is static across layers.
    \item \textbf{Adjacency matrix} $\mathbf{A} \in \{0,1\}^{K \times K}$ - specifies the sparsity structure of the network.
    \item \textbf{Weight matrix} $\mathbf{W}^{(l)} \in \mathbb{R}_{+}^{K \times K}$ - defines the strength of the interactions in the network via  edge-specific intensities whose dynamics depend on node-specific features. Corresponds to expected number of events induced via the channel.
\end{itemize}
\fi

%For the background rates $\lambda_k^{(0)}$, immigrant events are those not attributed to preceding events; in our application, $\lambda_k^{(0)}$ is constant in time. %$\lambda^{0}_{k}(t) = \lambda^{0}_{k}$,
%and can be modelled as gamma-distributed variables, i.e.
%\begin{align}
%\lambda_{k}^{0} \sim \mbox{Gamma}(\alpha^0_{\lambda},\beta^0_{\lambda}).
%\end{align}

For the interaction kernel $g_{k'\rightarrow k}$, we follow \cite{linderman2014discovering} and use a logistic-normal density with parameters $\theta_{k'\rightarrow k}=\{\mu_{k'\rightarrow k},\tau_{k'\rightarrow k}\}$. Writing $\Delta t=t-s_{k',n}$:
\begin{align}
g_{k'\rightarrow k}(\Delta t|\mu_{k'\rightarrow k},\tau_{k'\rightarrow k})&=\frac{1}{Z}\exp\left\{\frac{-\tau_{k'\rightarrow k}}{2}\left(\sigma^{-1}\left(\frac{\Delta t}{\Delta t_{max}}\right)-\mu_{k'\rightarrow k}\right)^{2}\right\}.
\end{align}
Here, $\sigma^{-1}(x)=\ln\left(\frac{x}{1-x}\right)$ and $Z=\frac{\Delta(\Delta t_{max}-\Delta t)}{\Delta t_{max}}\left(\frac{\tau_{k'\rightarrow k}}{2\pi}\right)^{-1/2}$.

This choice of kernel facilitates conjugacy with a Normal-Gamma prior. Namely, since $g_{k'\rightarrow k}$ is a probability density, it integrates to one when $s_{k,n}<T-\Delta t_{max}$; thus, when $\Delta t_{max}\ll T$, the spikes at the end of the dataset can be ignored, leaving a likelihood proportional to a product of logistic-normal densities. Because the logistic transform is invertible, we can work with logit-transformed intervals, for which the likelihood becomes a product of normal densities, conjugate with a Normal-Gamma prior.
%Weak Normal-Gamma prior is used for the parameters of the logistic normal distribution $\mu$ and $\tau$, i.e.: $$\mu, \tau \sim \mathcal{NG}(\mu, \tau|\mu_{\mu},k_{\mu},\alpha_{\tau},\beta_{\tau}).$$

%The usual link function for shape regression structure is the logarithm function.

For all $k'$ and $k$ and all layers $l=1,\ldots,L$, the prior hierarchy is summarised below.
\begin{align}
A_{k'\rightarrow k}&\sim\mbox{Bern}(\rho),~ \rho \sim \mbox{Beta}(\alpha_{\rho},\beta_{\rho}),~ W^{(l)}_{k'\rightarrow k} \sim  \mbox{Gamma}\left(\frac{1}{\kappa^{(l)}},\frac{1}{\kappa^{(l)} \mu_{k'\rightarrow k}^{(l)}}\right), \notag\\
\kappa^{(l)} &\sim \mbox{InvGamma}({a^0}_{\kappa},{b^0}_{\kappa}),~b_{\kappa}^{(l)} \sim \mbox{InvGamma}({a^0}_{\kappa_b},~{b^0}_{\kappa_b})\footnotemark,\notag\\
\eta_{k'\rightarrow k}^{(l)}&=\bm{x_{k'\rightarrow k}}^{(l)}\bm{\beta}^{(l)}, \mu_{k'\rightarrow k}^{(l)}=\exp\left(\eta_{k'\rightarrow k}^{(l)}\right),~ \beta^{(l)}  \sim \mbox{MultivariateNormal}(\mu^0_{\beta},\Sigma^0),  \notag\\ 
% \mu_{k'\rightarrow k}&=\exp\left(\eta_{k'\rightarrow l}\right) \notag\\
 g_{k'\rightarrow k}(\Delta t|\mu, \tau)&=\frac{1}{Z}\exp\left\{\frac{-\tau}{2}\left(\sigma^{-1}\left(\frac{\Delta t}{\Delta t_{max}}\right)-\mu\right)^{2}\right\},~ Z=\frac{\Delta(\Delta t_{max}-\Delta t)}{\Delta t_{max}}\left(\frac{\tau}{2\pi}\right)^{-1/2},~ \notag\\
 \sigma^{-1}(x)&=ln\left(\frac{x}{1-x}\right), \mu, \tau \sim \mathcal{NG}(\mu, \tau|\mu^0_{\mu},k^0_{\mu},\alpha^0_{\tau},\beta^0_{\tau}), \lambda_{k}^{(0)} \sim \mbox{Gamma}(\alpha^0_{\lambda},\beta^0_{\lambda}), \notag
\end{align}
\footnotetext{Other options explored are Gamma and Half-Cauchy, with results presented in relevant sections.}
The list of the hyperparameters required to be specified and their default values are as follows:
$\alpha^0_{\lambda}=1,
\beta^0_{\lambda}=1,
\alpha^0_{\rho}=10,
\beta^0_{\rho}=10,
a^0_{\kappa}=0.001,
b^0_{\kappa}=10,
a^0_{\kappa_b}=0.001,
b^0_{\kappa_b}=10,
\mu^0_{\mu}=-1,
k^0_{\mu}=10,
\alpha^0_{\tau}=10.5,
\beta^0_{\tau}=1,
\mu^0_{\beta}=0,
\Sigma^0=10*\bm{I},
\Delta t_{max}=10.$ The parameters are chosen such that the priors are weakly informative or uninformative. $\Delta t_{max}$ is chosen to reflect the value used in the empirical application. The influence of chosen $\Delta t_{max}$ and priors for $\beta$ and its hyper-parameters will be assessed in Section \ref{sec:scenarioanalysis}.

\section{Bayesian Inference}\label{sec:bayesianinference}

We extend the Gibbs sampler of \citet{linderman2014discovering} to the multiplex setting ($L>1$), for which many full conditional distributions have standard forms and can be sampled directly. This section describes the key extensions needed for the multiplex model; the full algorithm is provided in Appendix~\ref{sec:componentinference}. These extensions include: (i) layer-specific excitation strengths $W^{(l)}_{k'\rightarrow k}$ via gamma regression on covariates, necessitating Metropolis-Hastings updates for $\bm{\beta}^{(l)}$, $\kappa^{(l)}$, and $b_{\kappa}^{(l)}$; (ii) an augmented parent-allocation step using latent indicators $z_{k,n}$ to identify both the triggering event and responsible layer; (iii) an updated adjacency-matrix step that samples $\bm{A}$ under the layered intensity decomposition prior to parent resampling consistent with layer-specific intensities; (iv) stabilisation of adaptive Metropolis-Hastings updates for sparse networks; and (v) label-switching post-processing. Non-conjugate components are sampled using Metropolis-within-Gibbs steps.

We implement gamma regression for edge-specific intensity parameters using adaptive random-walk Metropolis-Hastings with adaptive scale (ASM; \citealp{atchade2005adaptive}) followed by adaptive scaling within adaptive Metropolis-Hastings (ASWAM; \citealp{andrieu2008tutorial,atchade2010limit}). Inference relies on 20,500 MCMC samples, discarding a 10\% burn-in. The first 500 updates use ASM to estimate the initial covariance for the subsequent ASWAM sampler. See \citet{griffin2013advances} for a comprehensive algorithmic review.

In sparse financial networks, overconditioning on zero-weight edges ($A_{k'\rightarrow k}=0$) can lead to slow mixing of the regression parameters. For these edges, the posterior $p(W_{k'\rightarrow k}^{(l)}|\ldots)$ equals the prior, offering no information about weight sizes while still influencing $\kappa$ and $\bm{\beta}$. Specifically, very small $\mu$ values on uninformative edges artificially inflate the $1/(\kappa\mu)$ term, risking overflow and destabilising the $\bm{W}$ updates. Appendices~\ref{sec:adaptiveRWMH} and \ref{sec:algorithmstabilisation} detail the Adaptive Metropolis-Hastings algorithms and their stabilisation.

Label switching occurs when a model's likelihood is invariant to component label permutations, rendering parameters non-identifiable. In our multiplex model, the likelihood is exactly invariant to permutations of the layer labels only when there are no covariates or when covariates are equal across layers. With finite samples and weak layer-specific information, the sampler may still display label-switching-like behaviour because the data do not clearly separate the layer contributions. Our framework uses latent indicators $z_{k,n}=(k',n',l)$ to assign each event to a specific trigger and layer. Where switching is evident in the sampler output, we apply the Equivalence Classes Representatives (ECR) post-processing algorithm \citep{papastamoulis2014handling}, which relabels draws by matching to a representative allocation. See Appendix~\ref{sec:labelswitching} for details.

\subsection{Network Measures} \label{sec:networkmeasures}

The graph $G$ inferred by the model is a weighted, directed graph representing estimated transmission of extreme credit-spread jumps. Consequently, the relevant network summaries are those that answer substantive questions about transmission \citep{Billio2012,Diebold2014,Bianchi2019}: whether stress is diffuse or concentrated, whether it flows predominantly in one direction or through feedback relationships, whether it clusters within groups of firms, and which institutions transmit or receive the greatest estimated excitation.

At the graph level, the sparsity parameter $\rho$ summarises how broadly links are supported across the network: a sparse network indicates that estimated transmission is concentrated on relatively few directed relationships. Reciprocity distinguishes predominantly directional propagation from mutual feedback links. Transitivity and weighted clustering measure whether transmission is locally concentrated among connected groups of firms. Small-world networks have two primary characteristics: a short average shortest path length and high clustering. We use transitivity, average clustering and weighted average clustering metrics to assess small-world phenomena. Such a structure matters for systemic risk because short routes combined with local clustering may allow stress to spread rapidly between institutions.

Degree-based summaries address concentration in particular institutions. Weighted out-degree measures the total estimated excitation transmitted by an institution, whereas weighted in-degree measures the excitation it receives. Comparing their distributions identifies whether transmission is spread relatively evenly or concentrated in a small number of outwardly influential hubs. Additionally, we analyse degree distributions to determine whether the network exhibits scale-free properties.\footnote{The classic definition of a scale-free network is that its degree distribution $P(k)$ follows a power law, $k^{-\alpha}$. Some more recent research, e.g.\ \citet{broido2019scale}, suggests that strict scale-free networks are empirically rare.} Degree concentration is relevant for systemic fragility: a scale-free structure can facilitate widespread contagion because shocks to highly connected hub nodes can distribute stress across a large part of the network.\footnote{\citet{das2022bankregulation} study the correspondent-bank network of all U.S. banks on the eve of the Great Depression. They find that its pyramid-shaped topology was inherently fragile and systemically risky, and that a bank's network position and the risk of its neighbours are strong predictors of bank survivorship. This provides related motivation for examining concentration of transmission in systemically consequential institutions.}

Preferential attachment and homophily provide related, but distinct, interpretations of network formation. Concentration in hub-like institutions can be consistent with preferential attachment, whereby already well-connected institutions attract additional links over time. However, testing this mechanism would require a dynamic analysis of link formation, which is outside the scope of the present application. Homophily concerns whether similar institutions are more strongly linked; in our application it is examined through the asset-similarity channel, including the industry-similarity covariate, rather than inferred from topology alone. Since the proposed model separates the excitation network into layers, the same degree-based summaries can also be calculated by layer to assess how the application-specific contagion channels contribute to these conclusions.

To analyse node-level influence, we focus on degree connectedness and eigenvector centrality applied to the inferred weighted adjacency matrix. Following \citet{Diebold2014}, total connectedness maps to mean node degree. We interpret weighted out-degree as a firm's systemic contribution, analogous to CoVaR \citep{tobias2016covar}, and weighted in-degree as its vulnerability to systemic shocks, analogous to expected shortfall \citep{Acharya2017}.

We supplement direct degree analysis with weighted eigenvector centrality (ECW). As advocated by \citet{Bianchi2019}, ECW better captures influence in complex financial webs by reflecting both direct connectivity and the global recursive influence of a node's neighbours. This uniquely differentiates systemic hubs from highly active but locally isolated nodes, bringing a nuanced understanding to shock propagation. Full network property analyses are detailed in Appendix \ref{sec:empirical_results_app}.

\section{Simulation Study}\label{sec:simulationstudy}

In this section, we present simulated scenarios to introduce the model via a simple example (Section \ref{sec:illustrativeexample}) and explore estimation behaviour under varying conditions (Section \ref{sec:scenarioanalysis}). Each generated process is termed a ``scenario,'' while estimation variations are described as hyperparameter and layer settings. Common parameters across analyses are summarised in Appendix \ref{sec:illustrative_example_app}, Table \ref{tab:common_params}.

%\input{tab/simulation_study/common_params}

 %The differences between the generated scenarios are shown in Table \ref{tab:scenarios}. %The differences in estimation of these scenarios will be described in Section \ref{sec:scenarioanalysis} as required.

\subsection{Illustrative Example}\label{sec:illustrativeexample}

To demonstrate that the model successfully recovers the true generative structures, we first simulate an illustrative two-layer network (the ``small'' model) following the generative model in Appendix \ref{sec:generative_model}. The specification includes: $K=10$ nodes, sparsity $\rho=0.06$, fixed covariates $\bm{x}^{(l)}=(1,0.1,0.01)$ and equal linear predictor coefficients $\bm{\beta}^{(l)}=(0,1,1)$ across layers $l \in \{0, 1\}$, constant background rates $\lambda^{(0)}_k=0.2$, a two-week maximum excitation window $\Delta t_{max}=14/365$, and an observation period of $T=100$ years.

\begin{figure}[htbp!]
\centering
\includegraphics[width=0.5\textwidth, angle=0, trim= 150 25 150 50]{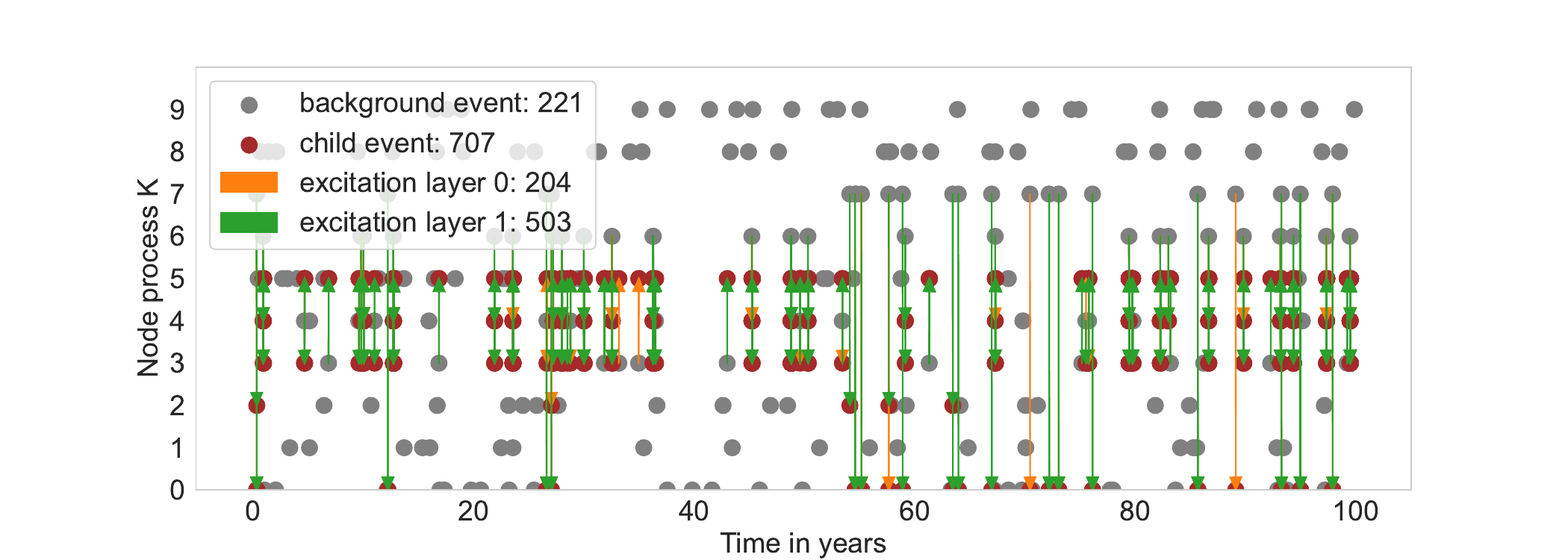}
\caption{Generated process. Events in time with corresponding parent-child relationship.}
\label{fig:process}
\end{figure}

This artificially sparse scenario forces events into short bursts relative to the long observation window ($\Delta t_{max} \ll T$). Fixing the covariates further isolates structural behaviors and highlights label-switching dynamics. The resulting adjacency matrix has $6$ edges, and the simulated process is visualised in Figure \ref{fig:process}.

\iffalse
\begin{figure}[htbp!]
\centering
\begin{subfigure}[b]{0.6\textwidth}
\includegraphics[width=1\textwidth, angle=0, trim= 150 25 150 50]{img/simulation_study/2l_sigma10/process.pdf}
\caption{Events in time with corresponding parent-child relationship.}
\label{fig:process_disabled}
\end{subfigure}
\centering
\begin{subfigure}[b]{0.6\textwidth}
\includegraphics[width=1\textwidth, angle=0, trim= 250 100 250 100]{img/simulation_study/2l_sigma10/gen_events.pdf}
\caption{Event count matrices: 928 total events, of which 204 are generated via first layer, 503 via second layer and 221 events are background events.}
\label{fig:count}
\end{subfigure}
\caption{Generated process}
\end{figure}
\fi

\begin{figure}[htbp!]
\centering
\begin{minipage}[b]{0.38\textwidth}
\centering
\includegraphics[width=1\textwidth, angle=0, trim= 40 30 25 28,clip]{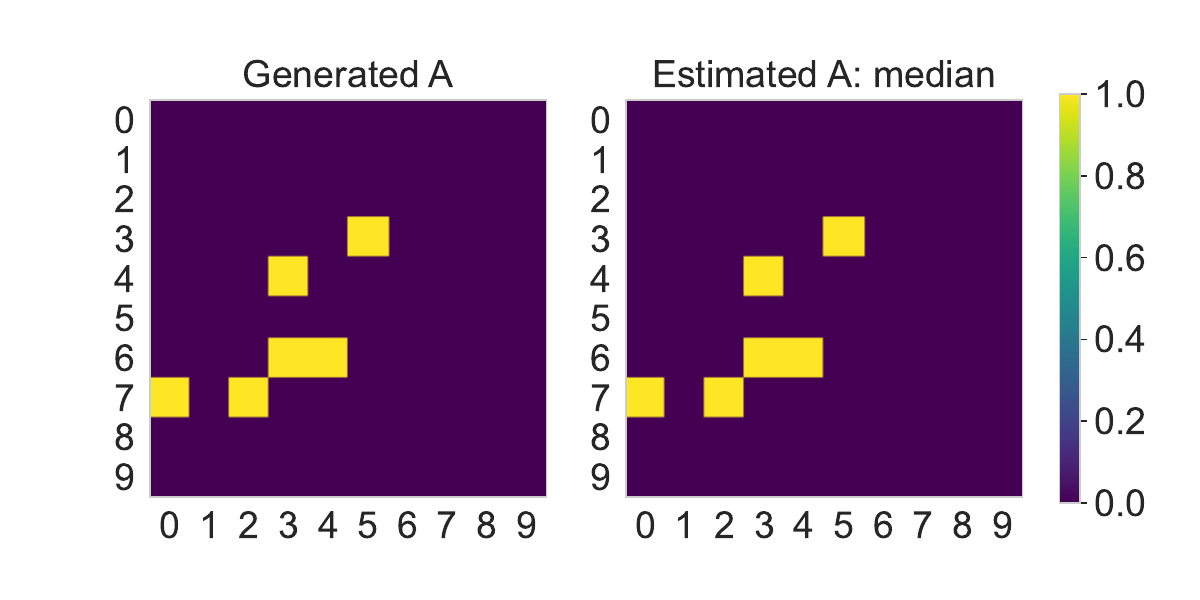}
\end{minipage}
\begin{minipage}[b]{0.38\textwidth}
\centering
\includegraphics[width=1\textwidth, angle=0, trim= 40 30 25 28,clip]{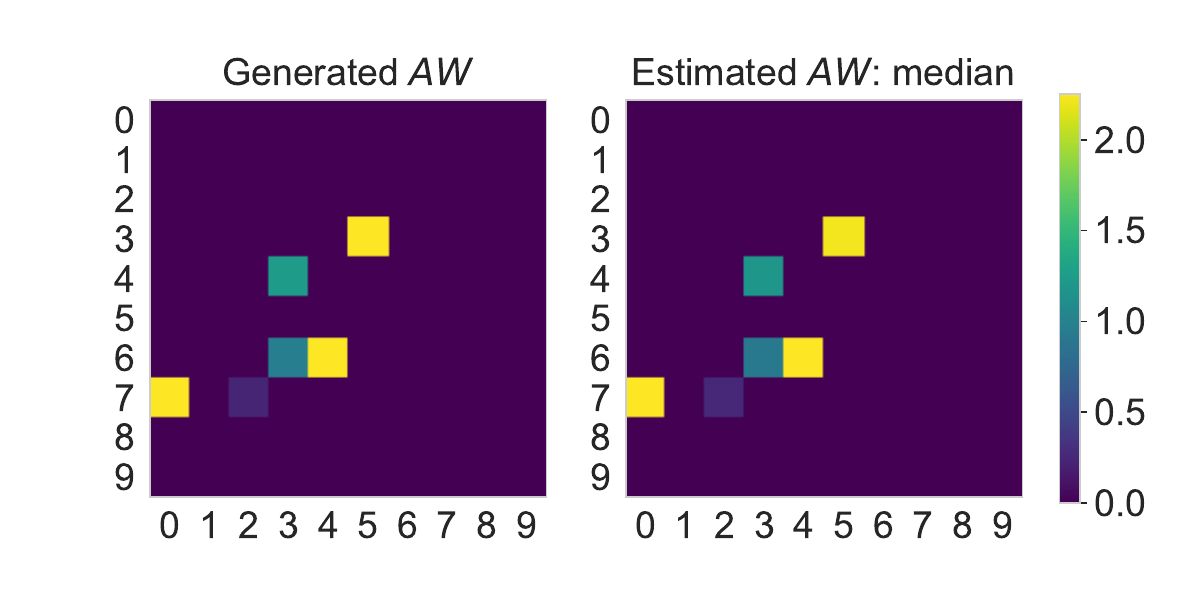}
\end{minipage}
\caption{Generated versus estimated (posterior median) components. Left: adjacency matrix (sample $\rho=0.06$, posterior median $\hat\rho=0.067$). Right: excitation network (sample $\bm{AW}=0.109$, posterior median $\bm{\hat{AW}}=0.107$).}
\label{fig:adjacency}
\label{fig:weights}
\end{figure}

In Figure \ref{fig:adjacency} we show the generated versus posterior medians of the estimated network components. The results indicate that both background intensity and excitation network ($\bm{AW}$) are recovered to high accuracy. The posterior medians of the average of the matrix $\hat{\mathbf{AW}}$ ($0.107$) and $\hat\rho$ ($0.066$) are nearly identical to the generated values (average of $\mathbf{AW}=0.109$ and $\rho=0.06$). Parameters related to the processes with the most events are very close to the ground truth, while those with small numbers of events are subject to sampling error, as illustrated by edge $A_{7,2}$ with only $5$ generated events and by the background rates $\lambda^{(0)}_k$. Posterior estimates and HDIs for background rates and weights are included in the Appendix \ref{sec:illustrative_example_app}.

%\input{tab/simulation_study/weights}

\iffalse
\begin{figure}[htbp!]
     \centering
     \begin{subfigure}{0.49\textwidth}
         \includegraphics[width=\textwidth, angle=0, trim= 75 25 75 0,clip]{img/simulation_study/2l_sigma10/w_ci.pdf}
     \end{subfigure}
  %   \hfill
     \begin{subfigure}{0.49\textwidth}
         \includegraphics[width=\textwidth, angle=0, trim= 75 25 75 0,clip]{img/simulation_study/2l_sigma10/par_ci.pdf}
     \end{subfigure}
\caption{95\% Credible Intervals for edge weights $W_{i,j}$, sparsity parameters $\rho$ and background rates $\lambda^0_{k}$.}
\label{fig:95ci_weights_lambda}
\end{figure}
\fi

% \begin{figure}[htbp!]
% \centering
% \includegraphics[width=0.5\textwidth, angle=0, trim= 50 0 0 0]{img/simulation_study/estimated_g.png}
% \caption{Generated vs estimated interaction kernel $g_{\theta_{k'\rightarrow k}}$ (posterior median) for elements where $A_{k'\rightarrow k}=1$.}
% \label{fig:interaction_kernel}
% \end{figure}

%Similarly, interaction kernel estimates are shown in Figure \ref{fig:interaction_kernel}. As expected, the estimated weight parameter and the interaction kernel are most accurate and have the narrowest HDIs for an edge $A_{3,5}$ with most of the generated events ($386$).

Table \ref{tab:mutau_gen} reports posterior medians and HDIs for the interaction-kernel parameters. Estimates of $\mu$ are close to the generated value on most active edges, especially edge $A_{3,5}$ with $386$ generated events, while estimates of $\tau$ are less precise. The resulting function estimate is more informative than either parameter alone: Figure \ref{fig:interaction_kernel_app} in Appendix \ref{sec:illustrative_example_app} shows that the generated kernel lies within the 95\% HDI for most lags on most active edges.

\begin{table}[htbp!]
\centering
\resizebox{0.61\textwidth}{!}{%
\begin{tabular}{cccccccc}
\hline
\textbf{$k'$} & \textbf{$k$} & \textbf{$\mu_{g}$} & \textbf{$\tau_{g}$} & \textbf{$\mu$} & \textbf{$\tau$} & \textbf{95\% HDI $\mu$} & \textbf{95\% HDI $\tau$} \\ \hline
3 & 5 & -1 & 10 & -1.05 & 5.87  & $[-1.10,-1.00]$ & $[4.77,6.98]$ \\
4 & 3 & -1 & 10 & -0.83 & 11.23 & $[-0.88,-0.77]$ & $[8.20,14.23]$ \\
6 & 3 & -1 & 10 & -1.06 & 8.06  & $[-1.19,-0.94]$ & $[4.63,11.85]$ \\
6 & 4 & -1 & 10 & -0.95 & 13.53 & $[-1.00,-0.90]$ & $[10.09,17.08]$ \\
7 & 0 & -1 & 10 & -0.90 & 8.49  & $[-0.98,-0.82]$ & $[5.97,11.27]$ \\
7 & 2 & -1 & 10 & -0.95 & 8.52  & $[-1.14,-0.78]$ & $[4.21,13.76]$ \\ \hline
\end{tabular}%
}
\caption{Posterior median estimates and HDIs for the interaction-kernel parameters $\mu_{k'\rightarrow k}$ and $\tau_{k'\rightarrow k}$ on active edges. Subscript $g$ denotes the generated value; edge $A_{3,5}$ has 386 generated events and edge $A_{7,2}$ has 5 generated events.}
\label{tab:mutau_gen}
\end{table}

\iffalse
\begin{figure}[htbp!]
    \centering
    % First part of the image
\resizebox{0.75\textwidth}{!}{%
    \begin{subfigure}{0.2\textwidth}
        \centering
        \includegraphics[width=\textwidth, angle=0, trim= 15 1077 30 205,clip]{img/simulation_study/2l_sigma10/estimated_g.pdf}
    \end{subfigure}\hfill
    % Second part of the image
    \begin{subfigure}{0.2\textwidth}
        \centering
        \includegraphics[width=\textwidth, angle=0, trim= 15 625 30 650,clip]{img/simulation_study/2l_sigma10/estimated_g.pdf}
  \end{subfigure}
    \begin{subfigure}{0.2\textwidth}
            
        \centering
        \includegraphics[width=\textwidth, angle=0, trim= 15 175 30 1105,clip]{img/simulation_study/2l_sigma10/estimated_g.pdf}
    \end{subfigure}
    }
    \caption{Generated vs estimated interaction kernel $g_{\theta_{k'\rightarrow k}}$ (posterior median) for elements where $A_{k'\rightarrow k}=1$.}
    \label{fig:interaction_kernel}
\end{figure}
\fi

\begin{figure}[htbp!]
     \centering
     \begin{subfigure}{0.35\textwidth}
         \includegraphics[width=\textwidth, angle=0, trim= 36 15 55 34,clip]{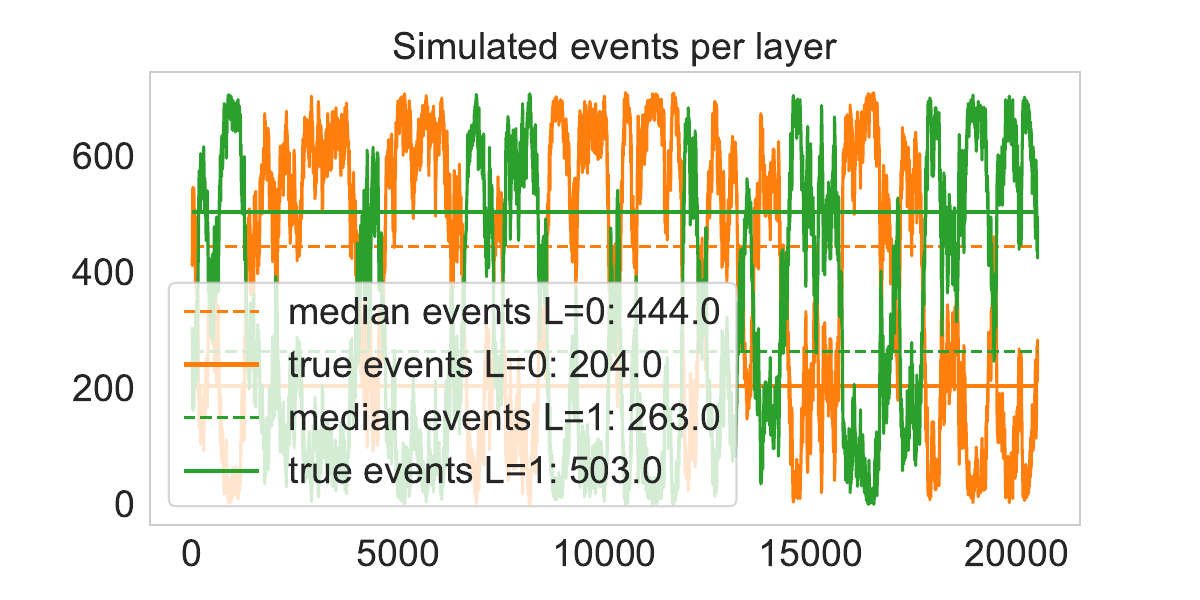}
     \end{subfigure}    
     \begin{subfigure}{0.35\textwidth}
         \includegraphics[width=\textwidth, angle=0, trim= 36 15 55 34,clip]{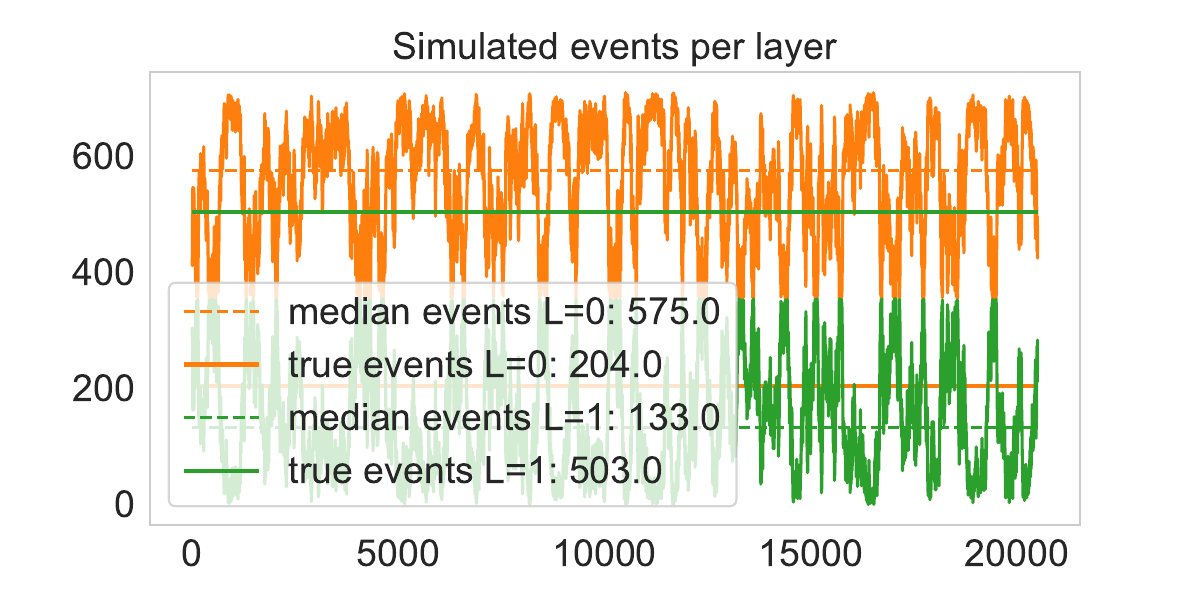}
     \end{subfigure}
\caption{Sampler output for the number of allocated events per layer before and after ECR relabelling (iterative version 2; background allocations excluded).}
\label{fig:ecr_samples}
\end{figure}

Although the aggregate excitation network is recovered accurately, the illustrative scenario has identical covariate specifications in its two layers and therefore admits label switching. Figure \ref{fig:ecr_samples} displays sampled layer event allocations before and after ECR relabelling. ECR is able to separate two layer contributions with different posterior event allocations; the numerical layer labels themselves have no interpretation in this symmetric scenario. Appendix Figure \ref{fig:weights_layer_ecr} shows the corresponding relabelled layer-level posterior medians. Section \ref{sec:scenarioanalysis} next studies estimation when covariates and coefficients differ across layers and under alternative hyperparameter and layer settings.

\subsection{Analysis of Simulated Scenarios}\label{sec:scenarioanalysis}

We simulate a number of event processes and estimate them under different hyperparameter and layer settings to explore accuracy and stability. The study varies network size, event counts, covariate effects, the number of fitted layers, $\Delta t_{max}$, label switching and hyperparameter choices. Table \ref{tab:scenarios} describes each generated scenario.

\iffalse
\begin{itemize}
\item the size of the network,
\item number of events per unit of time,
\item covariate influence,
\item impact due to misspecification of layers or $\Delta t_{max}$ parameters,
\item label switching,
\item sensitivity to prior specification of hyper parameters.

\fi
\begin{table}[htbp!]
\centering
% Slightly larger font and more padding for readability
\resizebox{\textwidth}{!}{%
{\Large\setlength{\tabcolsep}{7pt}\renewcommand{\arraystretch}{1.12}
\begin{tabular}{cccccccccccccc}
\hline
\textbf{Scen.}&\textbf{$K$} & \textbf{$L$} & \textbf{$\rho$} & \textbf{$\Delta t_{\text{max}}$} & \textbf{$T$} & \textbf{$\beta_0$} & \textbf{$\beta$} & \textbf{$\lambda^{(0)}_k$} & \textbf{$N$} & \textbf{$N_{\text{backg}}$} & \multicolumn{2}{c}{\textbf{$N_{\text{layer}}$}} & \textbf{$\bm{x}$} \\
\hline
small & 10 & 2 & 0.06 & 0.04 & 100 & $[0\ 0]$ & $\left[\begin{array}{cc} 1 & 1 \\ 1 & 1 \end{array}\right]$ & 0.2 & 928 & 221 & 204 & 503 & $\bm{x}^{(0)}=\bm{x}^{(1)}=(1,0.1,0.01)$ \\
small-beta & 10 & 2 & 0.06 & 0.04 & 100 & $[-1\ -1]$ & $\left[\begin{array}{cc} 2.0 & 1.0 \\ 0.2 & 0.1 \end{array}\right]$ & 0.2 & 855 & 181 & 263 & 411 & $\bm{x}^{(0)}= (1, x_1, x_2), ~ \forall x_1, x_2 \in [0, 2] \cap \mathbb{Z}$ \\ & & & & & & & & & & & & & $\bm{x}^{(1)}= (1, x_1, x_2), ~ \forall x_1, x_2 \in [1, 10] \cap \mathbb{Z}$ \\
medium & 20 & 2 & 0.1 & 10.5 & 20000 & $[-2\ -2]$ & $\left[\begin{array}{cc} 0.2 & -0.2 \\ 0.2 & -0.2 \end{array}\right]$ & 0.005 & 4590 & 1980 & 1143 & 1467 &$\bm{x}^{(0)}= (1, x_1, x_2), ~ \forall x_1, x_2 \in [0, 2] \cap \mathbb{Z}$ \\ & & & & & & & & & & & & & $\bm{x}^{(1)}= (1, x_1, x_2), ~ \forall x_1, x_2 \in [1, 10] \cap \mathbb{Z}$ \\
medium-beta & 20 & 2 & 0.1 & 10.5 & 10000 & $[-2\ 1]$ & $\left[\begin{array}{cc} 0.5 & 0.0 \\ 0.0 & -1.0 \end{array}\right]$ & 0.01 & 5182 & 1992 & 1969 & 1221 & $\bm{x}^{(0)}= (1, x_1, x_2), ~ \forall x_1, x_2 \in [0, 2] \cap \mathbb{Z}$ \\ & &  & & & & & & & & & & & $\bm{x}^{(1)}= (1, x_1, x_2), ~ \forall x_1, x_2 \in [1, 10] \cap \mathbb{Z}$  \\
large-beta & 30 & 2 & 0.085 & 10.5 & 7500 & $[-3.05\ -1.70]$ & $\left[\begin{array}{cc} 0.05 & 0.0 \\ 0.0 & -0.035 \end{array}\right]$ & 0.0058 & 4710 & 1264 & 879 & 2567 & $\bm{x}^{(0)}= (1, x_1, x_2), ~ \forall x_1, x_2 \in [0, 2] \cap \mathbb{Z}$ \\ & & & & & & & & && & & & $\bm{x}^{(1)}= (1, x_1, x_2), ~ \forall x_1, x_2 \in [1, 10] \cap \mathbb{Z}$  \\
\hline
\end{tabular}%
}% end large/resizebox group
}
\caption{Generated scenarios.}
\label{tab:scenarios}
\end{table}

The ``small'' scenario is the illustrative two-layer example from Section \ref{sec:illustrativeexample} with identical fixed covariates and coefficients across layers. The ``small-beta'' scenario uses different coefficients for each layer and layer-specific covariates, with covariates taking values in $\{0,1,2\}$ in layer 0 and $\{1,\ldots,10\}$ in layer 1. The ``medium'' and ``medium-beta'' scenarios increase the network size to $K=20$ and generate approximately 5,000 events, with only the latter specifying distinct coefficients across layers. The ``large-beta'' scenario uses $K=30$ nodes, $\rho=0.085$, distinct layer coefficients and a larger set of active edges than the earlier sparse 30-node design. It is deliberately closer to the stability boundary, with a combined spectral radius of $0.92$ for the integrated excitation matrix $G=\sum_l A\odot W^{(l)}$ (Appendix \ref{sec:stability}), but remains non-explosive in the generated sample.

Common estimation parameters and the fitted hyperparameter and layer settings are summarised in Appendix \ref{sec:illustrative_example_app}, Tables \ref{tab:common_params} and \ref{tab:scenarios_estimation}. In brief, the ``base'' setting uses two layers and the reference priors; ``invGamma'', ``sigma100'' and ``kappahypers100'' alter prior specifications; ``estKappaHyper'' and ``estKappa'' fix selected hyperparameters; the ``dt'' suffixes alter $\Delta t_{max}$; and ``3layer'' deliberately fits one additional layer.

\paragraph{Observation 1. Hawkes process estimation quality}
Graphical residual summaries indicate that the fitted intensities describe all generated processes reasonably well. Appendix \ref{sec:scenario_analysis_app} presents the ``medium'' scenario as an example and describes how these summaries are generated and interpreted.

% See for e.g. the compensators for processes in scenario 8 in Figure \ref{fig:comp_multi_8} and corresponding step functions for processes 0 to 8 (all the remaining ones looks similar thus ommitted for space reasons) Figures \ref{fig:step1_8}, \ref{fig:step2_8} and \ref{fig:step3_8}.

\iffalse
\begin{figure}[htbp!]
\centering
\includegraphics[width=1\textwidth, angle=0, trim= 110 100 120 100,clip]{img/simulation_study/8l_base/comp_multi.pdf}
\caption{Scenario ``medium''. Rate at event time and compensator values at event times for processes $K$.}
\label{fig:comp_multi_8}
\end{figure}
\fi

\paragraph{Observation 2. Adjacency matrix}
The adjacency matrix and the sparsity parameter $\rho$ are typically estimated with high precision (Table \ref{tab:rho}). Accuracy of the adjacency matrix increases when the network has more non-zero edges, that is, a higher $\rho$, and when more processes or nodes are observed while keeping the sparsity level roughly similar. In the five scenarios considered here, the posterior medians for $\rho$ are close to their generated values. HDIs are wider in the two small scenarios and narrower in the medium and large scenarios, reflecting the greater information available about network sparsity. The ``large-beta'' scenario illustrates this effect: with $88$ generated edges, the posterior median for $\rho$ is $0.094$ against a generated value of $0.085$.

\begin{table}[htbp!]
\centering
\resizebox{0.85\textwidth}{!}{%
\begin{tabular}{ccccccccccc}
\hline
\textbf{Scenario} & \textbf{K} & \textbf{T} & \(\Delta t_{\text{max}}\) & \(\lambda_0\) & \textbf{N} & \textbf{N backg} & \textbf{N layer} & \(\rho_{\text{gen}}\) & \(\rho\) & \textbf{95\% HDI}  \\ \hline
small & 10 & 100 & 0.04 & 0.2 & 928 & 221 & [204 503] & 0.06 & 0.067 & [0.03,0.12]  \\
small-beta & 10 & 100 & 0.04 & 0.2 & 855 & 181 & [263 411] & 0.06 & 0.066 & [0.02,0.12]  \\
medium & 20 & 20000 & 10.5 & 0.005 & 4590 & 1980 & [1143 1467] & 0.1 & 0.097 & [0.07,0.13]  \\
medium-beta & 20 & 10000 & 10.5 & 0.01 & 5182 & 1992 & [1969 1221] & 0.1 & 0.087 & [0.06,0.12]  \\
large-beta & 30 & 7500 & 10.5 & 0.0058 & 4710 & 1264 & [879 2567] & 0.085 & 0.094 & [0.075,0.115]  \\ \hline
\end{tabular}%
}
\caption{Adjacency matrix estimation accuracy with decreasing sparsity/increasing number of processes. Posterior median estimates are presented as $\rho$.}
\label{tab:rho}
\end{table}

\iffalse
\begin{figure}[htbp!]
\centering
\includegraphics[width=0.9\textwidth, angle=0, trim= 250 150 250 150]{img/simulation_study/8l_base/estimated_network_layer.pdf}
\caption{Scenario 8. Generated vs estimated network per layer (posterior median).}
\label{fig:weights_layer8}
\end{figure}
\fi

\begin{figure}[htbp!]
\centering
\includegraphics[width=0.50\textwidth, angle=0, trim= 5 5 5 45, clip]{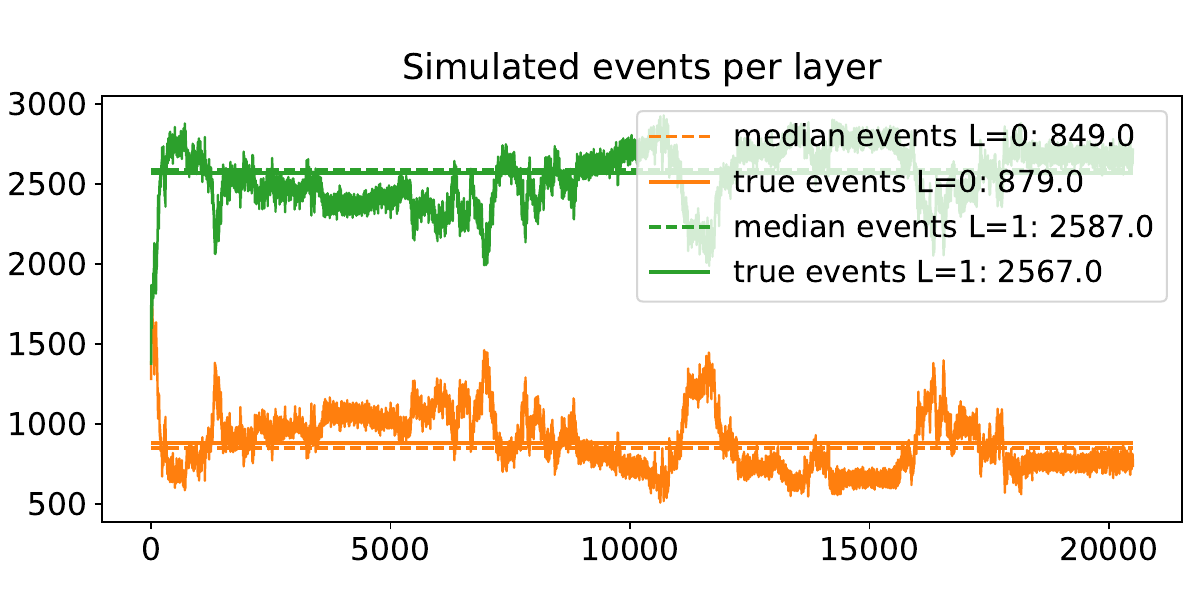}
\caption{Scenario ``large-beta''. Allocated events per layer without ECR relabelling. After burn-in, the sampler traces remain separated around the generated layer event counts.}
\label{fig:labelswitching11}
\end{figure}

\paragraph{Observation 3. Layer-Level Weights}
Layer-level weight recovery depends on whether layers are distinguishable in the observed data. Label switching occurs in the ``small'' scenario because its layer covariates and coefficients are identical, and label-switching behaviour is also more evident in the smaller fitted cases where layer-specific information is limited. In contrast, Figure \ref{fig:labelswitching11} shows the sampler output for ``large-beta'' without ECR relabelling: after burn-in, the allocated-event traces for the two layers remain separated and their posterior medians are close to the generated layer totals.

\paragraph{Observation 4. Covariate Coefficients and Prior Sensitivity}
The scenarios with a ``beta'' suffix test recovery when layer coefficients differ. Supporting tables are reported in Appendix \ref{sec:scenario_analysis_app}.

Coefficient estimation depends mainly on the amount of active edge-level information, since active edges provide the effective sample for covariate estimation. Coefficients are therefore harder to estimate in small or sparse scenarios, especially when layer-specific signal is weak or layers are not clearly separated. In the largest scenario, with $88$ generated edges and $4710$ events, the coefficients are estimated more accurately than in the smaller sparse scenarios (Appendix Table \ref{tab:beta_11}).

Prior sensitivity is most visible in the lower-information settings. The inverse-Gamma prior and fixed-$\kappa$ variants can move coefficient estimates further from the generated values, while wider Gaussian priors can shift intercept estimates in some scenarios. This does not make coefficient estimation impossible, but it indicates that coefficient interpretation should be framed in terms of stronger or weaker evidence across network size, sparsity and prior choices.

\paragraph{Observation 5. Layer misspecification}
To study robustness to misspecification of the number of layers, we estimate scenarios ``small-beta'' and ``medium'' with 3 layers as opposed to 2 used in scenario generation. Third-layer covariates are populated from a standard normal distribution.

\iffalse
\begin{figure}[htbp!]
\centering
\includegraphics[width=0.9\textwidth, angle=0, trim= 250 150 250 150]{img/simulation_study/8l_layer/estimated_network_layer.pdf}
\caption{Scenario ``medium''. Generated vs estimated network per layer (posterior median).}
\label{fig:weights_layer8_3l}
\end{figure}

\begin{figure}[htbp!]
\centering
\includegraphics[width=0.9\textwidth, angle=0, trim= 250 150 250 150]{img/simulation_study/4l_layer/estimated_network_layer.pdf}
\caption{Scenario ``small-beta''. Generated vs estimated network per layer (posterior median).}
\label{fig:weights_layer4_3l}
\end{figure}
\fi

We fit the two-layer ``small-beta'' and ``medium'' generated processes with three layers to examine whether an unnecessary layer can be separated from supported layers. In the ``small-beta'' setting, label ambiguity materially affects fitted layer weights: iterative version 1 of ECR\footnote{ECR iterative version 1 is a non-probability-based post-processing algorithm that updates the pivot allocation using relabelled allocations from preceding iterations. Further details of the algorithms are provided in Appendix \ref{sec:labelswitching}.} reduces total layer-weight RMSE from $4.64$ without relabelling to $2.21$ (Appendix Figure \ref{fig:ecr_samples_4_layer}). Following this relabelling, the additional fitted layer has aggregate posterior-median weight $9$, below the dominant fitted layer with weight $17$, although it is not below the weaker fitted layer with weight $1$. In ``medium'', the additional fitted layer has weight $2$, compared with $8$ for the dominant fitted layer and $2$ for the weaker fitted layer; ECR variants also produce similar total RMSE, indicating little practical label-switching effect. Thus, in these displayed cases, an additional fitted layer is distinguishable from the dominant layer structure, but is not necessarily cleanly separated from every supported layer. Table \ref{tab:layer_misp_ecr} reports the ECR post-processing comparison supporting this interpretation.
%This can be seen in Figure \ref{fig:weights_layer4_3l_ecr} where the estimated $\bm{AW}$ network for layer 1 is near zero.

\begin{table}[htbp!]
\centering
\resizebox{1.\textwidth}{!}{%
\begin{tabular}{llllllllllllll}
\hline
\textbf{Scenario} & \textbf{Setting} & \textbf{ECR} & \textbf{$\sum AW_{gen}$} & \textbf{$\sum AW^{(0)}_{gen}$} & \textbf{$\sum AW^{(1)}_{gen}$}  & \textbf{ $\sum AW$} & \textbf{$\sum AW^{(0)}$} & \textbf{$\sum AW^{(1)}$}  & \textbf{$\sum AW^{(2)}$}& \textbf{$RMSE^{(0)}$} & \textbf{$RMSE^{(1)}$} & \textbf{$RMSE^{(2)}$} & \textbf{$RMSE^{total}$} \\ \hline
small-beta                 & 3layer        & n/a                 & 32                     & 8                       & 24                                             & 24                 & 2                   & 11                  & 10                  & 0.39              & 2.92              & 1.34              & 4.64                \\
small-beta                & 3layer        & iterative1          & 32                     & 8                       & 24                                             & 27                 & 1                   & 17                  & 9                   & 0.19              & 1.29              & 0.73              & 2.21                \\
small-beta                 & 3layer        & iterative2          & 32                     & 8                       & 24                                            & 25                 & 7                   & 1                   & 16                  & 1.81              & 0.19              & 2.68              & 4.68                \\
medium                 & 3layer        & n/a                 & 14                     & 5                       & 9                                             & 12                 & 2                   & 8                   & 2                   & 0.28              & 0.20              & 0.11              & 0.59                \\
medium                & 3layer        & iterative1          & 14                     & 5                       & 9                                              & 12                 & 2                   & 8                   & 2                   & 0.28              & 0.21              & 0.08              & 0.57                \\
medium                 & 3layer        & iterative2          & 14                     & 5                       & 9                                             & 12                 & 8                   & 1                   & 2                   & 0.21              & 0.08              & 0.27              & 0.56                \\ \hline
\end{tabular}%
}
\caption{Effect of ECR relabelling for ``small-beta'' and ``medium'' processes fitted with three layers. The reported sums use posterior medians; the generated third-layer sum is zero and is omitted.}
\label{tab:layer_misp_ecr}
\end{table}

\iffalse
\begin{figure}[htbp!]
\centering
\includegraphics[width=0.9\textwidth, angle=0, trim= 250 150 250 150]{img/simulation_study/4l_layer/estimated_network_layer_ecr_iterative1_.pdf}
\caption{Scenario ``small-beta''. Generated vs estimated network per layer (posterior median) post adoption of ECR algorithm (iterative 1 ECR algorithm)}
\label{fig:weights_layer4_3l_ecr}
\end{figure}
\fi

\begin{figure}[htbp!]
    \centering
        % First part of the image
    \begin{subfigure}{0.36\textwidth}
        \centering
        \includegraphics[width=\textwidth, angle=0, trim= 12 2 40 25,clip]{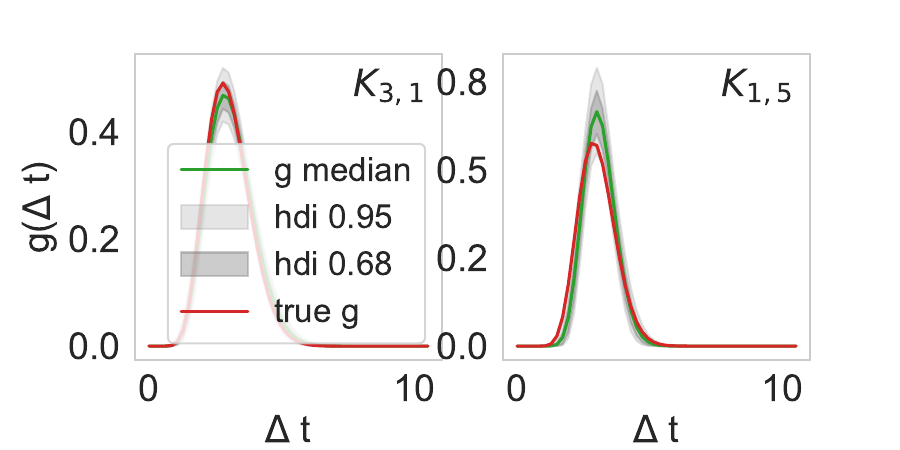}
        \caption{$\Delta t_{max}=10.5$}
    \end{subfigure}%
    % First part of the image
    \begin{subfigure}{0.345\textwidth}
        \centering
       \includegraphics[width=\textwidth, angle=0, trim= 28 4 40 25,clip]{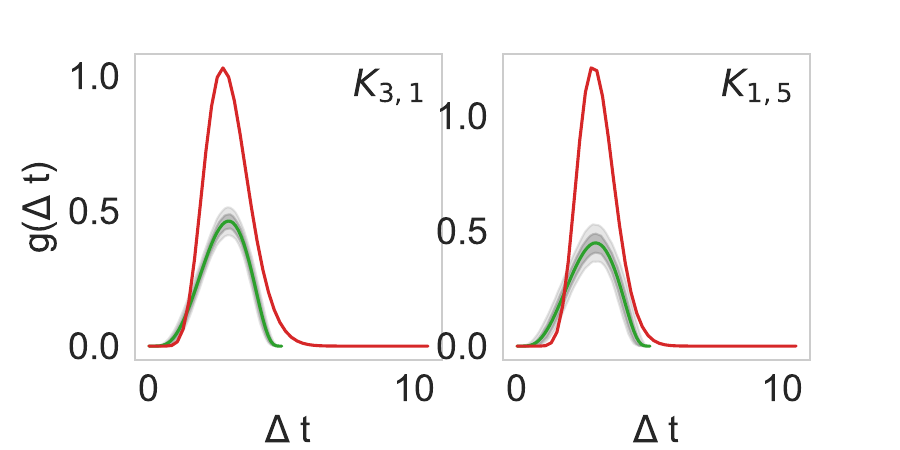}
     \caption{$\Delta t_{max}=5.5$}
    \end{subfigure}%
    % Second part of the image
    \begin{subfigure}{0.345\textwidth}
        \centering
          \includegraphics[width=\textwidth, angle=0, trim= 28 2 40 25,clip]{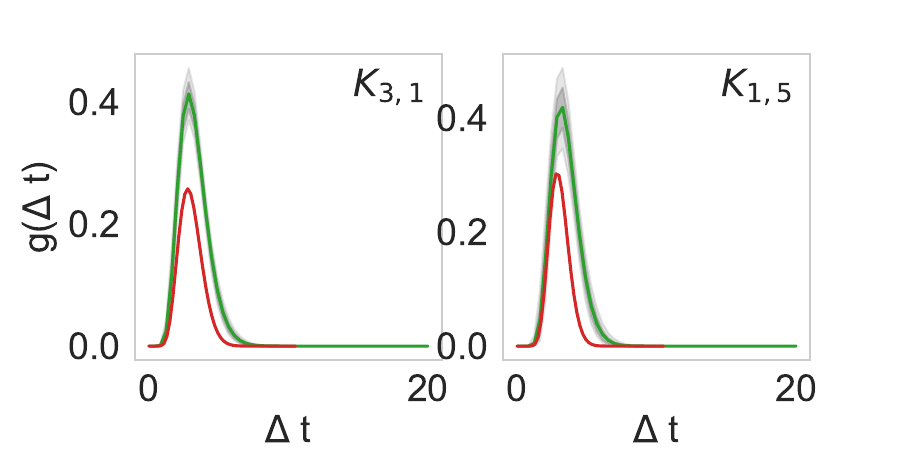}
        \caption{$\Delta t_{max}=20.5$}
    \end{subfigure}
    \caption{Scenario ``medium''. Generated vs estimated interaction kernel $g_{\theta_{k'\rightarrow k}}$ (posterior median) under $\Delta t_{max}$ misspecification.}
    \label{fig:interaction_kernel_misspecification}
\end{figure}

\paragraph{Observation 6. Misspecification of $\Delta t_{max}$}
We fit the ``medium'' process using lower and higher values of $\Delta t_{max}$. Figure \ref{fig:interaction_kernel_misspecification} shows that this choice most directly changes the inferred interaction kernel: an interval that is too short forces the estimated kernel to zero before the generated support ends, while a longer interval permits excitation at longer delays. This supports treating $\Delta t_{max}$ as a sensitivity choice in the empirical application; these simulations do not establish that the largest horizon is uniformly preferable.

\paragraph{Observation 7. Sensitivity to parameters $\alpha_{\rho}$ and $\beta_{\rho}$}
In settings with an ``ab10'' suffix, $\alpha_{\rho}$ and $\beta_{\rho}$ are set to $10$ rather than $1$. The principal visible difference is in posterior uncertainty for $\rho$ (Appendix \ref{sec:scenario_analysis_app}, Table \ref{tab:rho_accuracy_8}); the corresponding aggregate $\bm{AW}$ comparison is reported in Table \ref{tab:aw_accuracy_8}.

\paragraph{Observation 8. $\kappa$ estimation stability dependency on priors}
We compare alternative $\kappa$ prior settings for the ``medium'' scenario in Appendix \ref{sec:scenario_analysis_app}. Coefficient and $\kappa$ inference can be prior-sensitive when layer-specific information is weak. This is expected because layer-specific regression coefficients depend on how clearly the data separate the channel contributions. The empirical analysis therefore interprets covariate effects cautiously, placing emphasis on channel-level summaries, network structure, and the strongest and most consistent coefficient signals across runs under different prior specifications.

\FloatBarrier

\section{Empirical Application}\label{sec:empirical}

For our empirical application we infer the network through which extreme credit-spread jumps transmit systemic risk and investigate the drivers behind this transmission using the data described in Section \ref{sec:empiricaldataset}.

\subsection{Results}
\label{sec:empiricalresults}

Based on the findings in the analysis of simulated scenarios in Section \ref{sec:scenarioanalysis}, we evaluated several model set-ups in an application to modelling systemic risk. The common/reference parameters used across all model set-ups are provided in Appendix \ref{sec:empirical_results_app}, Table \ref{tab:common_params_rd}.

In particular, the following structural variations of the models are evaluated: (i) varying numbers of model layers (denoted as ``1l'', ``2l'' and ``3l''); and (ii) different combinations of node-specific covariates, spanning from base models without covariates (``base'', where covariates are set to zero) to permutations incorporating asset similarity, solvency and profitability metrics.

To provide a focused analysis, we report on selected covariate-regression model set-ups within each architectural class, guided by WAIC \citep{watanabe2010asymptotic} computed using quarterly blocks of the observed event data\footnote{Quarterly blocks are used as the pointwise units so that extreme credit-spread jumps occurring close together during market-stress periods are not treated as independent prediction units.} (see Appendix \ref{sec:empirical_results_app}, Table \ref{tab:waic_all_scenarios_app}). Cross-layer WAIC differences are small relative to their standard errors, so we do not interpret them as evidence that one architecture is predictively dominant. These configurations are the single-layer model with asset similarity (``1l asset''), the two-layer model incorporating asset similarity and solvency (``2l asset solv''), and the three-layer model using all three covariates (``3l all''). The base set-up for each architecture is retained as a benchmark to assess the incremental value of adding covariates. We also retain two alternatives testing different maximum influence horizons ($\Delta t_{max}=5$ and $20$ days) for the three-layer set-up as key robustness checks (``3l all 5dt'' and ``3l all 20dt''). Results for the remaining fitted configurations are available in Appendix \ref{sec:empirical_results_app}. We first interpret the inferred networks, before considering channel covariates and model suitability.

\subsubsection{Estimated Network and Network Measures}

The inferred networks indicate sparse and predominantly directional transmission of extreme credit-spread jumps. We use the posterior of $\rho$ as the model-based summary of network sparsity. Low reciprocity indicates that estimated excitation generally flows from transmitting institutions to recipients rather than through mutual feedback links. Appendix Tables \ref{tab:scenarios_reciprocity_rd_app}--\ref{tab:scenarios_acw_layer_rd_app} report reciprocity, degree correlation, transitivity and clustering summaries. In particular, the low weighted clustering estimates provide little support for strongly clustered or small-world-like transmission in the fitted network. Layer-level weighted average clustering shows only minor differences across covariate set-ups, with clustering generally lower than in the corresponding base benchmarks (Appendix Figure \ref{tab:scenarios_acw_layer_rd_app}).

To check the degree-distribution behaviour of the inferred networks, we fit kernel density estimate (KDE), log-normal, exponential and power-law distributions to the degree data and compare them using log-likelihood ratios (LRs) and Kolmogorov--Smirnov (KS) divergence tests. The results are summarised in Appendix Table \ref{tab:degree_tests_app}. For weighted degree, the LR comparisons favour a power-law fit over the KDE, log-normal and exponential alternatives in every reported model set-up. This suggests scale-free properties in estimated transmission strength: outward transmission is concentrated among a small number of hub-like institutions, making weighted metrics useful for identifying potential systemic hubs.

Node-level differences are most distinctive between the single-layer base benchmark and the ``3l all'' set-up. We therefore use ``3l all'', which represents all three application channels, for the node-level importance analysis. Table \ref{tab:topnodes} reports the top tickers for each metric, Table \ref{tab:topnodes_names} gives the matching node numbers, ticker names and frequencies, and Appendix Section \ref{sec:empirical_app_network_measures}, Table \ref{tab:topnodes_list}, gives full firm and covariate details. We find that: (i) metrics frequently identify the same institutions (only 29 unique nodes reach any top-10 list, and 13 reach any top-5 list); (ii) weighting alters rankings and marginally shifts selected nodes, except for in-degree; and (iii) layer-specific rankings mostly reshuffle the full weighted lists. The most frequently flagged institutions are Citigroup (19 top-10 lists), Lincoln National (15), Intesa Sanpaolo (14), and Allianz, American International Group, Marriott and HeidelbergCement (11 each).

\begin{table}[htbp!]
\centering
\resizebox{1\textwidth}{!}{%
{\large\setlength{\tabcolsep}{6pt}\renewcommand{\arraystretch}{1.06}
\begin{tabular}{llllllllllllllllllll}
\hline
\textbf{Rank} & \textbf{BC}                & \textbf{CC}                & \textbf{EC}                & \textbf{ECW}               & \textbf{D}                 & \textbf{DW}                & \textbf{DW L0}             & \textbf{DW L1}             & \textbf{DW L2}             & \textbf{ID}                & \textbf{IDW}               & \textbf{IDW L0}            & \textbf{IDW L1}            & \textbf{IDW L2}            & \textbf{OD}                & \textbf{ODW}               & \textbf{ODW L1}            & \textbf{ODW L2}            & \textbf{ODW L3}            \\ \hline
1                       & \cellcolor[HTML]{63BE7B}TSN & \cellcolor[HTML]{FBA275}CB & \cellcolor[HTML]{FFEB84}ISP & \cellcolor[HTML]{FFEB84}ISP & \cellcolor[HTML]{F86F6C}AIG  & \cellcolor[HTML]{F86F6C}AIG  & \cellcolor[HTML]{F8706C}AIG  & \cellcolor[HTML]{F8706C}AIG  & \cellcolor[HTML]{F8706C}AIG  & \cellcolor[HTML]{FFEB84}ISP & \cellcolor[HTML]{FFEB84}ISP & \cellcolor[HTML]{F7E984}ISP & \cellcolor[HTML]{F7E984}ISP & \cellcolor[HTML]{FBA476}CB & \cellcolor[HTML]{F86F6C}AIG  & \cellcolor[HTML]{F86F6C}AIG  & \cellcolor[HTML]{F8706C}AIG  & \cellcolor[HTML]{F8706C}AIG  & \cellcolor[HTML]{F8706C}AIG  \\
2                       & \cellcolor[HTML]{D2DE82}MAR & \cellcolor[HTML]{FCBD7B}DHR & \cellcolor[HTML]{78C47D}SPG & \cellcolor[HTML]{FBA275}CB & \cellcolor[HTML]{63BE7B}TSN & \cellcolor[HTML]{63BE7B}TSN & \cellcolor[HTML]{71C27C}TSN & \cellcolor[HTML]{71C27C}TSN & \cellcolor[HTML]{71C27C}TSN & \cellcolor[HTML]{FBA275}CB & \cellcolor[HTML]{FBA275}CB & \cellcolor[HTML]{82C77D}SPG & \cellcolor[HTML]{FBA476}CB & \cellcolor[HTML]{FA9D75}C & \cellcolor[HTML]{63BE7B}TSN & \cellcolor[HTML]{63BE7B}TSN & \cellcolor[HTML]{71C27C}TSN & \cellcolor[HTML]{71C27C}TSN & \cellcolor[HTML]{71C27C}TSN \\
3                       & \cellcolor[HTML]{F8766D}ALV  & \cellcolor[HTML]{FFEB84}ISP & \cellcolor[HTML]{FBA275}CB & \cellcolor[HTML]{FA9B74}C & \cellcolor[HTML]{FA9B74}C & \cellcolor[HTML]{D2DE82}MAR & \cellcolor[HTML]{FA9D75}C & \cellcolor[HTML]{D0DE82}MAR & \cellcolor[HTML]{D0DE82}MAR & \cellcolor[HTML]{FA9B74}C & \cellcolor[HTML]{FA9B74}C & \cellcolor[HTML]{FA9D75}C & \cellcolor[HTML]{FA9D75}C & \cellcolor[HTML]{82C77D}SPG & \cellcolor[HTML]{FA9B74}C & \cellcolor[HTML]{D2DE82}MAR & \cellcolor[HTML]{D0DE82}MAR & \cellcolor[HTML]{D0DE82}MAR & \cellcolor[HTML]{D0DE82}MAR \\
4                       & \cellcolor[HTML]{FCBD7B}DHR & \cellcolor[HTML]{FDCD7E}GE & \cellcolor[HTML]{FA9B74}C & \cellcolor[HTML]{78C47D}SPG & \cellcolor[HTML]{D2DE82}MAR & \cellcolor[HTML]{FA9B74}C & \cellcolor[HTML]{D0DE82}MAR & \cellcolor[HTML]{FA9D75}C & \cellcolor[HTML]{FA9D75}C & \cellcolor[HTML]{78C47D}SPG & \cellcolor[HTML]{78C47D}SPG & \cellcolor[HTML]{FBA476}CB & \cellcolor[HTML]{E5E483}L & \cellcolor[HTML]{E5E483}L & \cellcolor[HTML]{D2DE82}MAR & \cellcolor[HTML]{FA9B74}C & \cellcolor[HTML]{FA9D75}C & \cellcolor[HTML]{FEDA80}HEI & \cellcolor[HTML]{FA9D75}C \\
5                       & \cellcolor[HTML]{E1E383}LNC & \cellcolor[HTML]{78C47D}SPG & \cellcolor[HTML]{FDCD7E}GE & \cellcolor[HTML]{EBE683}L & \cellcolor[HTML]{FDD680}HEI & \cellcolor[HTML]{FDD680}HEI & \cellcolor[HTML]{FEDA80}HEI & \cellcolor[HTML]{FEDA80}HEI & \cellcolor[HTML]{FEDA80}HEI & \cellcolor[HTML]{EBE683}L & \cellcolor[HTML]{EBE683}L & \cellcolor[HTML]{FDD17F}GE & \cellcolor[HTML]{82C77D}SPG & \cellcolor[HTML]{FDD17F}GE & \cellcolor[HTML]{FDD680}HEI & \cellcolor[HTML]{FDD680}HEI & \cellcolor[HTML]{FEDA80}HEI & \cellcolor[HTML]{FA9D75}C & \cellcolor[HTML]{FEDA80}HEI \\
6                       & \cellcolor[HTML]{F86F6C}AIG  & \cellcolor[HTML]{F87B6E}AV  & \cellcolor[HTML]{EBE683}L & \cellcolor[HTML]{FDCD7E}GE & \cellcolor[HTML]{F8766D}ALV  & \cellcolor[HTML]{E1E383}LNC & \cellcolor[HTML]{DDE182}LNC & \cellcolor[HTML]{F7E984}ISP & \cellcolor[HTML]{F7E984}ISP & \cellcolor[HTML]{FDCD7E}GE & \cellcolor[HTML]{FDCD7E}GE & \cellcolor[HTML]{DDE182}LNC & \cellcolor[HTML]{DDE182}LNC & \cellcolor[HTML]{F7E984}ISP & \cellcolor[HTML]{F8766D}ALV  & \cellcolor[HTML]{FAEA84}JPM & \cellcolor[HTML]{F3E884}JPM & \cellcolor[HTML]{F3E884}JPM & \cellcolor[HTML]{F8696B}AAL  \\
7                       & \cellcolor[HTML]{FA9B74}C & \cellcolor[HTML]{FA9F75}CAT & \cellcolor[HTML]{B9D780}MUV & \cellcolor[HTML]{E1E383}LNC & \cellcolor[HTML]{FFEB84}ISP & \cellcolor[HTML]{F8766D}ALV  & \cellcolor[HTML]{F8776D}ALV  & \cellcolor[HTML]{DDE182}LNC & \cellcolor[HTML]{F8776D}ALV  & \cellcolor[HTML]{E1E383}LNC & \cellcolor[HTML]{E1E383}LNC & \cellcolor[HTML]{E5E483}L & \cellcolor[HTML]{BAD780}MUV & \cellcolor[HTML]{BAD780}MUV & \cellcolor[HTML]{FAEA84}JPM & \cellcolor[HTML]{F8766D}ALV  & \cellcolor[HTML]{F8776D}ALV  & \cellcolor[HTML]{FEE683}HPQ & \cellcolor[HTML]{F8776D}ALV  \\
8                       & \cellcolor[HTML]{F97F6F}BA & \cellcolor[HTML]{C3DA81}MET & \cellcolor[HTML]{E1E383}LNC & \cellcolor[HTML]{B9D780}MUV & \cellcolor[HTML]{E1E383}LNC & \cellcolor[HTML]{FFEB84}ISP & \cellcolor[HTML]{F8696B}AAL  & \cellcolor[HTML]{F8776D}ALV  & \cellcolor[HTML]{DDE182}LNC & \cellcolor[HTML]{B9D780}MUV & \cellcolor[HTML]{B9D780}MUV & \cellcolor[HTML]{BAD780}MUV & \cellcolor[HTML]{C3DA81}MET & \cellcolor[HTML]{C3DA81}MET & \cellcolor[HTML]{F8696B}AAL  & \cellcolor[HTML]{F8696B}AAL  & \cellcolor[HTML]{F8696B}AAL  & \cellcolor[HTML]{63BE7B}VIA & \cellcolor[HTML]{F3E884}JPM \\
9                       & \cellcolor[HTML]{FDD680}HEI & \cellcolor[HTML]{FBAB77}CS & \cellcolor[HTML]{FCBD7B}DHR & \cellcolor[HTML]{C3DA81}MET & \cellcolor[HTML]{6EC17C}TGT & \cellcolor[HTML]{FAEA84}JPM & \cellcolor[HTML]{F3E884}JPM & \cellcolor[HTML]{F3E884}JPM & \cellcolor[HTML]{F3E884}JPM & \cellcolor[HTML]{C3DA81}MET & \cellcolor[HTML]{C3DA81}MET & \cellcolor[HTML]{FBA676}CON & \cellcolor[HTML]{F97E6F}AZN  & \cellcolor[HTML]{DDE182}LNC & \cellcolor[HTML]{E1E383}LNC & \cellcolor[HTML]{6EC17C}TGT & \cellcolor[HTML]{DDE182}LNC & \cellcolor[HTML]{F8696B}AAL  & \cellcolor[HTML]{D4DF82}LUV \\
10                      & \cellcolor[HTML]{FFEB84}ISP & \cellcolor[HTML]{FBA476}CON & \cellcolor[HTML]{C3DA81}MET & \cellcolor[HTML]{FBA476}CON & \cellcolor[HTML]{FAEA84}JPM & \cellcolor[HTML]{F8696B}AAL  & \cellcolor[HTML]{F7E984}ISP & \cellcolor[HTML]{FEE683}HPQ & \cellcolor[HTML]{FA9573}BP & \cellcolor[HTML]{FBA476}CON & \cellcolor[HTML]{FBA476}CON & \cellcolor[HTML]{FDC77D}ENE & \cellcolor[HTML]{FDD17F}GE & \cellcolor[HTML]{FBA676}CON & \cellcolor[HTML]{6EC17C}TGT & \cellcolor[HTML]{FEE182}HPQ & \cellcolor[HTML]{79C57D}TGT & \cellcolor[HTML]{F8776D}ALV  & \cellcolor[HTML]{FA9573}BP \\ \hline
\end{tabular}%
}% end large/resizebox group
}
\caption{Top 10 tickers per metric: betweenness (BC), closeness (CC), eigenvector (EC), and degree (D/ID/OD) centralities, including weighted versions (ECW/DW/IDW/ODW). See Table \ref{tab:topnodes_names} for full firm names.}
\label{tab:topnodes}
\end{table}

\begin{table}[htbp!]
\centering
\scriptsize
\setlength{\tabcolsep}{3pt}
\resizebox{\textwidth}{!}{%
\begin{tabular}{rcccllr@{\hspace{8pt}}rcccllr}
\hline
\textbf{Node} & \textbf{Ticker} & \textbf{F10} & \textbf{F5} & \textbf{Name} & \textbf{Ind.} & \textbf{Reg.} & \textbf{Node} & \textbf{Ticker} & \textbf{F10} & \textbf{F5} & \textbf{Name} & \textbf{Ind.} & \textbf{Reg.} \\ \hline
22 & C & 19 & 18 & Citigroup & Bank. & Amer. & 0 & AAL & 7 & 0 & Anglo American & Metals/Mining & APAC \\
63 & LNC & 15 & 1 & Lincoln National & Insur. & Amer. & 26 & CON & 6 & 0 & Continental & Auto. & EMEA \\
57 & ISP & 14 & 7 & Intesa Sanpaolo & Bank. & EMEA & 86 & TGT & 4 & 0 & Target & Retail & Amer. \\
6 & ALV & 11 & 1 & Allianz & Insur. & -- & 37 & DHR & 3 & 2 & Danaher & Med. Equip. & -- \\
3 & AIG & 11 & 10 & American Intl. Grp & Insur. & -- & 53 & HPQ & 3 & 0 & HP & Tech. Hardw. & -- \\
66 & MAR & 11 & 11 & Marriott Intl. & Leisure & Amer. & 19 & BP & 2 & 0 & BP & Oil/Gas & -- \\
48 & HEI & 11 & 10 & HeidelbergCement & Constr. Mat. & EMEA & 65 & LUV & 1 & 0 & Southwest Air. & Transp. & Amer. \\
88 & TSN & 10 & 10 & Tyson Foods & Food & -- & 9 & AZN & 1 & 0 & AstraZeneca & Biotech & APAC \\
58 & JPM & 10 & 0 & JPMorgan Chase & Bank. & Amer. & 10 & BA & 1 & 0 & Boeing & Aero/Def. & Amer. \\
44 & GE & 8 & 4 & General Electric & Divers. Ind. & Amer. & 29 & CS & 1 & 0 & AXA & Insur. & EMEA \\
84 & SPG & 8 & 8 & Simon Property & REIT & -- & 91 & VIA & 1 & 0 & ViacomCBS & Entertain. & -- \\
25 & CB & 8 & 8 & Chubb & Insur. & -- & 24 & CAT & 1 & 0 & Caterpillar & Machinery & Amer. \\
69 & MET & 7 & 0 & MetLife & Insur. & Amer. & 8 & AV & 1 & 0 & Aviva & Insur. & EMEA \\
61 & L & 7 & 5 & Loews & Insur. & -- & 40 & ENE & 1 & 0 & Enel & Utilities & EMEA \\
71 & MUV & 7 & 0 & Munich Re & Insur. & -- &  &  &  &  &  &  & \\ \hline
\end{tabular}%
}
\caption{Compact top nodes by frequency, showing node number, ticker, frequency in the top-10 and top-5 lists, name, industry and region. Full firm and covariate details are reported in Appendix Table \ref{tab:topnodes_list}.}
\label{tab:topnodes_names}
\end{table}

\begin{figure}[htbp!]
\centering
\includegraphics[width=\linewidth, trim=0 18 0 0, clip]{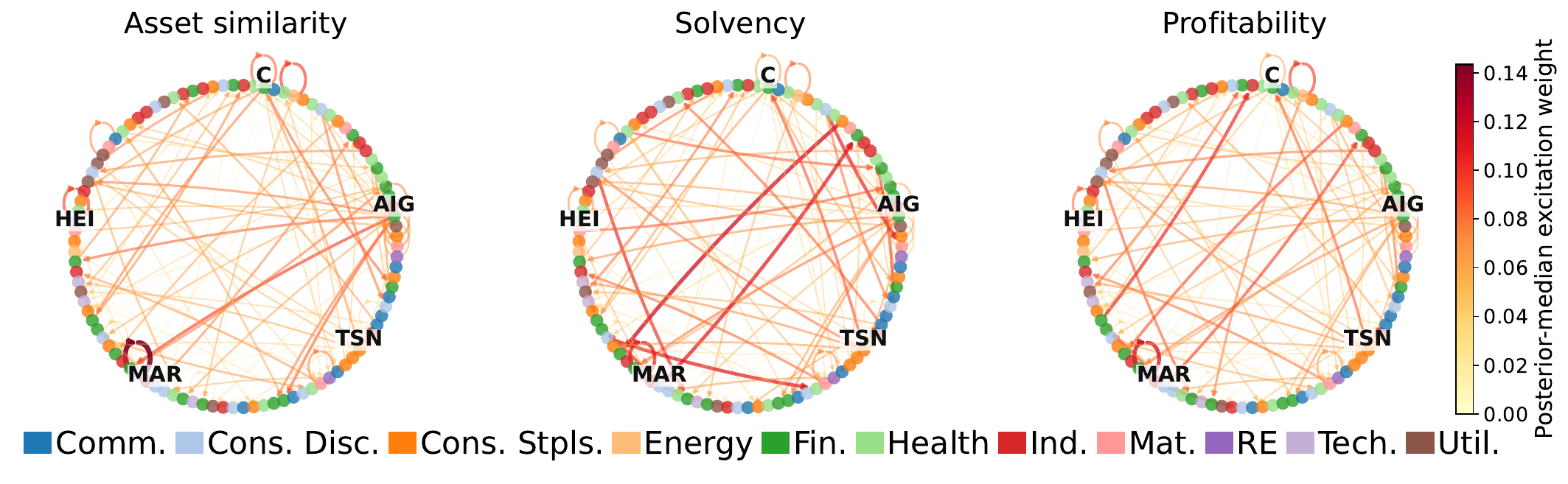}
\caption{Posterior-median directed excitation networks for the three candidate contagion channels in the ``3l all'' model set-up. Arrow direction shows transmission direction, darker and wider edges indicate larger excitation weights, and node positions are fixed across panels. Nodes are coloured by sector. Ticker labels identify the five largest weighted net transmitters (AIG: American International Group; TSN: Tyson Foods; MAR: Marriott International; HEI: HeidelbergCement; C: Citigroup).}
\label{fig:network_rd_3lall}
\end{figure}

Weighted out-degree is more dispersed than weighted in-degree, so differences in outward transmission drive much of the node-level variation (Figure \ref{fig:network_rd_3lall} and Appendix Figure~\ref{fig:3l_all_l}). The top five weighted net transmitters are American International Group, Tyson Foods, Marriott International, HeidelbergCement and Citigroup. These nodes also appear among the largest outward transmitters in the directed network, indicating that their outgoing excitation exceeds their incoming excitation rather than merely reflecting high overall connectivity. Citigroup and Lincoln National appear among the relatively small set of firms with notable positions in both incoming and outgoing summaries, consistent with potential feedback roles. Weighted eigenvector centrality identifies institutions whose importance reflects connections to other influential nodes rather than direct outward transmission alone; its three highest-ranked institutions are Intesa Sanpaolo, Chubb and Citigroup.

Only a minimal set of nodes appear in both the principal incoming and outgoing rankings. At the layer level, the estimates for the asset similarity, solvency and profitability channels remain broadly similar across model set-ups, although the strongest contrast appears between asset similarity and solvency. Among the top nodes flagged by the importance metrics, a large proportion originate from the U.S. banking and insurance sectors and have high profitability and solvency ratios. Notable examples include American International Group and Citigroup, both widely recognised as central contributors to risk spread during the 2007--2008 financial crisis. The separate directed channel networks in Figure \ref{fig:network_rd_3lall} make the sparse common edge set visible and show prominent transmitters through numerous darker and wider outgoing edges; variation in those edges across panels displays channel-specific contributions.

\FloatBarrier

\subsubsection{Contagion Channels and Covariates}

We consider three candidate contagion channels: asset similarity, solvency and profitability. Table \ref{tab:scenarios_weights_layer_rd} reports excitation weights, which are interpreted as channel-specific contributions for the covariate-regression model set-ups. In the ``3l all'' set-up, the summed weights for asset similarity, solvency and profitability are $6.00$, $6.28$ and $5.99$, respectively. Thus, solvency has the largest posterior-median contribution in this set-up, but the difference is small and does not support a claim that one channel dominates. This differs from the single-layer and two-layer set-ups, where the asset similarity channel is influential when considered alone and has a slightly greater summed contribution than solvency in ``2l asset solv'' ($8.72$ compared with $8.33$). As additional channels are introduced, the influence is distributed more evenly. The shorter excitation period in ``3l all 5dt'' also gives comparatively even channel totals, which may reflect short-term jumps being driven more by common market stress than by slower-moving fundamentals such as profitability or solvency.

%\enlargethispage{2\baselineskip}
Table \ref{tab:scenarios_covariates_rd} reports the gamma-regression coefficients within each channel. Sender covariates describe the institution from which excitation is transmitted; recipient covariates describe the counterparty receiving it. In the asset similarity channel, industry categorisation ($\beta_2$) has a clearly positive estimate in ``1l asset'' and remains positive, though less precisely estimated, in ``2l asset solv''. Its posterior median also remains positive in ``3l all'', although its interval includes zero after the other channels are included. These estimates provide signs that extreme credit-spread jumps transmit more strongly within similar industries, with the effect less precisely isolated when more channels are fitted. For solvency, the transmitting-firm coefficient remains negative in the three-channel set-up, whereas the recipient-firm effect is weaker and its interval includes zero. Profitability gives a similar direction for the sender effect, but with weaker evidence. The empirical interpretation is therefore that poorer firm fundamentals are more clearly associated with transmitting than receiving stress, while the sizes of these covariate effects are modest. The transformed solvency and profitability variables are moderately positively associated across firms (Spearman's $\rho=0.36$; Pearson $r=0.36$, $n=99$), so overlapping channel information may contribute to coefficient attenuation when both channels are included. This motivates considering model selection or shrinkage on layers when several channel specifications carry related information.

\subsubsection{Time Variation in the Empirical Network}
\label{sec:empirical_time_variation_3all}

We next compare the full-period network with the three regime-specific fits introduced in Figure \ref{fig:process_rd2} and discussed in Section \ref{sec:empiricaldataset}: pre-GFC/GFC, euro crisis/calm, and COVID-19. The aim is simply to check whether the full-period network is an average of materially different transmission regimes.

\begin{table}[htbp!]
\centering
\footnotesize
\setlength{\tabcolsep}{1.2pt}
\renewcommand{\arraystretch}{1.03}
\setlength{\arrayrulewidth}{0.6pt}
\resizebox{0.93\textwidth}{!}{%
\begin{tabular}{llrrrrlrrrr}
\arrayrulecolor{black}\hline
\textbf{Period} & \textbf{Years} & \textbf{$T$} & \textbf{$N$} & \textbf{$N/T$} & \textbf{$\rho$} & \textbf{95\%} & \textbf{$\sum A$} & \textbf{$\sum AW$} & \textbf{$\max AW$} & \textbf{WAIC/$N$} \\
\arrayrulecolor{black}\hline
Full & 2004--2022 & 4,696 & 1,905 & 0.406 & \cellcolor[HTML]{F8696B}0.025 & {[}0.021,0.029{]} & 113 & \cellcolor[HTML]{F8696B}25.00 & \cellcolor[HTML]{5A8AC6}0.37 & \cellcolor[HTML]{FBCACD}12.02 \\
Pre-GFC/GFC & 2004--2009 & 1,549 & 1,118 & 0.722 & \cellcolor[HTML]{FBE8EB}0.014 & {[}0.011,0.018{]} & 57 & \cellcolor[HTML]{FABDC0}17.83 & \cellcolor[HTML]{608EC8}0.39 & \cellcolor[HTML]{F99FA1}11.36 \\
Euro/calm & 2010--2019 & 2,608 & 397 & 0.152 & \cellcolor[HTML]{5A8AC6}0.000 & {[}0.000,0.001{]} & 0 & \cellcolor[HTML]{5A8AC6}0.00 & -- & \cellcolor[HTML]{5A8AC6}15.00 \\
COVID-19 & 2020--2022 & 539 & 390 & 0.724 & \cellcolor[HTML]{D5E0F1}0.010 & {[}0.007,0.012{]} & 10 & \cellcolor[HTML]{CEDBEF}8.91 & \cellcolor[HTML]{F8696B}1.06 & \cellcolor[HTML]{F8696B}10.55 \\
\arrayrulecolor{black}\hline
\end{tabular}%
}
\caption{Time-variation comparison for the selected full-period fit and the three regime-specific refits. Sum $A$, Sum $AW$ and Max $AW$ use posterior-median active excitation entries, including self-excitation as in Table \ref{tab:scenarios_weights_layer_rd}; 95\% is the posterior HDI for $\rho$.}
\label{tab:timevariation_params}
\end{table}

Table \ref{tab:timevariation_params}, with supporting summaries in Appendix \ref{sec:empirical_results_app} (Tables \ref{tab:timevariation_covariates}--\ref{tab:timevariation_layers}), shows that the contrast is sharp. The full-period fit has $\rho=0.025$ and Sum $AW=25.00$, while the pre-GFC/GFC fit has lower $\rho=0.014$ but still retains substantial aggregate excitation, Sum $AW=17.83$. The euro-crisis/calm window has no posterior-median excitation network; this is an artefact of defining extreme credit-spread jumps using the 99th-percentile threshold estimated over the full period, which leaves few common-threshold events in the calmer middle window but keeps the regimes directly comparable. COVID-19 is sparse, with $\rho=0.010$ and only 10 active posterior-median entries, but its largest edge is much stronger than in the full-period fit (Max $AW=1.06$ versus 0.37). Reciprocity and clustering remain low across the non-empty networks, while weighted-degree skewness is highest in COVID-19, indicating a small number of intense transmitters rather than a broad network.

\begin{figure}[htbp!]
\centering
{\captionsetup{font=normalsize}
\makebox[\textwidth][c]{\includegraphics[width=0.95\textwidth, trim=0 34 0 0, clip]{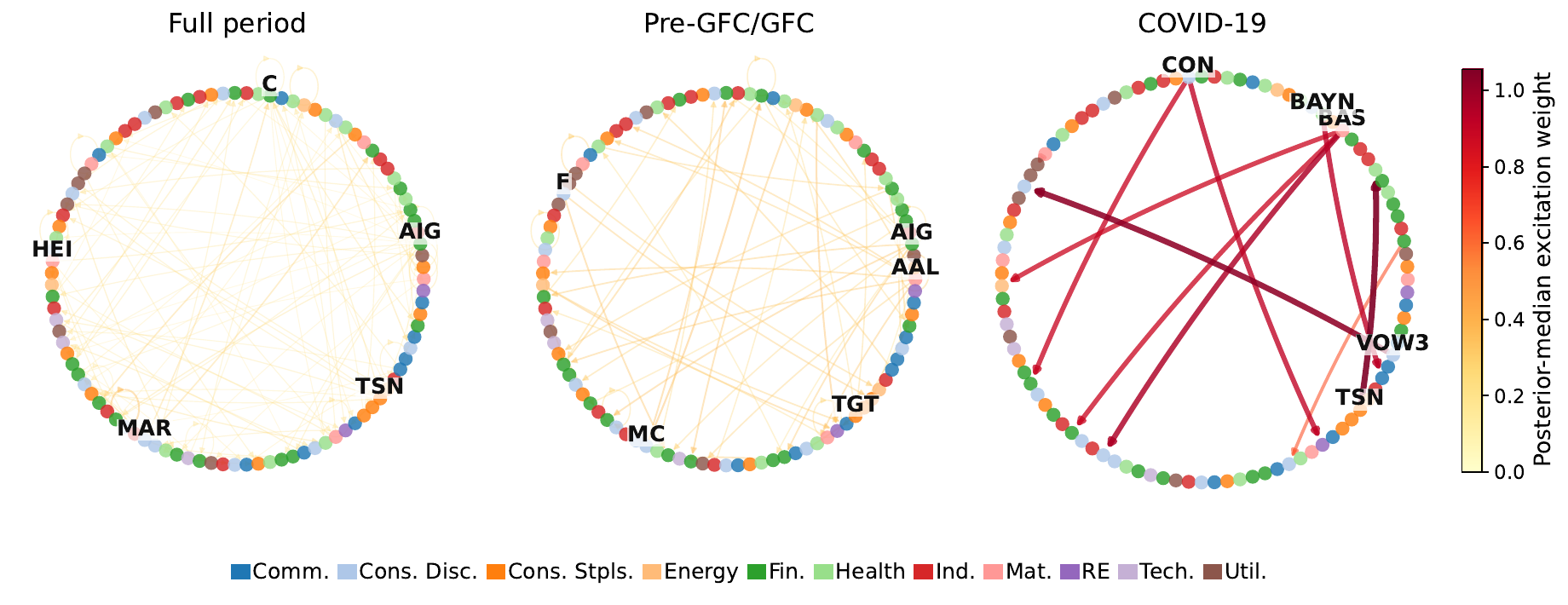}}
\vspace{-0.6em}

{\footnotesize
\setlength{\tabcolsep}{1.2pt}
\begin{tabular}{ccccccccccc}
\textcolor[HTML]{1F77B4}{\rule{0.75em}{0.75em}} Comm. &
\textcolor[HTML]{AEC7E8}{\rule{0.75em}{0.75em}} Cons. Disc. &
\textcolor[HTML]{FF7F0E}{\rule{0.75em}{0.75em}} Cons. Stpls. &
\textcolor[HTML]{FFBB78}{\rule{0.75em}{0.75em}} Energy &
\textcolor[HTML]{2CA02C}{\rule{0.75em}{0.75em}} Fin. &
\textcolor[HTML]{98DF8A}{\rule{0.75em}{0.75em}} Health &
\textcolor[HTML]{D62728}{\rule{0.75em}{0.75em}} Ind. &
\textcolor[HTML]{FF9896}{\rule{0.75em}{0.75em}} Mat. &
\textcolor[HTML]{9467BD}{\rule{0.75em}{0.75em}} RE &
\textcolor[HTML]{C5B0D5}{\rule{0.75em}{0.75em}} Tech. &
\textcolor[HTML]{8C564B}{\rule{0.75em}{0.75em}} Util.
\end{tabular}}
\caption{Overall posterior-median excitation networks for the selected full-period fit, the pre-GFC/GFC refit and the COVID-19 refit. The euro-crisis/calm refit is omitted because it has no positive posterior-median excitation edges. Node positions, sector colours and edge scaling follow Figure \ref{fig:network_rd_3lall}; ticker labels identify the top five weighted net transmitters in each displayed period (Full period: AIG, MAR, TSN, HEI and C; Pre-GFC/GFC: AAL, AIG, TGT, MC and F; COVID-19: BAS, CON, TSN, VOW3 and BAYN).}
\label{fig:timevariation_overall_networks}
}
\end{figure}

Figure \ref{fig:timevariation_overall_networks} reinforces this interpretation: the full-period and pre-GFC/GFC networks contain broader crisis transmission structure, whereas COVID-19 is concentrated in a few high-weight edges. Appendix Figure \ref{fig:timevariation_layer_networks} gives the corresponding layer-level view, and Appendix Table \ref{tab:timevariation_topnodes} shows how the leading institutions change by regime. AIG and Citigroup remain prominent in the pre-GFC/GFC weighted-degree and ECW rankings, while Tyson Foods and Marriott reappear during COVID-19. This shift from financial-sector names toward non-financial firms such as BASF, Continental, Tyson Foods, AstraZeneca and Marriott is consistent with a broader real-economy stress episode. Appendix Table \ref{tab:timevariation_degree_tests} reports the degree-distribution checks, with the scale-free evidence again clearest for weighted degree.

The subperiod analysis, together with the residual-process diagnostics in Appendix \ref{sec:goodnessoffit_emp}, indicates remaining time variation in event activity. Modelling this time variation directly is a natural extension, while the present analysis focuses on the interpretable multiplex transmission network and its stability across economically meaningful regimes.

\FloatBarrier

\subsubsection{Comparison to Previous Work}

The most comparable inferred-network approaches to measuring systemic node importance include the Granger-causality networks of \citet{Billio2012}, the vector autoregressive connectedness model of \citet{Diebold2014}, and the Markov-switching graphical SUR model of \citet{Bianchi2019}. The studies compared above focus on samples dominated by large financial institutions over periods heavily dominated by the 2007--2008 financial crisis, whereas our application also includes insurance and non-financial firms over 2004--2022. Table \ref{tab:comparison_ranks} presents ranks of institutions across these studies and our selected network summaries.

\citet{Diebold2014} flag Citigroup, Wells Fargo and JPMorgan Chase as among the most net-connected institutions; \citet{Billio2012} use principal components analysis and Granger-causality networks to reach related conclusions, emphasising out-to-other rather than in-from-other measures. In contrast, our broader application tracks large-cap non-financial and insurance entities as well as banks. American International Group has the highest weighted net-degree, reflecting numerous outward edges to the broader network (Appendix Figure \ref{fig:topedges_combined}). Chubb, Lincoln National and Hartford Financial Services Group also rank highly in weighted net-degree and weighted eigenvector centrality, consistent with the importance of insurance firms reported by \citet{Billio2012}.

\citet{Bianchi2019} examine the S\&P 100 using a Markov-switching graphical SUR model, measuring centrality using weighted eigenvector centrality variants based on Fama--French and I-CAPM approaches. Whereas they isolate regimes to compare network topologies, we estimate a single network over the full period. Our wider dataset\footnote{Unlike the S\&P 100 framework of \citet{Bianchi2019}, our dataset includes firms from the S\&P 500 and European indices across a wider range of sectors.} covers calmer periods and multiple crises, including COVID-19, placing the systemic influence of banks and financial firms in a broader context. While we confirm that a small number of institutions stand out, expanding the sample also highlights insurance and non-financial firms ranked highly by weighted eigenvector centrality.

The appearance of non-financial firms among the top five weighted net transmitters is plausibly connected to the broader sample period, particularly the COVID-19 stress episode, but should not be read as evidence of direct balance-sheet contagion from these firms. In the posterior-median aggregate $AW$ network, positive net transmission means that weighted outgoing excitation exceeds weighted incoming excitation. Marriott International, Tyson Foods and HeidelbergCement may therefore act as real-economy stress indicators for travel and discretionary demand, food and supply-chain pressures, and construction or energy-intensive industrial activity. Their displayed edges in Appendix Figure \ref{fig:topedges_combined} are predictive excitation relationships in credit-risk events, not structural claims about contractual exposures or default cascades.

Finally, the studies compared above do not explicitly embed covariates. \citet{Longstaff2010} uses a vector autoregressive framework to show that contagion propagates primarily through liquidity and risk-premium channels using aggregate market variables, whereas our model relies on firm-specific indicators rather than market-wide shocks. \citet{adelfio2021financial} apply an extension of a space-time ETAS model in which the spatial component captures similarities between European financial institutions through CDS price distances. This formulation represents institutional proximity in a way related to our weight-regression channels, which use node-level covariates averaged across time. Their evidence that non-crisis spatial distances remain stable, together with \citet{avdjiev2019measuring}, motivates our use of structural firm characteristics in dependency weights to provide a more granular evaluation of extreme credit-spread jump transmission than reliance on endogenous price reactions alone.

\begin{landscape}

\begin{table}[htbp!]
\centering
\resizebox{1\linewidth}{!} {%
\begin{tabular}{lllllllllllllllllllllllllllll}
\hline
\multicolumn{1}{c}{}                                & \multicolumn{1}{c}{}                                & \multicolumn{1}{c}{}                                    & \multicolumn{1}{c}{}                                & \multicolumn{1}{c}{}                                & \multicolumn{1}{c}{}                                  & \multicolumn{9}{c}{\textbf{Relative ranks}}                                                                                                                                                                                                                                                   & \multicolumn{4}{c}{\footnotesize\textbf{ECW}}                                                                                              & \multicolumn{1}{c}{\footnotesize\textbf{IDW}} & \multicolumn{1}{c}{\footnotesize\textbf{ODW}} & \multicolumn{4}{c}{\footnotesize\textbf{NDW}}                                                                                                  & \multicolumn{4}{c}{\footnotesize\textbf{DW}}                                                                                                                \\
\hline
\end{tabular}%
}\caption{Relative rank comparison of our model's weighted (net-)degree (NDW/DW) and eigenvector centrality (ECW) versus alternative systemic risk measures: OTO, AM and MPC~\citep{Billio2012}, net-connectedness (NC)~\citep{Diebold2014}, and ECW under FF/I-CAPM~\citep{Bianchi2019}. Blanks denote absent firms or values outside reported top thresholds.
}
\label{tab:comparison_ranks}
\end{table}

\begin{table}[htbp!]
\centering
\resizebox{1.5\textwidth}{!}{%
\begin{tabular}{llllllllllllllllllllllllllllllll}
\hline
\textbf{Model set-up} & \textbf{Avg$AW^{(0)}$} & \textbf{Avg$AW^{(1)}$} & \textbf{Avg$AW^{(2)}$} & \textbf{Sum$AW^{(0)}$} & \textbf{Sum$AW^{(1)}$} & \textbf{Sum$AW^{(2)}$} & \textbf{Min$AW^{(0)}$} & \textbf{Max$AW^{(0)}$} & \textbf{Min$AW^{(1)}$} & \textbf{Max$AW^{(1)}$} & \textbf{Min$AW^{(2)}$} & \textbf{Max$AW^{(2)}$} & \textbf{Len95\% $AW_0$} & \textbf{Len95\% $AW_1$} & \textbf{Len95\% $AW_2$} & \textbf{95\%/$AW^{(0)}$}  & \textbf{95\%/$AW^{(1)}$}  & \textbf{95\%/$AW^{(2)}$} \\ \hline
%1l nocovs & 143 & \cellcolor[HTML]{FCAB78}0.08 &  &  & \cellcolor[HTML]{FB9F76}11.55 &  &  & \cellcolor[HTML]{6CC183}0.00 & \cellcolor[HTML]{F9787A}0.41 &  &  &  &  & {[}0.01,0.24{]}  &                  &                  & {[}0.02,0.13{]}  &                  &                  & \cellcolor[HTML]{FB9975}0.23 &                              &                              & \cellcolor[HTML]{FBA076}0.11 &                              &                              & \cellcolor[HTML]{FFDD82}2.90 &                              &                              & \cellcolor[HTML]{FFDF82}1.42 &                              & \\
1l base & \cellcolor[HTML]{FA8471}0.10 &  &  & \cellcolor[HTML]{FA8471}13.31 &  &  & \cellcolor[HTML]{B5DFC2}0.04 & \cellcolor[HTML]{FAACAE}0.28 &  &  &  &  & \cellcolor[HTML]{FFDA81}0.18 &                              &                              & \cellcolor[HTML]{A2D07E}1.91 &                              & \\
1l asset & \cellcolor[HTML]{FA7F70}0.10 &  &  & \cellcolor[HTML]{F9786E}14.07 &  &  & \cellcolor[HTML]{A9DAB7}0.03 & \cellcolor[HTML]{FA9496}0.34 &  &  &  &  & \cellcolor[HTML]{FED981}0.18 &                              &                              & \cellcolor[HTML]{9FCF7E}1.89 &                              & \\
\hline
\end{tabular}%
}
\caption{Model-layer excitation-weight estimates for edges where $A_{i,j}=1$. Averages, sums, minima and maxima are across all edges.}
\label{tab:scenarios_weights_layer_rd}
\end{table}

%\end{landscape}

\begin{table}[htbp!]
\centering
\resizebox{1\textwidth}{!}{%
\begin{tabular}{lcccccccccccccccccccccc}
\hline
\textbf{Model set-up}& \textbf{Cov}&\textbf{L}&\textbf{$\beta_0$}&\textbf{95\% HDI}&\textbf{68\% HDI}&\textbf{$\beta_1$}&\textbf{95\% HDI}&\textbf{68\% HDI}&\textbf{$\beta_2$}&\textbf{95\%HDI}&\textbf{68\% HDI}&\textbf{$\beta_3$}&\textbf{95\% HDI}&\textbf{68\% HDI}\\
\hline
%1l nocovs                                    & none                                   & 0                              &                                                      &                                       &                                       &                                                      &                                                                  &                                                                  &                                                      &                                                                 &                                                                 &                                                      &                                                                 &                                                                 \\
1l base                                      & 0                                      & 0                              & \cellcolor[HTML]{FAAEB0}-2.19                        & {[}-2.29,-2.1{]}                      & {[}-2.24,-2.15{]}                     & \cellcolor[HTML]{F97F82}0.51                         & {[}-20.2,19.18{]}                                                & {[}-9.85,10.19{]}                                                &                                                      &                                                                 &                                                                 &                                                      &                                                                 &                                                                 \\
\hline
\end{tabular}%
}
\caption{Covariate coefficients. Green cells indicate the HDI excludes 0, and yellow indicates that the interval contains 0 but is skewed ($\geq$75\% on one side).}
\label{tab:scenarios_covariates_rd}
\end{table}
%\end{landscape}

\end{landscape}

\FloatBarrier

\section{Conclusion}\label{sec:conclusion}

Our empirical application demonstrates how the Multiplex Network Hawkes model can integrate information associated with asset similarity, solvency and profitability into a structured multi-layer framework for modelling transmission of extreme credit-spread jumps. By separating candidate contagion channels, the framework relates event transmission to firm-level information and enables the influence of institutions to be assessed using topological metrics. In particular, weighted degree and weighted eigenvector centrality identify institutions such as American International Group and Citigroup as important conduits for risk transmission.

The inferred transmission network is sparse and predominantly directional, with only a few reciprocal feedback links and outward transmission concentrated among a small number of institutions. The preference for a power-law fit in the weighted-degree comparisons is consistent with scale-free properties in transmission strength, in line with the disproportionate role of influential financial institutions emphasised by \citet{Bianchi2019}, \citet{Billio2012} and \citet{Diebold2014}. Conversely, low reciprocity and low weighted clustering provide little support for strongly reciprocal or small-world-like propagation. This differs from the stronger small-world features reported by \citet{Diebold2014} for a narrower network centred on financial institutions during the crisis period, and may reflect the wider range of institutions and years considered here.

The channel decomposition is informative but should be interpreted cautiously. The asset-similarity estimates provide signs that extreme credit-spread jumps transmit more strongly within similar industries, consistent with a homophily interpretation and with the role of similarity-based information considered by \citet{adelfio2021financial}. Solvency and profitability provide additional candidate channels through firm fundamentals. However, the channel contributions are similar in the three-channel set-up and the estimated covariate effects are modest, so these results do not support a claim that any single channel dominates.

Collectively, these findings show how multiplex representation and covariate-driven link weights can provide a more granular assessment of systemic risk transmission. Compared with prior work, the application broadens the institution set and time period by considering banks, insurance firms and non-financial firms over 2004--2022. The resulting network recovers several institutions highlighted previously while also showing that a wider cross-industry sample changes the visible composition of influential transmitters.

The regime-specific refits show that the full-period network averages over materially different periods. The pre-GFC/GFC network retains broader crisis transmission structure with a financial core, the euro-crisis/calm window has no posterior-median excitation network under the common full-period event threshold, and COVID-19 is sparse but contains a few much stronger edges involving more non-financial firms. These contrasts motivate extending the framework to allow time variation in background intensities or network structure, while preserving the multiplex decomposition used here to compare candidate transmission channels.

\printindex 

\newpage
%\singlespacing
\setlength{\bibsep}{0pt plus 0.3ex}
\addcontentsline{toc}{section}{References} % to add references to table of contents
\bibliographystyle{agsm}                % file to determine the style of references

\bibliography{references}                     % read references from example.bib

\if1\mnhincludeappendix
\clearpage
\begin{center}
{\Large\bf Supplementary Material}\\[0.5em]
{\large A Multiplex Network Hawkes Model for Systemic Risk Measurement}
\end{center}
\appendix 

\section{Empirical Data}\label{app:data}

The potential contagion channels and candidate variables considered for the excitation network are described below.

\paragraph*{Asset Similarity Contagion Channel}
For this channel, we focus on interconnections between companies based on shared characteristics. Asset-similarity covariates are categorical variables associated with each edge (pair of nodes). That is, if both nodes share the selected category value, the variable is $1$, otherwise it is $0$. The following characteristics were considered for this channel, with those selected highlighted in bold:
\begin{itemize}
\item financial index (of which the company is part of),
\item \textbf{sector},
\item industry group/\textbf{industry}/sub-industry (BICS levels 2-4), 
\item segment (BICS levels 5-6),
\item country of risk,
\item entity region,
\item \textbf{region of largest revenue}.
\end{itemize}

Unlike asset-similarity covariates, liquidity, solvency and profitability covariates are node specific. For each channel/metric, the edge-weight regression includes the values for sender node $k'$ and recipient node $k$. Below we describe the financial measures considered for these candidate contagion channels.

\paragraph*{Liquidity Candidate Variables}
Liquidity was initially considered as a possible contagion channel because the availability of liquid assets may affect shock absorption. The candidate variables below were evaluated during exploratory data analysis, but no liquidity variable was retained in the final empirical model:

\begin{itemize}
    \item Current Ratio: $\text{Current Ratio}=\frac{\text{Total Current Assets}}{\text{Current Liabilities}}$.
    
    \item Quick Ratio: $\text{Quick Ratio}=\frac{\text{Total Current Assets} - \text{Inventories}}{\text{Current Liabilities}}$.
    
    \item CDS Volume: $\text{CDS Volume}=\text{Total trading volume of credit default swaps}$.
    
    \item Equity Shares Volume: $\text{Equity Shares Volume}=\text{Total trading volume of equity shares}$.
\end{itemize}

\paragraph*{Solvency Contagion Channel}
For this channel, we consider solvency metrics that assess a company's long-term financial stability and ability to meet debt obligations. The candidate indicators are listed below, with the one selected for our model highlighted in bold:

\begin{itemize}
    \item Debt-to-Assets Ratio: $\text{Debt-to-Assets Ratio}=\frac{\text{Total Debt}}{\text{Total Assets}}$.
    
    \item Debt-to-Equity Ratio: $\text{Debt-to-Equity Ratio}=\frac{\text{Total Debt}}{\text{Total Equity}}$.
    
    \item Financial Leverage: $\text{Financial Leverage}=\frac{\text{Average Total Assets}}{\text{Average Total Equity}}$.
    
    \item \textbf{Asset Coverage Ratio}: $\text{Asset Coverage Ratio}=\frac{\text{Total Tangible Assets} - \text{Intangible Assets}}{\text{Total Debt}}$.
    
    \item Interest Coverage Ratio: $\text{Interest Coverage Ratio}=\frac{\text{Earnings Before Interest and Taxes}}{\text{Interest Expense}}$.
    
    \item Long-term debt-to-common equity ratio: $\text{Long-term Debt to Common Equity Ratio}=\frac{\text{Long-term Debt}}{\text{Common Equity}}$.
\end{itemize}

\paragraph*{Profitability Contagion Channel}
The profitability channel can be defined by indicators that gauge how effectively a company translates sales into profits. The indicators we considered are listed below, with the selected one highlighted in bold:

\begin{itemize}
    \item \textbf{Book Value Per Share}: $\text{Book Value Per Share}=\frac{\text{Total Equity}}{\text{Number of Outstanding Shares}}$.
    
    \item Equity Price: $\text{Equity Price}=\text{Current market price of a company's stock}$.
    
    \item Price-to-earnings ratio: $\text{Price to Earnings Ratio}=\frac{\text{Market Value per Share}}{\text{Earnings Per Share}}$.
    
    \item Market Capitalisation: $\text{Market Capitalisation}=\text{Total number of shares} \times \text{Current share price}$.
\end{itemize}

\begin{figure}[htbp!]
    \centering
    % First part of the image
    \begin{subfigure}{0.25\textwidth}
        \centering
        \includegraphics[width=\textwidth, angle=0, trim= 50 50 50 50,clip]{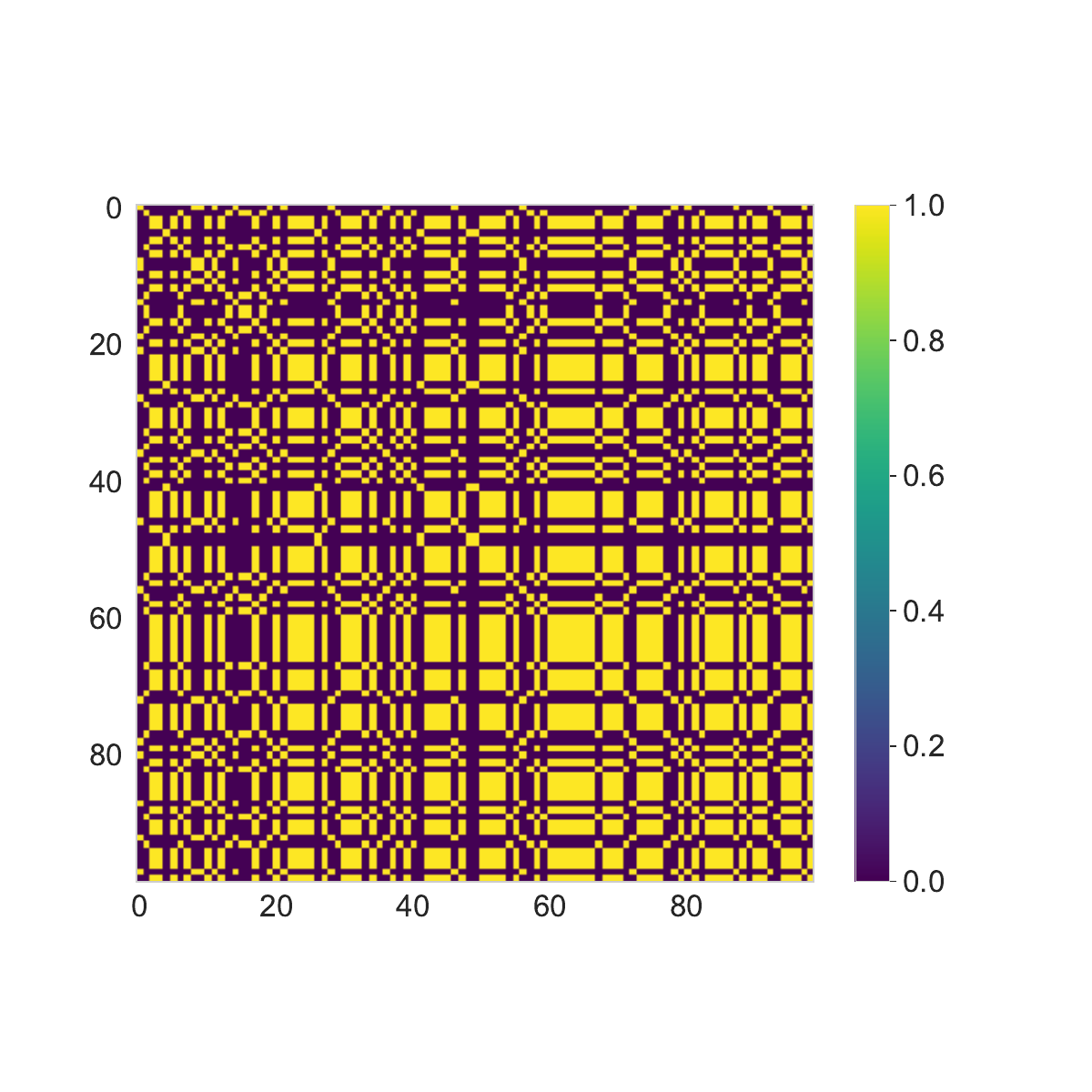}
    \subcaption{Index}
    \end{subfigure}
    % the image
    \begin{subfigure}{0.25\textwidth}
        \centering
        \includegraphics[width=\textwidth, angle=0, trim= 50 50 50 50,clip]{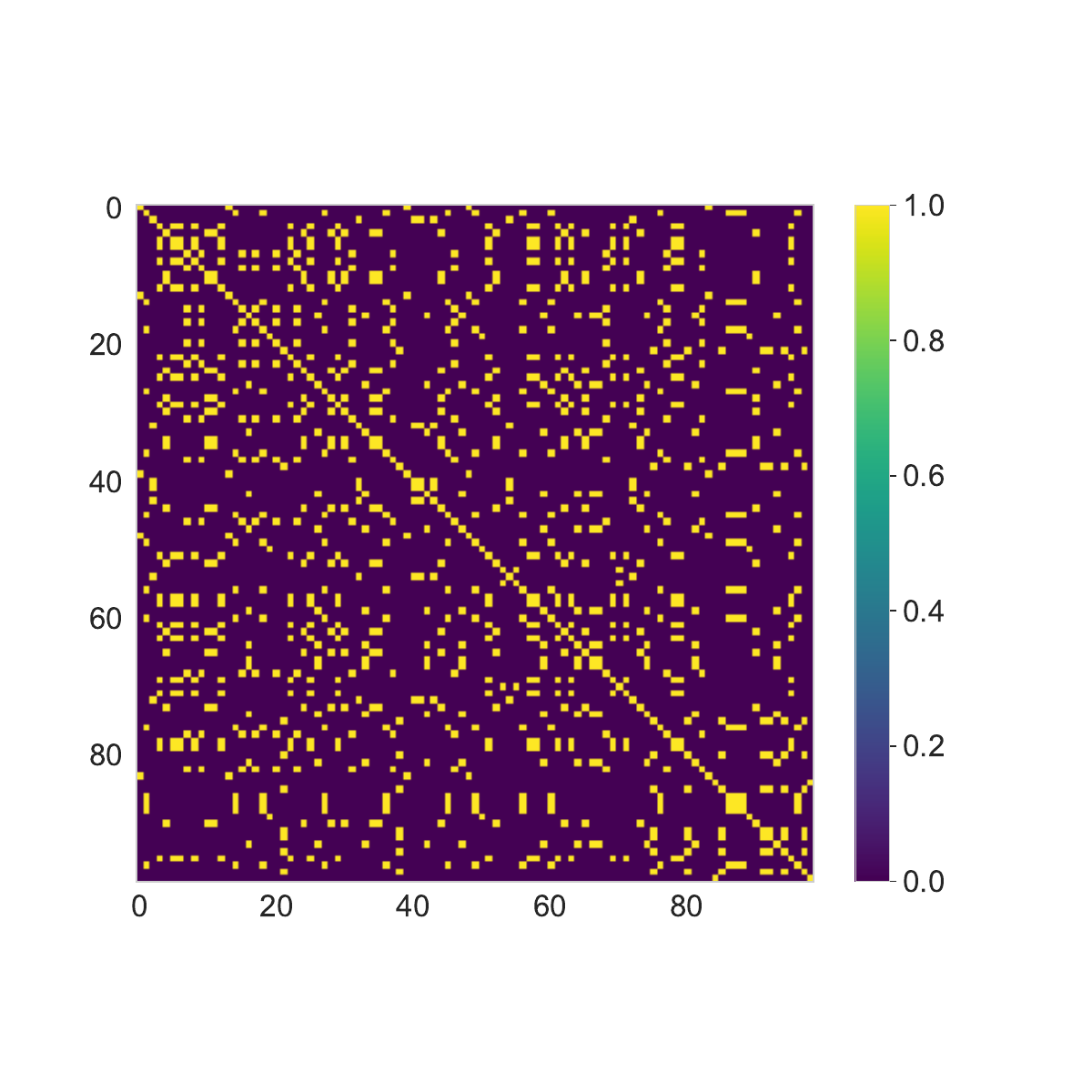}
    \subcaption{Sector (BICS level 1)}
    \end{subfigure}
    \begin{subfigure}{0.25\textwidth}
        \centering
        \includegraphics[width=\textwidth, angle=0, trim= 50 50 50 50,clip]{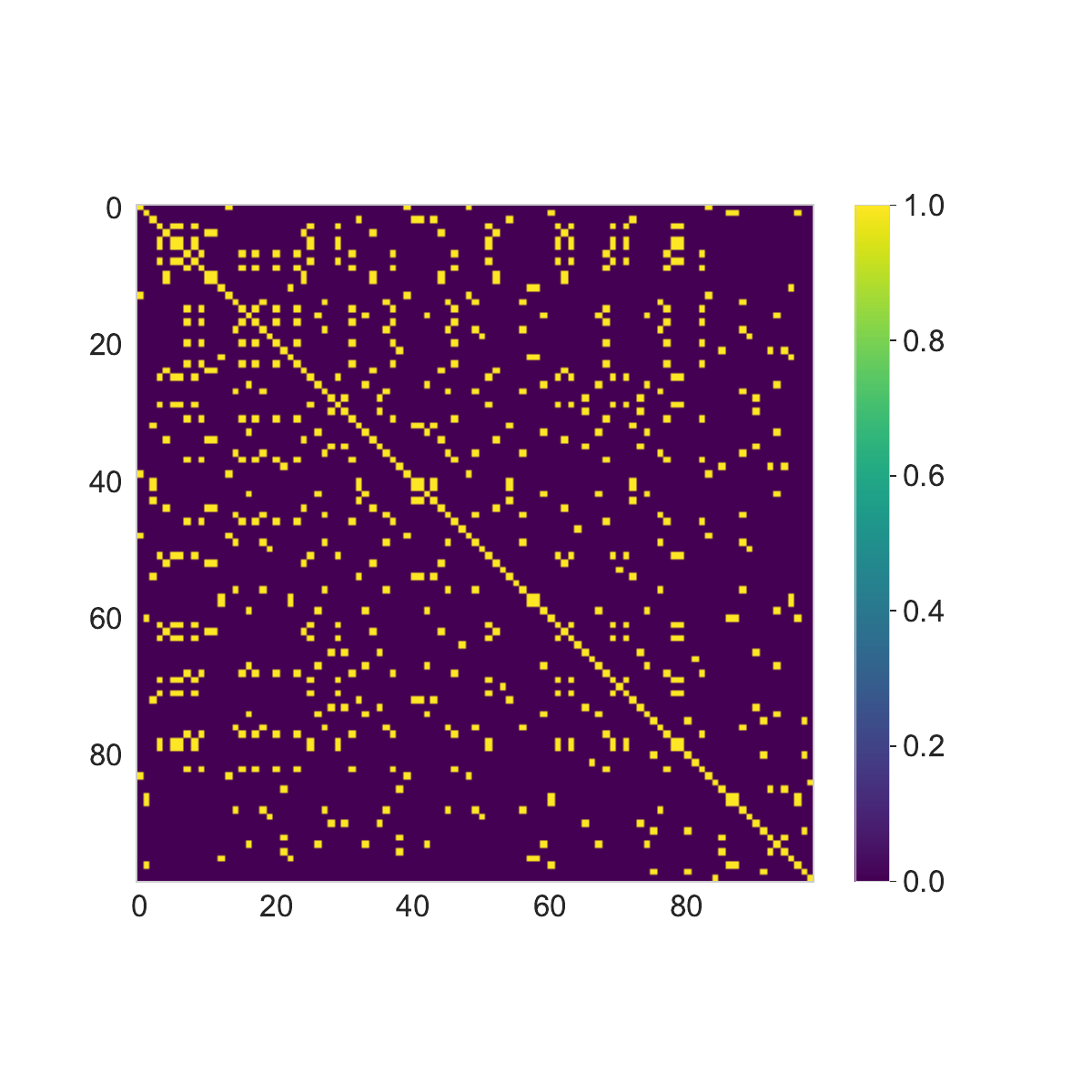}
               \subcaption{Industry group (BICS lvl 2)}
    \end{subfigure}\hfill
    % Second part of the image
    \begin{subfigure}{0.25\textwidth}
        \centering
        \includegraphics[width=\textwidth, angle=0, trim= 50 50 50 50,clip]{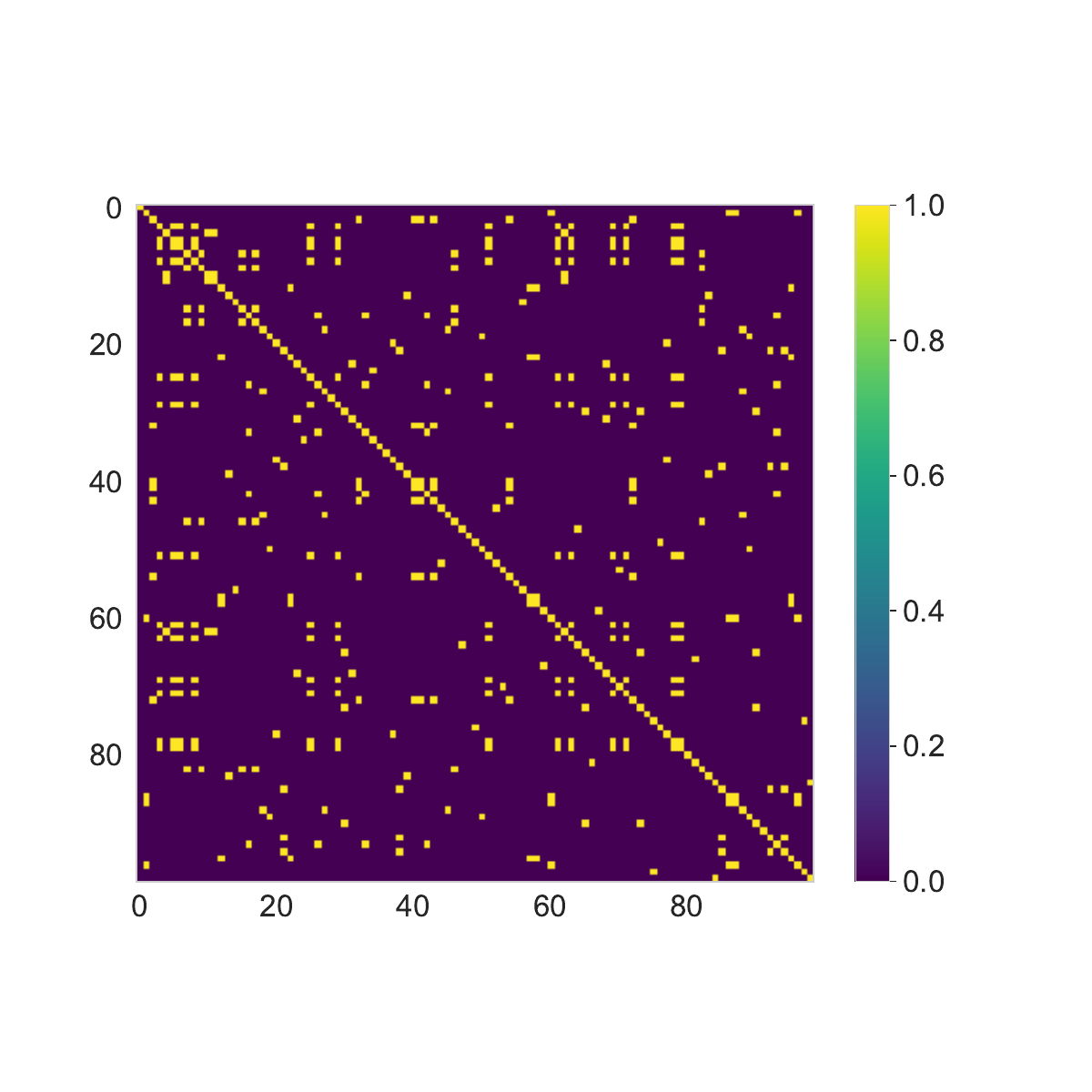}
               \subcaption{Industry (BICS level 3)}
    \end{subfigure}
        % First part of the image
    \begin{subfigure}{0.25\textwidth}
        \centering
        \includegraphics[width=\textwidth, angle=0, trim= 50 50 50 50,clip]{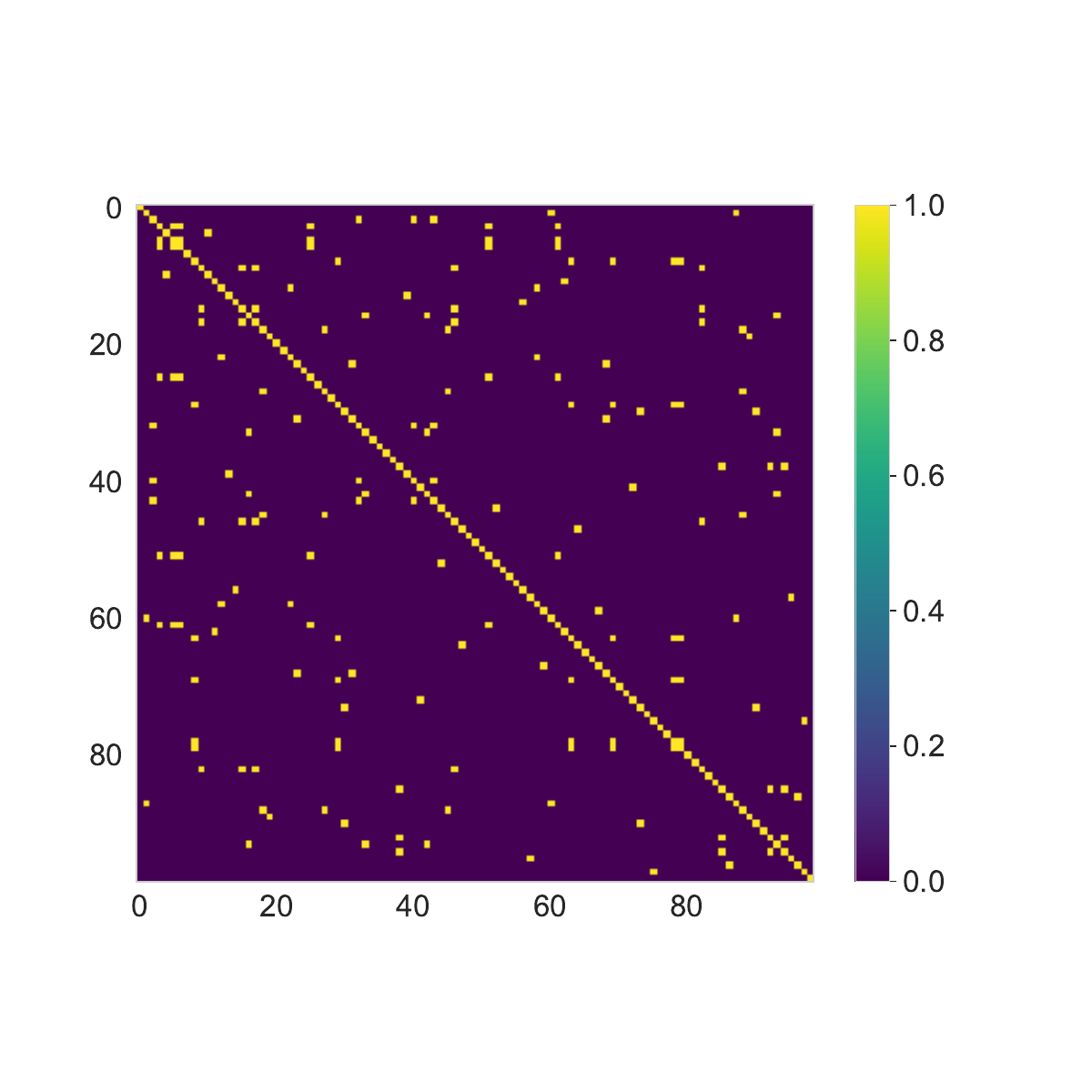}
               \subcaption{Sub-industry (BICS level 4)}
    \end{subfigure}
    % Second part of the image
    \begin{subfigure}{0.25\textwidth}
        \centering
        \includegraphics[width=\textwidth, angle=0, trim= 50 50 50 50,clip]{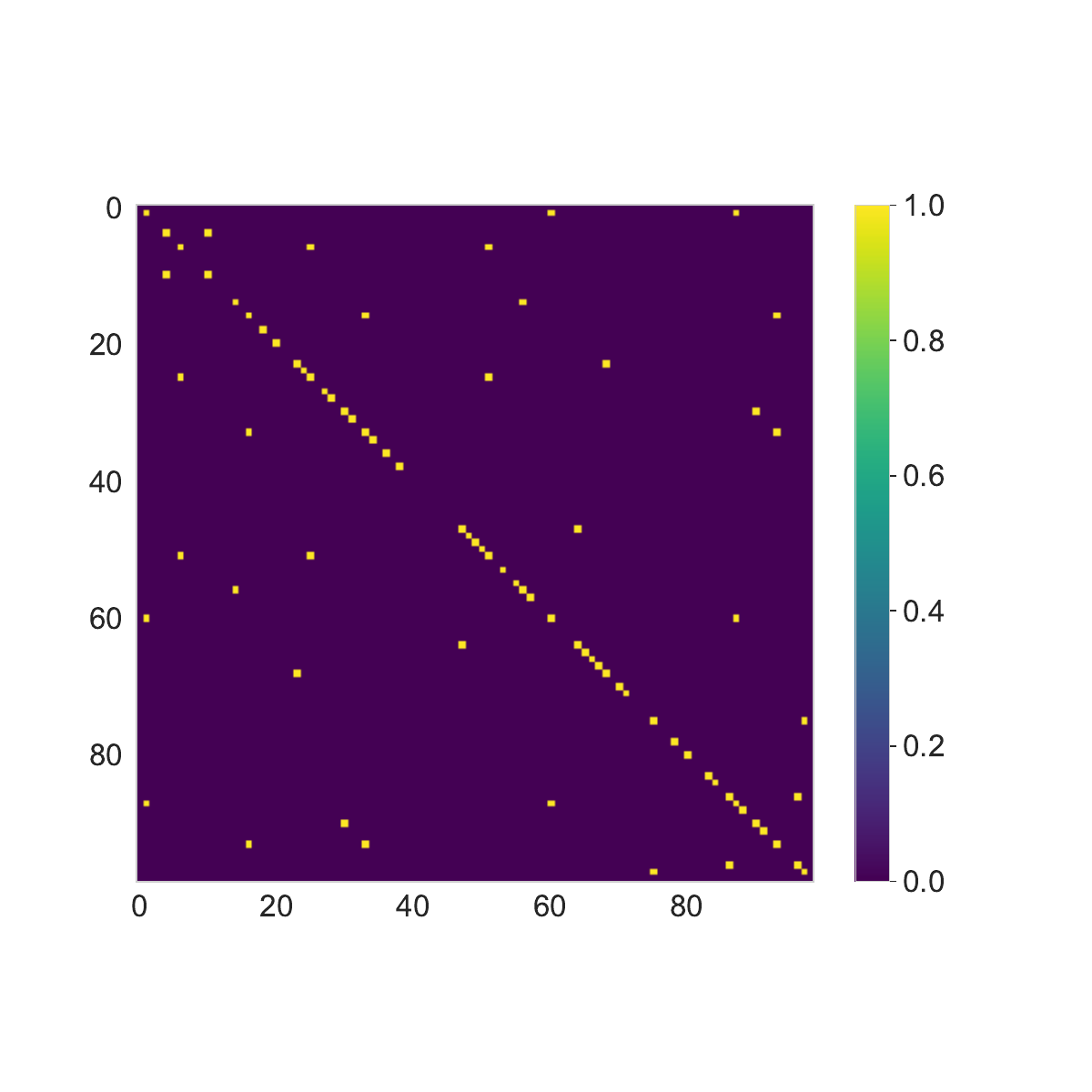}
               \subcaption{Segment (BICS level 5)}
    \end{subfigure}\hfill
    %\begin{subfigure}{0.25\textwidth}
     %   \centering
      %  \includegraphics[width=\textwidth, angle=0, trim= 50 50 50 50,clip]{img/empirical_application/data/covariate_6_BICS_LEVEL_6_SEGMENT_NAME.pdf}
       %        \caption{Segment}
    %\end{subfigure}
    % Second part of the image
   
    %\begin{subfigure}{0.48\textwidth}
     %   \centering
     %   \includegraphics[width=\textwidth, angle=0, trim= 50 50 50 50,clip]{img/empirical_application/data/covariate_7_BICS_LEVEL_7_SEGMENT_NAME.pdf}
    %\end{subfigure}\vfill
    %        \begin{subfigure}{0.48\textwidth}
    %    \centering
    %    \includegraphics[width=\textwidth, angle=0, trim=50 50 50 50,clip]{img/empirical_application/data/covariate_8_BICS_LEVEL_8_SEGMENT_NAME.pdf}
    %\end{subfigure}\hfill
    % Second part of the image
    \begin{subfigure}{0.25\textwidth}
        \centering
        \includegraphics[width=\textwidth, angle=0, trim= 50 50 50 50,clip]{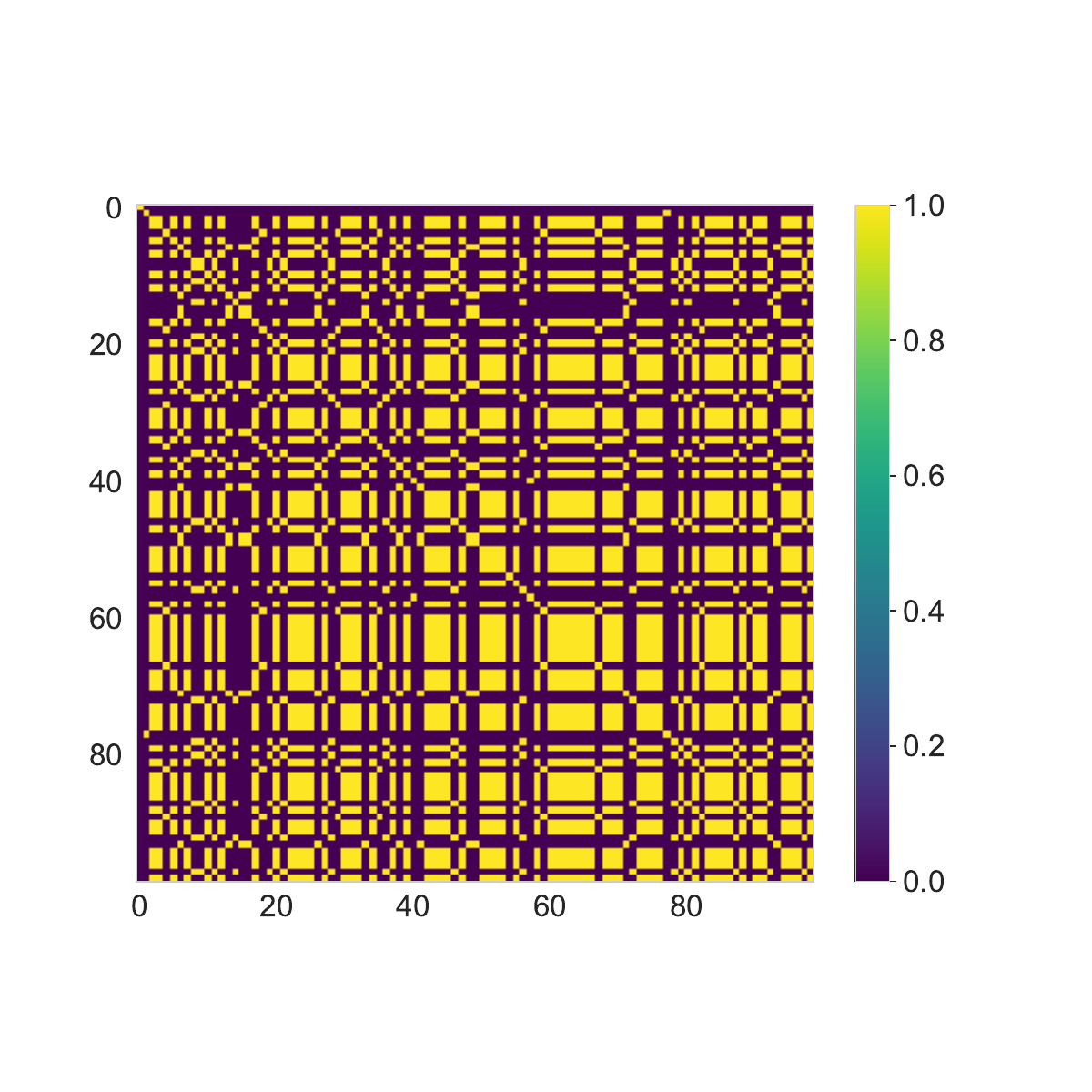}
               \subcaption{Country of risk}
    \end{subfigure}
        % First part of the image
    \begin{subfigure}{0.25\textwidth}
        \centering
        \includegraphics[width=\textwidth, angle=0, trim= 50 50 50 50,clip]{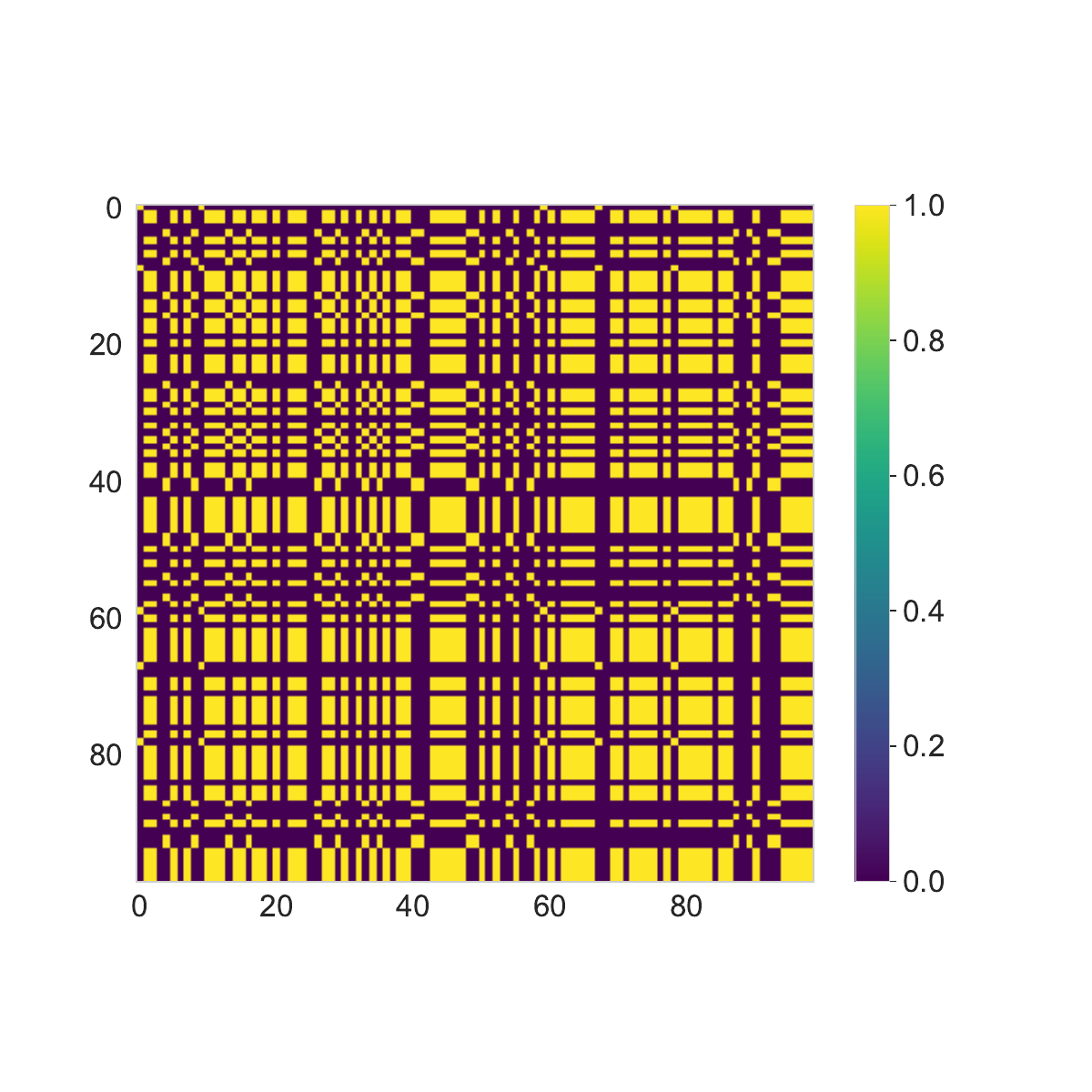}
        \subcaption{Region of largest revenue}
    \end{subfigure}
    % Second part of the image
    \begin{subfigure}{0.25\textwidth}
        \centering
        \includegraphics[width=\textwidth, angle=0, trim=  50 50 50 50,clip]{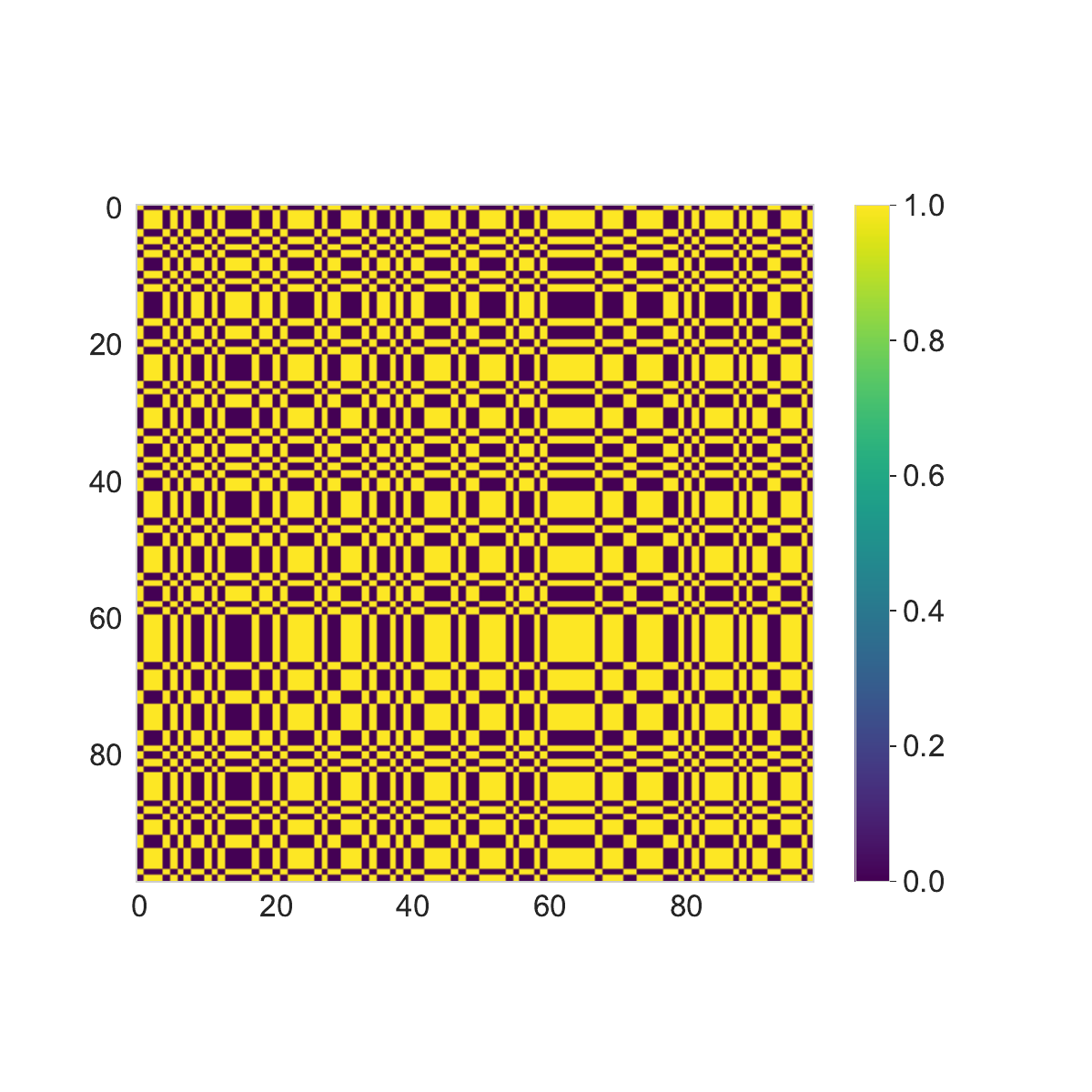}
        \subcaption{Entity region}
    \end{subfigure}
    \caption{Selection of categorical covariates.}
    \label{fig:cat_covariates}
\end{figure}

We select firms with sufficient CDS data between 25 January 2004 and 25 January 2022 from constituents of the largest indices: S\&P 500 (SPX), Deutscher Aktien Index (DAX), Euro STOXX 50 (SX5E), and FTSE 100 (UKX). To select categorical variables, we look at the most distinct indicators (that is, those with lower correlation with other candidates). The categorical covariates considered are shown in Figure \ref{fig:cat_covariates}. For categorical data, we assign an edge a value of $1$ if the node categories match (i.e. if both companies share the same value) and $0$ otherwise. Each plot is a binary matrix, where the row-column pair corresponds to the edge $W_{k' \rightarrow k}$ of the network. The value $1$ indicates that the categorical variable of node $k'$ matches that of node $k$, and $0$ indicates the opposite. Each covariate results in a single binary value for an edge in the asset similarity channel. Selected covariates were sector, industry and region of largest revenue.

\begin{figure}[htbp!]
    \centering
    \includegraphics[width=0.55\textwidth, angle=0, trim=150 450 80 450, clip]{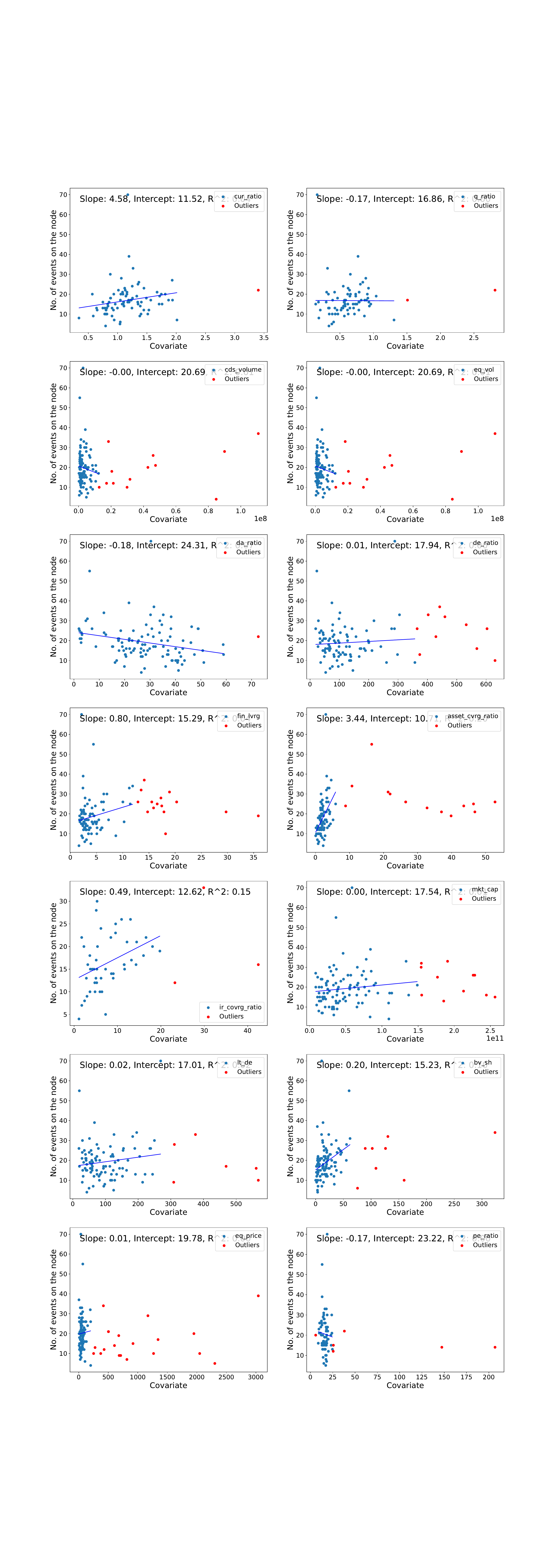}
    \caption{Linear regression of events on covariates prior to transformation.}
    \label{fig:regression_pre_outliers}
\end{figure}

\begin{figure}[htbp!]
    \centering
    \includegraphics[width=0.55\textwidth, angle=0, trim=150 450 80 450, clip]{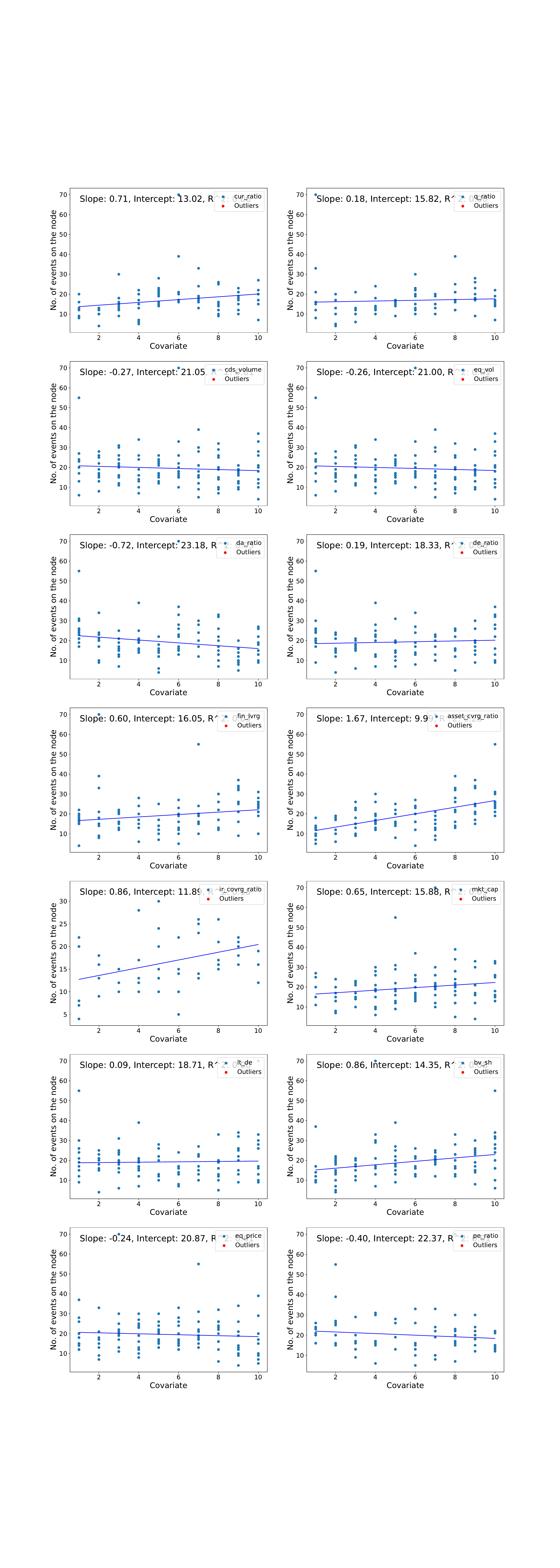}
    \caption{Linear regression of events on covariates post transformation.}
    \label{fig:regression_outliers}
\end{figure}

To identify variables for the other channel-specific weight regressions, we consider data availability throughout the sample period and examine descriptive relationships between event counts and candidate variables. Because financial ratios contain substantial outliers (Figure \ref{fig:regression_pre_outliers}), we transform them into values from $1$ to $10$ according to decile intervals. Figure \ref{fig:regression_outliers} displays the transformed-variable comparisons considered during exploratory selection.

For the selected transformed variables used in the final application, solvency and profitability are moderately positively associated across firms (Spearman's $\rho=0.36$; Pearson $r=0.36$, $n=99$). This relationship is relevant when interpreting attenuation of channel-specific coefficient estimates in joint model set-ups.

The full set of nodes with corresponding selected covariates is provided in Table \ref{tab:scenarios_nodes_rd}.

\begin{table}[htbp!]
\resizebox{\textwidth}{!}{%
\begin{tabular}{lllllllll}
\hline
\textbf{Node} & \textbf{Index} & \textbf{Name}                            & \textbf{SECTOR}        & \textbf{INDUSTRY}              & \textbf{REGN OF LARGEST REV}   & \textbf{\# Events} & \textbf{Solv} & \textbf{Prof} \\ \hline
0             & UKX            & Anglo American PLC                       & Materials              & Metals \& Mining               & Asia Pacific                   & 20                 & 9             & 7             \\
1             & SX5E           & Koninklijke Ahold Delhaize NV            & Consumer Staples       & Retail - Consumer Staples      & The Americas                   & 15                 & 7             & 3             \\
2             & SPX            & American Electric Power Co Inc           & Utilities              & Electric Utilities             & The Americas                   & 13                 & 7             & 8             \\
3             & SPX            & American International Group Inc         & Financials             & Insurance                      &                                & 34                 & 9             & 10            \\
4             & DAX            & Airbus SE                                & Industrials            & Aerospace \& Defense           & EMEA - Europe, Middle East, Af & 17                 & 9             & 4             \\
5             & SPX            & Allstate Corp/The                        & Financials             & Insurance                      & The Americas                   & 30                 & 10            & 9             \\
6             & SX5E           & Allianz SE                               & Financials             & Insurance                      &                                & 26                 & 10            & 10            \\
\hline
\end{tabular}%
}
\caption{Nodes and selected covariate values in the empirical application.}
\label{tab:scenarios_nodes_rd}
\end{table}

\paragraph*{Data limitations}
\begin{itemize}
\item The model uses static covariates and therefore does not represent material changes in firms' balance sheets over the sample period.
\item The fixed population produces a closed-system analysis and excludes companies that defaulted during the sample period.
\item Firms without adequate CDS or covariate data are excluded, which may introduce selection bias.
\item Where a categorical value is unavailable, the corresponding match covariate is set to $0$.
\end{itemize}

\section{Stability of the Multiplex Network Hawkes process}\label{sec:stability}
This appendix records the stability condition referenced in Section \ref{sec:multiplexnetworkhawkes}. As noted by \cite{linderman2014discovering} and \cite{daley2003introduction}, Hawkes processes must be constrained to ensure their positive feedback does not lead to an infinite number of expected events. A stable single-layer system must satisfy $\lambda_{\max}=\max |\mathrm{eig}(A\odot W)|<1$. In the multiplex case used in the simulation study, the same condition is applied to the combined integrated excitation matrix, $G=\sum_{l=1}^{L} A\odot W^{(l)}$, so that $\lambda_{\max}=\max |\mathrm{eig}(G)|<1$. Here $\lambda_{\max}$ is a spectral/eigenvalue quantity and is distinct from the background intensities $\lambda^{(0)}_k$ reported in the simulation tables. When conditioning on finite datasets, placing weakly informative priors on the network parameters is generally sufficient to satisfy this constraint.

\section{Component inference}\label{sec:componentinference}

The Metropolis-within-Gibbs sampling algorithm for the Multiplex Network Hawkes model is summarized in Algorithm \ref{alg:gibbs}. The underlying component inference leverages the Poisson superposition theorem, which decomposes the Hawkes process into independent additive components by treating the unknown generative origins of each event as latent parent variables.

\begin{algorithm} [htbp!]
\caption{Metropolis-within-Gibbs sampler for the Multiplex Network Hawkes model}
\begin{algorithmic} % The number [1] indicates that lines are numbered
\State 1. Introduce auxiliary parent variables $z_{k,n}$ for each $s_{k,n}$
\State 2. Sample background rates $\lambda^{(0)}_k|\{z_{k,n}=(0,0,0)\}_{n=1}^{N_k}$
\State 3. Sample impulse response parameters $\theta_{k'\rightarrow k}|\{s_{k,n},z_{k,n}\}$
\State 4. Sample $\bm{W}^{(l)}|\{s_{k,n},z_{k,n}\}, \theta_{k'\rightarrow k},\bm{\beta^{(l)}}$
\State 5. Sample gamma-regression parameters $\bm{\beta}^{(l)}|\bm{W}^{(l)}$, $\kappa^{(l)}|\bm{W}^{(l)}$, and scale hyperparameter $b_{\kappa}^{(l)}$ using an adaptive random-walk Metropolis-Hastings sampler (see Section \ref{sec:adaptiveRWMH}).
\State 6. Sample $\bm{A}|\{s_{k,n}\},\bm{W}^{(l)},\theta_{k'\rightarrow k}$ from the marginal distribution after integrating out the parents.
\State 7. Sample parents $z_{k,n}|\{s_{k,n}\},\bm{A},\bm{W}^{(l)},\theta_{k'\rightarrow k}$
\end{algorithmic}
\label{alg:gibbs}
\end{algorithm}

To operationalize this, each observed event time $s_{k,n}$ on sub-process $k$ is augmented with a latent parent indicator $z_{k,n}$. We set $z_{k,n}=(0,0,0)$ if the event is exogenous (caused by the background rate $\lambda^{(0)}_k$), and $z_{k,n}=(k',n',l)$ if it was triggered by the $n'$-th event on sub-process $k'$ via network layer $l$. The complete-data likelihood is then:
\begin{align*}
&p\left(\{\{s_{k,n},z_{k,n}\}_{n=1}^{N_k}\}_{k=1}^K \mid \{\lambda^{(0)}_k\},\{\{h_{k' \rightarrow k}^{(l)}(\Delta t)\}_{k'\rightarrow k}\}_{l=1}^L\right) \\
&\quad = \prod_{k=1}^K p\left(\{s_{k,n}: z_{k,n}=(0,0,0)\} \mid \lambda^{(0)}_k\right) \times \\
&\qquad \prod_{k'=1}^K\prod_{n'=1}^{N_{k'}} \prod_{l=1}^L\prod_{k=1}^K p\left(\{s_{k,n}: z_{k,n}=(k',n',l)\} \mid h_{k' \rightarrow k}^{(l)}(t-s_{k',n'})\right),
\end{align*}
where expressing the components in terms of densities yields the following likelihood:

\begin{align*}
&p\left(\{\{s_{k,n},z_{k,n}\}_{n=1}^{N_k}\}_{k=1}^K \mid \{\lambda^{(0)}_k\}_{k=1}^{K},\{\{h_{k' \rightarrow k}^{(l)}(\Delta t)\}_{k'\rightarrow k}\}_{l=1}^{L}\right)=\\
&\quad \prod_{k=1}^K\left[\exp\left\lbrace-\int_{0}^T\lambda^{(0)}_k\text{d}\tau\right\rbrace \times \prod_{n=1}^{N_k}\left({\lambda_k^{(0)}}\right)^{\delta_{z_{k,n},(0,0,0)}}\right] \times \\
&\quad \prod_{k'=1}^K\prod_{n'=1}^{N_{k'}} \prod_{l=1}^L\prod_{k=1}^K \left[\exp\left\lbrace-\int_{s_{k',n'}}^T h^{(l)}_{k'\rightarrow k}(\tau - s_{k',n'})\text{d}\tau\right\rbrace \times \right. \\
&\qquad\qquad\qquad\qquad \left. \prod_{n=1}^{N_k} h^{(l)}_{k'\rightarrow k}(s_{k,n} - s_{k',n'})^{\delta_{z_{k,n},(k',n',l)}}\right].
\end{align*}

Here and elsewhere, $\delta_{i,j}$ is the Kronecker delta function (i.e. $\delta_{i,j}=1$ if $i=j$). We now specify the marginal distributions required for Gibbs sampling, together with the prior specifications and posterior updates.

\subsection{Elements of Weights Matrices}

Noting that $W_{k'\rightarrow k}^{(l)}$ only appears in the impulse responses for which $z_{k,n}=(k',n',l)$, the conditional distribution for $W_{k'\rightarrow k}^{(l)}$ is proportional to:

\begin{align*}
&p(W_{k'\rightarrow k}^{(l)}|\{\{s_{k,n},z_{k,n}\}_{n=1}^{N_k}\}_{k=1}^K,\{\lambda^{(0)}_k\}_{k=1}^{K},\{\{h_{k' \rightarrow k}^{(l)}(\Delta t)\}_{k'\rightarrow k}\}_{l=1}^{L})\propto\\
&\prod_{n'=1}^{N_{k'}}\left[\exp\left\lbrace-\int_{s_{k',n'}}^T h^{(l)}_{k'\rightarrow k}(\tau - s_{k',n'})\text{d}\tau\right\rbrace \prod_{n=1}^{N_k} h^{(l)}_{k'\rightarrow k}(s_{k,n} - s_{k',n'})^{\delta_{z_{k,n},(k',n',l)}}\right]\\
&\times p(W_{k'\rightarrow k}^{(l)})=\\
&\prod_{n'=1}^{N_{k'}}\left[\exp\left\lbrace-\int_{s_{k',n'}}^T A_{k'\rightarrow k}W^{(l)}_{k'\rightarrow k}g_{k'\rightarrow k}(\tau - s_{k',n'})\text{d}\tau\right\rbrace \prod_{n=1}^{N_k} \left(A_{k'\rightarrow k}W^{(l)}_{k'\rightarrow k}g_{k'\rightarrow k}(s_{k,n} - s_{k',n'})\right)^{\delta_{z_{k,n},(k',n',l)}}\right]\\
&\times p(W_{k'\rightarrow k}^{(l)}).
\end{align*}
If $A_{k'\rightarrow k}=1$ and we ignore spikes after $T-\Delta t_{max}$ this is proportional to:
\begin{align*}
\exp\{-W^{(l)}_{k'\rightarrow k}N_{k'}\}(W^{(l)}_{k'\rightarrow k})^{N_{k'\rightarrow k}^{(l)}}p(W^{(l)}_{k'\rightarrow k})
\end{align*}
where $$N_{k'}$$ is the total number of events on node $k'$, and $$N_{k'\rightarrow k}^{(l)}=\sum_{n=1}^{N_k}\sum_{n'=1}^{N_{k'}} \delta_{z_{k,n},(k',n',l)}$$ is the number of events caused by an interaction on network layer $l$.
%\prod_{n=1}^N \prod_{l=1}^L\prod_{k=1}^K \left\langle\exp\left\lbrace-\int_{s_n}^T h^{(l)}_{c_n\rightarrow k'}(\tau - s_n)\text{d}\tau\right\rbrace \prod_{n'=1}^N h^{(l)}_{c_n\rightarrow c_{n'}}(s_{n'} - s_n)^{\delta_{c_{n'},k'}\delta_{z_{n'},(n,l)}}

When $p(W^{(l)}_{k'\rightarrow k})$ is a gamma distribution, the conditional distribution is also gamma. If $A_{k'\rightarrow k}=0$, the conditional distribution reduces to the prior.

\begin{align}
W^{(l)}_{k'\rightarrow k}|A_{k'\rightarrow k}=1, &\sim \mbox{Gamma}\left(\alpha_W^{(l)},\beta_W^{(l)}\right)\notag\\
p(\rho|\bm{A})&\propto \prod_{k'=1}^K\prod_{k=1}^K \rho^{A_{k'\rightarrow k}}(1-\rho)^{1-A_{k'\rightarrow k}} \rho^{\alpha^0_{\rho}-1}(1-\rho)^{\beta^0_{\rho}-1}\\
\beta_W^{(l)}&={\beta_W^0}^{(l)}+w^{(l)}\sum_{n'=1}^N \delta_{c_{n'},k'}\label{eq:beta_post2}
\end{align}

Because the weights depend on external covariates, we use gamma regression; instead of $W^{(l)}_{k'\rightarrow k}\sim \mbox{Gamma}(\alpha^0_{W},\beta^0_{W})$ we specify:

\begin{align*}
W^{(l)}_{k'\rightarrow k} &\sim  \mbox{Gamma}\left(\frac{1}{\kappa^{(l)}},\frac{1}{\kappa^{(l)} \mu_{k'\rightarrow k}^{(l)}}\right)\\
g(\mu_{k'\rightarrow k}^{(l)})&=\eta_{k'\rightarrow k}^{(l)}={\bm{x'}_{k'\rightarrow k}}^{(l)}\bm{\beta}^{(l)},
\end{align*}
where $\alpha_{W}^0 = \frac{1}{\kappa}$ and $\beta_{W}^0=\frac{1}{\kappa \mu_{k'\rightarrow k}}.$ The elements of the vector ${\bm{x'}_{k'\rightarrow k}}^{(l)}$ are functions of covariates corresponding to nodes $k'$ and $k$ in model layer $l$, whose form depends on the layer type; the elements of the vector $\bm{\beta^{(l)}}$ are the corresponding regression coefficients.

%\begin{align*}
%W^{(l)}_{k'\rightarrow k} &\sim  \mbox{Gamma}(\alpha_{W_{k'\rightarrow k}}^{(l)},\beta_{W}^{(l)})\\
%\mu_{k'\rightarrow k}^{(l)} &= \frac{\alpha_{W_{k' \rightarrow k}}^{(l)}}{\beta_W^{(l)}}\\
%g(\mu_{k'\rightarrow k}^{(l)})&=\eta_{k'\rightarrow k}^{(l)}={\bm{x'}_{k'\rightarrow k}}^{(l)}\bm{\gamma}^{(l)},
%\end{align*}

The gamma density in terms of $\kappa$ and $\mu$ can be written as:
$$f(y)=\frac{1}{\Gamma\left(\frac{1}{\kappa}\right)(\kappa\mu)^{1/\kappa}}y^{\frac{1}{\kappa}-1}e^{-\frac{1}{\kappa\mu}y}.$$
Thus, the likelihood is:
$$\prod_{i=1}^{K\times K}\frac{1}{\Gamma\left(\frac{1}{\kappa^{(l)}}\right)(\kappa^{(l)}\mu_{i}^{(l)})^{1/\kappa^{(l)}}}{W_{i}^{(l)}}^{\frac{1}{\kappa^{(l)}}-1}e^{-\frac{1}{\kappa^{(l)}\mu_{i}^{(l)}}W_{i}^{(l)}}$$

The posteriors $p\left(\bm{\beta^{(l)}}|\kappa^{(l)},\bm{W^{(l)}},\bm{x}^{(l)}\right)$ and $p\left(\kappa^{(l)}|\bm{\mu^{(l)}},\bm{W^{(l)}}\right)$ and their log-likelihoods $\ell(\bm{\beta^{(l)}})$ and $\ell(\kappa^{(l)})$ are then defined as follows:
\begin{align}
\label{eq:beta_post}
p\left(\bm{\beta^{(l)}}|\kappa^{(l)},\bm{W^{(l)}},\bm{x}^{(l)}\right)&=\prod_{i=1}^{K\times K}p\left(W_{i}^{(l)}|\kappa^{(l)},\bm{\beta^{(l)}},\bm{x}^{(l)}\right)p\left(\bm{\beta^{(l)}}\right)\\
&\propto \prod_{i=1}^{K\times K}\frac{1}{(\mu_{i}^{(l)})^{1/\kappa^{l}}}e^{-\frac{1}{\kappa^{(l)}\mu_{i}^{(l)}}W_{i}^{(l)}}p\left(\bm{\beta^{(l)}}\right) \\
\ell(\bm{\beta^{(l)}})&=-\frac{1}{\kappa^{(l)}}\sum_{i=1}^{K\times K}\left(ln(\mu_i^{(l)})+\frac{W_i^{(l)}}{\mu_{i}^{(l)}}\right)+ln\left(p\left(\bm{\beta^{(l)}}\right)\right)
\end{align}
\begin{align}
\label{eq:kappa_post}
p\left(\kappa^{(l)}|\bm{\mu^{(l)}},\bm{W^{(l)}}\right)&=\prod_{i=1}^{K\times K}p\left(W_{i}^{(l)}|\kappa^{(l)},\bm{\mu^{(l)}}\right)p\left(\kappa^{(l)}\right)\\
&\propto  \prod_{i=1}^{K\times K}\frac{1}{\Gamma\left(\frac{1}{(\kappa^{(l)})}\right){\mu_i^{(l)}\kappa^{(l)}}^{1/\kappa^{(l)}}}{W_{i}^{(l)}}^{\frac{1}{\kappa^{(l)}}-1}e^{-\frac{1}{\kappa^{(l)}\mu_{i}^{(l)}}W_{i}^{(l)}}p\left(\kappa^{(l)}\right)\\
\ell(\kappa^{(l)})&=-(K\times K)ln\left(\Gamma\left(\frac{1}{\kappa^{(l)}}\right)\right)-\frac{1}{\kappa^{(l)}}\sum_{i=1}^{K\times K}ln(\mu_i^{(l)}\kappa^{(l)})\\
&+\sum_{i=1}^{K\times K}\left( \left(\frac{1}{\kappa^{(l)}}-1\right)ln(W_i^{(l)})-\frac{1}{\kappa^{(l)}\mu_i^{(l)}}W_i^{(l)}\right)+ln\left( p\left(\kappa^{(l)}\right) \right)
\end{align}

In our application, we use adaptive random-walk Metropolis-Hastings methods with ASM by \citet{atchade2005adaptive} and ASWAM by \citet{andrieu2008tutorial,atchade2010limit}. %and Generalised Horseshoe (GHS) prior.

For adaptive scale (ASM) and adaptive scaling within adaptive Metropolis-Hastings (ASWAM) algorithms, we specify the priors as follows:

\begin{align*}
\kappa^{(l)} &\sim \mbox{HalfCauchy}(a_{\kappa}^{(l)},b_{\kappa}^{(l)}), \mbox{InvGamma}(a_{\kappa}^{(l)},b_{\kappa}^{(l)}),  \text{ or } \mbox{Gamma}(a_{\kappa}^{(l)},b_{\kappa}^{(l)})\\
\beta^{(l)} & \sim \mbox{MultivariateNormal}(\bm{\mu_{\beta}},\Sigma^{(l)}).
\end{align*}

Because the sampler can be sensitive to the scale parameter, we also allow $b_{\kappa}^{(l)}$ to be estimated using a Metropolis-Hastings step with analogous priors, i.e. $$b_{\kappa}^{(l)} \sim \mbox{HalfCauchy}(a_{\kappa_b},b_{\kappa_b}),\mbox{InvGamma}(a_{\kappa_b},b_{\kappa_b}),\mbox{Gamma}(a_{\kappa_b},b_{\kappa_b}).$$

We use the first 500 ASM updates to estimate the initial covariance matrices. In our application, un-normalised posteriors defined in Eq. \ref{eq:beta_post} and Eq. \ref{eq:kappa_post} are used in the acceptance ratio. Given the initial covariance estimated from these ASM updates, we use ASWAM for the remaining sampler updates.

\subsection{Elements of Adjacency Matrix}

With Aldous-Hoover graph priors, the entries in the binary adjacency matrix $\bm{A}$ are conditionally independent given the parameters of the prior. The likelihood introduces dependencies between the rows of $\bm{A}$, but each column can be sampled in parallel. Gibbs updates are complicated due to dependencies between graph and parent variables $z_{k,n}$, i.e. if $z_{k,n}=(k',n',l)$, then we must have $A_{k'\rightarrow k}=1$. Therefore, to improve mixing, we first update $\bm{A}|\{s_{k,n}\},\bm{\omega},\bm{W},\theta_{k'\rightarrow k}$ by marginalising the parent variables. The posterior is determined by the likelihood of the conditionally Poisson process $\lambda_{k}^{(l)}(t|\{s_{k,n}:s_{k,n}<t\})$ with and without the interaction $A_{k'\rightarrow k}$, together with the prior from the Aldous-Hoover graph model:
\begin{align*}
A_{k'\rightarrow k}=1 &\sim \mbox{Bern}(f).
\end{align*}
where in the simplest case, $f=\rho \sim \mbox{Beta}(\alpha^0_{\rho},\beta^0_{\rho})$, e.g. $\alpha^0_{\rho}=1$ and $\beta^0_{\rho}=1$.

The parameter $\rho$ is updated as follows:
\begin{align*}
\rho|\bm{A} &\sim \mbox{Beta}(\alpha_{\rho},\beta_{\rho})\\
p(\rho|\bm{A})&\propto \prod_{k'=1}^K\prod_{k=1}^K \rho^{A_{k'\rightarrow k}}(1-\rho)^{1-A_{k'\rightarrow k}} \rho^{\alpha^0_{\rho}-1}(1-\rho)^{\beta^0_{\rho}-1}\\
&=\rho^{\sum_{k'=1}^K\sum_{k=1}^K A_{k'\rightarrow k}+\alpha^0_{\rho}-1}(1-\rho)^{\sum_{k'=1}^K\sum_{k=1}^K(1- A_{k'\rightarrow k})+\beta^0_{\rho}-1}\\
\alpha_{\rho}&=\alpha^0_{\rho}+\sum_{k'=1}^K\sum_{k=1}^K A_{k'\rightarrow k}\\
\beta_{\rho}&=\beta^0_{\rho}+\sum_{k'=1}^K\sum_{k=1}^K(1- A_{k'\rightarrow k}).
\end{align*}

\subsection{Latent Indicator Variables}\label{sec:latentvar}
Given updates to $\bm{A}$, we then update parent variables $z_{k,n}|\{s_{k,n}\},\bm{A},\bm{W}^{(l)},\theta_{k'\rightarrow k}$ by sampling from a categorical distribution:
\begin{align*}
z_{k,n}| \bm{p} &\sim Cat((L+1),\bm{p})\\
\bm{p}&=(p_{k}^{(0,0,0)},\ldots,p_{k}^{(k',l)})\\
p_{k}^{(0,0,0)}&=\frac{\lambda_{k}^{(0)}}{\lambda_{k}}\\
p_{k}^{(k',l)}&=\frac{A_{k'\rightarrow k}W^{(l)}_{k'\rightarrow k}g_{k' \rightarrow k}(\Delta t)}{\lambda_{k}},
\end{align*}
where $z_{k,n}=(k',n',l)$ indicates the $n'$-th event on process $k'$ that caused event $s_{k,n}$ on process $k$ via layer $l$.

Thus:
\begin{align*}
p(\{\{z_{k,n}\}\}|\{\lambda_k^{(0)}\}_{k=1}^{K},\{\{h_{k' \rightarrow k}^{(l)}\}\}_{l=1}^L)\propto\prod_{k=1}^K\prod_{n=1}^{N_k}\left[ {p_{k}^{(0,0,0)}}^{\delta_{z_{k,n},(0,0,0)}}\prod_{l=1}^L \prod_{k'=1}^K{p_{k}^{(k',l)}}^{\delta_{z_{k,n},(k',n',l)}} \right].
\end{align*}

\subsubsection{Background Rates}
Conjugate Gamma priors can be used for constant background rates as follows:
\begin{align*}
\lambda_k^{(0)} &\sim \mbox{Gamma} (\alpha_{\lambda},\beta_{\lambda})\\
\lambda_k^{(0)}|\{z_{k,n}\} &\sim \mbox{Gamma}(\alpha_{\lambda}^{(k)},\beta_{\lambda}^{(k)})\\
\alpha_{\lambda}^{(k)}&=\alpha^0_{\lambda}+ \sum_{n=1}^{N_k}  \delta_{z_{k,n},{(0,0,0)}}\\
\beta_{\lambda}^{(k)}&=\beta^0_{\lambda}+ T_{k}
\end{align*}

\subsection{Interaction Kernel}

A normal-gamma prior is used for the parameters of the logistic normal distribution, $\mu$ and $\tau$:
$$\mu_{k'\rightarrow k}, \tau_{k'\rightarrow k} \sim \mathcal{NG}(\mu_{k'\rightarrow k}, \tau_{k'\rightarrow k}|\mu_{\mu}^0,k_{\mu}^0,\alpha_{\tau}^0,\beta_{\tau}^0).$$
This prior is conjugate with the likelihood. Note that $g_{k'\rightarrow k}(\Delta t)$ only appears in the term below:
\begin{align*}
\prod_{k'=1}^K \prod_{l=1}^L\prod_{n'=1}^{N_{k'}}\left[\exp\left\lbrace-\int_{s_{k',n'}}^T A_{k'\rightarrow k}W^{(l)}_{k'\rightarrow k}g_{k'\rightarrow k}(\tau - s_{k',n'})\text{d}\tau\right\rbrace \prod_{n=1}^{N_k} \left( A_{k'\rightarrow k}W^{(l)}_{k'\rightarrow k}g_{k' \rightarrow k}(s_{k,n} - s_{k',n'}) \right)^{\delta_{z_{k,n},(k',n',l)}}\right].
\end{align*}

Since each impulse response $g_{k' \rightarrow k}$ is a probability density function, its integral is one when the triggering event is earlier than $T-\Delta t_{max}$. Thus, when $\Delta t_{max} << T$, we can ignore spikes that occur at the very end of the dataset. The resulting likelihood is therefore proportional to a product of logistic normal densities. Because the logistic function is invertible, we work with logit-transformed intervals instead. The likelihood is then a product of normal densities, which is conjugate with the normal-gamma prior. Thus:
\begin{align*}
\mu_{k'\rightarrow k}, \tau_{k'\rightarrow k}| \{s_{k,n},z_{k,n}\}, \mu_{\mu}^0,\kappa_{\mu}^0,\alpha_{\tau}^0,\beta_{\tau}^0 \sim \mathcal{N}(\mu_{k'\rightarrow k},(\kappa_{\mu} \tau_{k'\rightarrow k})^{-1}) \times \mbox{Gamma}(\tau_{k'\rightarrow k}|\alpha_{\tau},\beta_{\tau})
\end{align*}
where
\begin{align*}
\mu_{\mu}&=\frac{\kappa_{\mu}^0\mu_{\mu}^0+m\bar{x}}{\kappa_{\mu}^0+m},\\
\kappa_{\mu}&=\kappa_{\mu}^0+m,\\
\alpha_{\tau}&=\alpha_{\tau}^0+\frac{m}{2},\\
x_{n'\rightarrow n}&=ln(s_{k,n}-s_{k',n'})-ln(\Delta t_{max}-(s_{k,n}-s_{k',n'})),\\
m&=\sum_{l=1}^{L}\sum_{n'=1}^{N_{k'}}\sum_{n=1}^{N_k}\delta_{z_{k,n},(k',n',l)},\\
\bar{x}&=\frac{1}{m}\sum_{l=1}^{L}\sum_{n'=1}^{N_{k'}}\sum_{n=1}^{N_k}\delta_{z_{k,n},(k',n',l)}x_{n' \rightarrow n},\\
\beta_{\tau}&=\beta_{\tau}^0+\frac{1}{2}\sum_{l=1}^{L}\sum_{n'=1}^{N_{k'}}\sum_{n=1}^{N_k}\delta_{z_{k,n},(k',n',l)}(x_{n' \rightarrow n}-\bar{x})^2+\frac{\kappa_{\mu}^0m(\bar{x}-\mu_{\mu}^0)^2}{2(\kappa_{\mu}^0+m)}.
\end{align*}

\section{Adaptive Random Walk Metropolis-Hastings Algorithms} \label{sec:adaptiveRWMH}

This appendix provides implementation details for the adaptive Metropolis-Hastings steps used to sample gamma-regression parameters for edge weights (see Section \ref{sec:bayesianinference}). We use the first 500 ASM updates to estimate the initial covariance matrices. The algorithm steps are given below.

\begin{enumerate}
\item Initialise starting value $\bm{\theta}$ (i.e. $\bm{\beta^{(l)}}$, $\kappa^{(l)}$ and $b_{\kappa}^{(l)}$).
\item For $g=1,2,\ldots,n$:
\begin{enumerate}
\item Sample $\bm{\theta}'=\bm{\theta}_g+\bm{\epsilon}_g$, where $\bm{\epsilon}_g\sim \mbox{N}(0,\bm{\zeta}^g)$
\item Let the next iterate be 
\begin{equation}
\bm{\theta}_{g+1}=\begin{cases} \bm{\theta}' \text{ with probability } \alpha_{\bm{\zeta}
_g}(\bm{\theta}_g,\bm{\theta'}|\bm{y})\\
\bm{\theta}_g \text{ with probability } 1-\alpha_{\bm{\zeta}
_g}(\bm{\theta}_g,\bm{\theta'}|\bm{y})\\
\end{cases}
\end{equation}
\item Update the proposal variance as
$$\bm{\zeta}_{g+1}=\rho\left(\zeta_g+w_g(\alpha_{\bm{\zeta}
_g}(\bm{\theta}_g,\bm{\theta'}|\bm{y})-\bm{\hat\tau})\right)$$
where  $w_g=O(g^{-\lambda})$ where $-1/2<\lambda<1$ (we choose $w_g=g^{-0.5}$)
\item The function $\rho$ is defined to stabilise the algorithm and defines a range of possible values for $\bm{\zeta}$.
\begin{equation}
\rho=\begin{cases}
\bm{\zeta}^{\min} \text{ if } \bm{\zeta}<\bm{\zeta}^{\min}\\
\bm{\zeta} \text{ if } \bm{\zeta}\in (\bm{\zeta}^{\min},\bm{\zeta}^{\max})\\
\bm{\zeta}^{\max} \text{ if } \bm{\zeta}>\bm{\zeta}^{\max}
\end{cases}
\end{equation}
where $\bm{\zeta}^{\max}$ and $\bm{\zeta}^{\min}$ are defined as 
$$\bm{\zeta}^{max}=\min(cap,\bm{\zeta}+\delta\bm{\zeta})$$  and $$\bm{\zeta}^{min}=\max(floor,\bm{\zeta}-\delta\bm{\zeta})$$ for appropriate cap/floor and $\delta$ values.
\item Acceptance ratio for MH is defined as follows: 
$$\alpha_{\bm{\zeta}
_g}(\bm{\theta}_g,\bm{\theta'}|\bm{y})=\min\left(1,\frac{\pi(\bm{\theta}'|\bm{y})q(\bm{\theta}',\bm{\theta}|\bm{y})}{\pi(\bm{\theta}|\bm{y})q(\bm{\theta},\bm{\theta}'|\bm{y})}\right)$$
In cases where we use normal random walk proposals, it is simply
$$\alpha_{\bm{\zeta}
_g}(\bm{\theta}^g,\bm{\theta'}|\bm{y})=\min\left(1,\frac{\pi(\bm{\theta}'|\bm{y})}{\pi(\bm{\theta}|\bm{y})}\right)$$
In cases where we use a log-normal proposal, we need to adjust the ratio using the Jacobian of the transformation, which simplifies to:
$$\alpha_{\bm{\zeta}
_g}(\bm{\theta}^g,\bm{\theta'}|\bm{y})=\min\left(1,\frac{\pi(\bm{\theta}'|\bm{y})\bm{\theta}}{\pi(\bm{\theta}|\bm{y})\bm{\theta}'}\right)$$
\end{enumerate}
\end{enumerate}

In our application, un-normalised posteriors defined in Eq. \ref{eq:beta_post} and Eq. \ref{eq:kappa_post} are used in the acceptance ratio.\\

Given the initial covariance estimated from the first 500 ASM updates, we use the ASWAM algorithm for the remaining updates as follows:

\begin{enumerate}
\item Initialise the starting value $\bm{\theta}$ (i.e. $\bm{\beta^{(l)}}$, $\kappa^{(l)}$ and $b_{\kappa}^{(l)}$) based on the final ASM update. $\bm{\zeta}_{0}$ is initialised using the variance of the ASM updates. ${\bm{s}_d}_{0}$ is initialised with a suitable small value.
\item For $g=n+1,n+2,\ldots,N$:
\begin{enumerate}
\item Sample $\bm{\theta}'=\bm{\theta}_g+\bm{\epsilon}_g$, where $\bm{\epsilon}_g\sim \mbox{N}(0,\bm{\zeta}_g)$
\item Let the next iterate be 
\begin{equation}
\bm{\theta}_{g+1}=\begin{cases} \bm{\theta}' \text{ with probability } \alpha_{\bm{\zeta}
_g}(\bm{\theta}_g,\bm{\theta'}|\bm{y})\\
\bm{\theta}_g \text{ with probability } 1-\alpha_{\bm{\zeta}
_g}(\bm{\theta}_g,\bm{\theta'}|\bm{y})\\
\end{cases}
\end{equation}
\item Compute new optimal variance in two steps
\begin{itemize}
\item Variance scalar multiplier:
$${\bm{s}_d}_{g+1}=\rho\left({\bm{s}_d}_g+w_g(\alpha_{{\bm{s}_d}
_g}(\bm{\theta}_g,\bm{\theta'}|\bm{y})-\bm{\hat\tau})\right)$$
where  $w_g=O(g^{-\lambda})$ where $1/2<=\lambda<=1$ (we choose $w_g=g^{-0.5}$)
\item Proposal variance:
$$\bm{\zeta}_{g+1}={\bm{s}_d}_{g+1}\frac{1}{g-1}\left(\sum_{j=1}^g\bm{\theta}_j,\bm{\theta}_j^T-\frac{\left(\sum_{j=1}^g\bm{\theta}_j\right)\left(\sum_{j=1}^g(\bm{\theta}_j\right)^T}{g}\right)+{\bm{s}_d}_{g+1}\epsilon\bm{I}_d$$
where $\epsilon$ is a small, positive constant and $\bm{I}_d$ is the $d$-dimensional identity matrix.
\end{itemize}

\item The variance updates are made more efficient using the one-step updates below, reusing quantities estimated previously. The updates use recursive relationships for the mean and variances as follows:
\begin{itemize}
\item The mean of samples $\bm{m}$ is updated as

$$\mathbf{m}_{g+1} = \frac{g}{g+1}\mathbf{m}_{g} + \frac{1}{g+1}\left(\mathbf{x}_{g+1}\right)= \mathbf{m}_{g}  +  \frac{1}{g+1}\left(\mathbf{x}_{g+1}-\mathbf{m}_{g}\right)$$

\item The covariance is then updated as follows
\begin{align}
\mathbf{\zeta}_{g+1} &= \frac{g}{g+1}\mathbf{\zeta}_{g} + \frac{1}{g+1}\left( 
\left(\mathbf{x}_{g+1} - \mathbf{m}_{g}\right)\left(\mathbf{x}_{g+1} - \mathbf{m}_{g+1}\right)^\top\right)\\
&=\mathbf{\zeta}_{g} + \frac{1}{g+1}\left( 
\left(\mathbf{x}_{g+1} - \mathbf{m}_{g}\right)\left(\mathbf{x}_{g+1} - \mathbf{m}_{g+1}\right)^\top-\mathbf{\zeta}_{g}\right)\\
s_{g+1} &= \frac{g-1}{g}s_{g} + \frac{1}{g+1}\left( 
\left(\mathbf{x}_{g+1} - \mathbf{m}_{g}\right)\left(\mathbf{x}_{g+1} - \mathbf{m}_{g+1}\right)^\top\right)\\
&=s_{g} + \frac{1}{g+1}\left( 
\left(\mathbf{x}_{g+1} - \mathbf{m}_{g}\right)\left(\mathbf{x}_{g+1} - \mathbf{m}_{g+1}\right)^\top-\frac{1}{g}s_{g}\right)
\end{align}

where $\mathbf{x}_{g+1}$ is the sample generated at step $g+1$, $s_g$ is the unbiased sample variance and $\bm{\zeta}_g$ is the biased sample variance.

\item However, to update covariances in a numerically stable way, we utilise Welford's algorithm:

$$M_g \triangleq \sum_{i=1}^{g} {(x_i - \overline{x}_g)}^2$$ 

or in other words

$$s_g = \frac{M_g}{g-1}$$

and thus an unbiased sample variance to be used in ASWAM can be calculated as:

$$M_{g+1} = M_{g} + (x_g - \overline{x}_{g+1})(x_g - \overline{x}_g)^\top$$
with 
$$s_{g+1}= \frac{M_{g+1}}{g}.$$

\end{itemize}
\end{enumerate}
\end{enumerate}

\subsection{Algorithm Stabilisation}\label{sec:algorithmstabilisation}
For edges where $A_{k'\rightarrow k}=0$, the posterior $p(W_{k'\rightarrow k}^{(l)}|\ldots)$ equals the prior: edges with no recorded events contain no information about weight magnitudes. However, under the specification above, they still contribute to estimation of the hyperparameters $\kappa$ and $\bm{\beta}$. Because the inferred contagion network is expected to be sparse, this can cause numerical instability. To stabilise hyperparameter estimation, we integrate out zero edges when estimating $\bm{\beta}$ and $\kappa$, giving the following log-posteriors:
\begin{align*}
\ell(\bm{\beta^{(l)}})&=-\frac{1}{\kappa^{(l)}}\sum_{i=1}^{K\times K}\left(ln(\mu_i^{(l)})+\frac{W_i^{(l)}}{\mu_{i}^{(l)}}\right)^{\delta_{A_{i},1}}+ln\left(p\left(\bm{\beta^{(l)}}\right)\right)
\end{align*}
and similarly for $\kappa$:
\begin{align*}
\ell(\kappa^{(l)})&=-\sum_{i=1}^{K\times K}\left(\delta_{A_{i},1}ln\Gamma\left(\frac{1}{\kappa^{(l)}}\right)\right)-\frac{1}{\kappa^{(l)}}\sum_{i=1}^{K\times K}ln(\mu_i^{(l)}\kappa^{(l)})^{\delta_{A_{i},1}}\\
&+\sum_{i=1}^{K\times K}\left( \left(\frac{1}{\kappa^{(l)}}-1\right)ln(W_i^{(l)})-\frac{1}{\kappa^{(l)}\mu_i^{(l)}}W_i^{(l)}\right)^{\delta_{A_{i},1}}+ln\left( p\left(\kappa^{(l)}\right) \right)
\end{align*}

\section{Label Switching} \label{sec:labelswitching}
Label switching occurs when the likelihood of a statistical model is invariant to permutations of component labels, making the component-specific parameters non-identifiable.

In our problem setting, the likelihood is invariant to permutations of layer labels only when layers have no covariates or equal/equivalent covariate specifications. When covariates differ, the likelihood is not invariant to these permutations, although label-switching behaviour can still occur with smaller samples. As in mixture models, the Multiplex Network Hawkes model uses a latent indicator $z_{k,n}=(k',n',l)$ that identifies the process and layer attributed to event $s_{k,n}$. We therefore use the ECR approach of \citet{papastamoulis2014handling} when relabelling is required, with implementation details following the R package documentation in \citet{papastamoulis2016label}.

ECR partitions allocation vectors into equivalence classes, where two allocations are equivalent if one can be obtained from the other by permuting labels. A representative is selected for each class, and at each sampler iteration a permutation reorders the allocation to its selected representative \citep{papastamoulis2016label}.

Two extensions dynamically adjust the pivot using previous iterations. We consider the three ECR versions documented by \citet{papastamoulis2016label}:
\begin{itemize}
    \item ECR default version, where the pivot is fixed at the start of the algorithm (see Algorithm \ref{alg:ecr}).
    \item ECR iterative version 1, which updates the pivot based on the mode/median of the permutations obtained in previous iterations (see Algorithm \ref{alg:ecr1}).
    \item ECR iterative version 2, which uses the classification probabilities from the sampler to update the pivot allocation vector upon each iteration (see Algorithm \ref{alg:ecr2}).
\end{itemize}

In our application, the pivot allocation vector is constructed from latent indicators $z_{n}$, which record the layer assigned to each event in each sampler update. The probabilities required for iterative version 2 are obtained in the parent-resampling step described in Section \ref{sec:latentvar}. In particular, from $p_{k}^{(k',l)}$ we obtain:
\begin{equation}
    p_{k}^{(l)}=\sum_{k'=0}^{K} p_{k}^{(k',l)}.
\end{equation}

\begin{algorithm}
\caption{ECR: Default version}
\begin{algorithmic}
\State Define a pivot allocation: $z^* = (z^*_1, \ldots, z^*_n)$.
\For{$t = 1, \ldots, m$}
    \State Find a permutation $\tau^{(t)} \in \mathcal{T}_{\bm{z}}$ that maximises $\sum_{i=1}^{n} I(\tau z^{(t)}_i = z^*_i)$.
\EndFor
\end{algorithmic}
\label{alg:ecr}
\end{algorithm}

\begin{algorithm}
\caption{ECR: Iterative version 1}
\begin{algorithmic}
\State Choose $m$ initial permutations $\tau^{(t)}$, $t = 1, \ldots, m$ (usually set to identity).
\State Update the pivot: $z^*_i = \text{mode}\{\tau z^{(t)}_i; t = 1, \ldots, m\}$, $i = 1, \ldots, n$ (mode replaced by median if mode is not defined). 
\For{$t = 1, \ldots, m$}
    \State Find a permutation $\tau^{(t)} \in \mathcal{T}_{K}$ that maximises $\sum_{i=1}^{n} I(\tau z^{(t)}_i = z^*_i)$.
\EndFor
\State If an improvement is made to $\sum_{t=1}^{m} \sum_{i=1}^{n} I(\tau z^{(t)}_i = z^*_i)$ go to step 2, finish otherwise.
\end{algorithmic}
\label{alg:ecr1}
\end{algorithm}

\begin{algorithm}
\caption{ECR algorithm: Iterative version 2}
\begin{algorithmic}
\State Choose $m$ initial permutations $\tau^{(t)}$, $t = 1, \ldots, m$ (usually set to identity).
\State Update the pivot: $z^*_i = \arg\max\{p^{(t)_i\tau_k}; t = 1, \ldots, m\}$, $i = 1, \ldots, n$.
\For{$t = 1, \ldots, m$}
    \State Find a permutation $\tau^{(t)} \in \mathcal{T}_{K}$ that maximises $\sum_{i=1}^{n} I(\tau z^{(t)}_i = z^*_i)$.
\EndFor
\State If an improvement is made to $\sum_{t=1}^{m} \sum_{i=1}^{n} I(\tau z^{(t)}_i = z^*_i)$ go to step 2, finish otherwise.
\end{algorithmic}
\label{alg:ecr2}
\end{algorithm}

\section{Network Measures}\label{sec:network_metric_defn}

The network measures used in our study follow definitions from the NetworkX Python library \cite{Hagberg2008}. For more comprehensive background on network measures and their interpretation, see \cite{newman2003structure} and \cite{holme2012structure}.

\subsection{Node Level Measures}

\textbf{Degree-Based Metrics.} For a node $v$, its \textbf{degree} is the number of incident edges. In directed networks, \textbf{in-degree} and \textbf{out-degree} count incoming and outgoing edges respectively. Weighted variants scale these counts by edge weights.

\textbf{Clustering Metrics.} The local \textbf{clustering} coefficient of $v$ is the proportion of actual connections between its neighbors out of all possible connections:
\[
\text{Clustering}(v) = \frac{2 T(v)}{\deg(v)(\deg(v) - 1)},
\]
where $T(v)$ is the number of triangles through $v$. \textbf{Weighted clustering} is defined as the geometric average of subgraph edge weights. \textbf{Square clustering} evaluates the probability that two neighbors of $v$ share a distinct common neighbor.

\textbf{Centrality Metrics.} \textbf{Closeness centrality} measures how quickly a node can reach all others:
\[
\text{Closeness}(v) = \frac{N - 1}{\sum_{u \neq v} d(v, u)},
\]
where $d(v, u)$ is the shortest-path distance. \textbf{Betweenness centrality} measures the fraction of all shortest paths ($\sigma_{st}$) that pass through $v$ ($\sigma_{st}(v)$):
\[
\text{Betweenness}(v) = \sum_{s \neq v \neq t} \frac{\sigma_{st}(v)}{\sigma_{st}}.
\]
\textbf{Eigenvector centrality} computes influence based on the centrality of a node's neighbors.

\subsection{Graph Level Measures}
Graph-level measures aggregate node structures to describe the entire network.

\begin{itemize}
    \item \textbf{Transitivity}: The global fraction of connected triads that form triangles.
    \item \textbf{Average Clustering}: The mean local clustering coefficient, $\frac{1}{N} \sum_{i} C_i$.
    \item \textbf{Density}: The fraction of present edges out of all possible edges.
    \item \textbf{Overall Reciprocity}: The fraction of directed edges that are bidirectional.
    \item \textbf{Degree Assortativity}: The Pearson correlation $r$ of degrees at both ends of an edge, measuring the tendency for nodes of similar degree to connect:
    \[
    r = \frac{\sum_{ij}(A_{ij} - k_i k_j / 2m) k_i k_j}{\sum_{ij}(k_i \delta_{ij} - k_i^2 / 2m) k_i k_j},
    \]
    where $A$ is the adjacency matrix, $k_i$ is the degree of node $i$, and $m$ is the total edge count.
\end{itemize}

\section{Goodness-of-Fit} \label{sec:goodnessoffit}
This appendix summarises the residual-process diagnostics used to assess fit in the simulation study and empirical application.
A standard method to assess the goodness-of-fit for Hawkes processes is residual process analysis, theoretically justified by \citet{daley2003introduction}. In particular, the random time transformation $\tau_i=\Lambda^*(t_i)=\int_0^t\lambda^*(u)\text{d} u$ takes the point process with conditional intensity $\lambda^*(t)$ and transforms it into a unit-rate Poisson process, i.e., $\tilde{N}(t)=N({\Lambda^*}^{-1}(t))$. This generalizes the well-known result that for a random variable $X$ with continuous distribution function $F$, the transformed variable $Y=F(X)$ is uniformly distributed.

Residual process analysis uses this transformation to test model fit \citep{ogata1988statistical}. If the compensator used for the time transformation perfectly matches the true generative model, the transformed residual process $\{\tau_i^{(k)}\}$ behaves exactly like a unit-rate Poisson process. Goodness-of-fit can thus be evaluated using established tests for unit-rate Poisson behavior.

Following \cite{daley2003introduction}, we assess model fit systematically:
\begin{enumerate}
\item For a given realization $\{t_1,\ldots,t_{N(T)}\}$ on a window $(0,T)$, specify a compensator $\Lambda_k^*(t)$ for process $k$.
\item Form the transformed time sequence $\{\tau_i^{(k)}\}=\Lambda_k^*(t_i)$.
\item Plot the cumulative step-function $Y_k(x)$ defined by the points $(x_i,y_i)=(\tau_i^{(k)}/\tau_T^{(k)},i/N_k(T))$ in the unit square $0\leq x,y\leq 1$.
\item Add confidence bands $y=x\pm Z_{1-\alpha}/\sqrt{\tau_T^{(k)}}$, where $\Phi(Z_p)=p$ denotes the standard normal CDF.
\item Implement an approximate $100(1-\alpha)\%$ test of the unit-rate Poisson hypothesis by checking whether the empirical process $Y_k(x)$ breaches the confidence bands from Step 4.
\end{enumerate}

Similar to a Kolmogorov-Smirnov test for probability distributions, this procedure identifies significant deviations from the expected unit-rate curve. It is an approximate test based on the large-sample Brownian motion approximation of the Poisson process, and it does not explicitly correct for the fact that parameters were estimated from the same sampled data.

A similar analysis can be performed on the transformed interarrival times, $Y_k=\tau_k-\tau_{k-1}$ \citep{ogata1988statistical}. Under perfect fit, these times are i.i.d.\ exponential random variables with unit mean. Consequently, the test statistics $U_k = 1 - \exp(-Y_k)$ should be i.i.d.\ uniform on $[0,1)$. Plotting $U_k$ against $U_{k+1}$ serves as a visual check for interval independence, helping to identify residual serial correlation. The primary advantage of residual analysis is its intuitive visual display, providing an excellent qualitative assessment of goodness-of-fit.

\section{Generative Model} \label{sec:generative_model}

This appendix records the simulation procedure consistent with the model specification in Section \ref{sec:multiplexnetworkhawkes}. Following \citet{linderman2014discovering}, the extended specification suggests an intuitive generative model. The Poisson superposition theorem implies that additive components of the process can be considered as independent processes, each giving rise to its own events; a sample path can therefore be generated as follows:
\begin{enumerate}
\item Draw the entries of adjacency matrix $\mathbf{A}$ conditional on the sparsity parameter: $A_{k'\rightarrow k}\sim\mbox{Bernoulli}\left(\rho\right)$.
\item Generate weight matrices $\mathbf{W}^{(l)}$ given layer/edge-level covariates and linear predictor coefficients.
\item For each process $k$, draw a number of background events $m_k\sim\mbox{Poisson}(\lambda_k^{(0)}T)$. 
\item Draw the $m_k$ event times i.i.d. from the uniform density $\mbox{Uniform}(0,T)$.
\item To generate a cascade of events, for each parent event and model layer, draw the number of induced child events per parent-child process: $m_{k'\rightarrow k}^{(l)}\sim\mbox{Poisson}\left(A_{k'\rightarrow k}W_{k'\rightarrow k}^{(l)}\right)$.
\item Draw the $m_{k'\rightarrow k}^{(l)}$ child event times i.i.d. from logistic-normal density $g_{k'\rightarrow k}\left(\Delta t\right)$.
\item Retain events where event times $\leq \Delta t_{max}$.
%\item For each impulse response, draw a number of induced events $m \sim Poisson(W_{k \rightarrow k'})$,
%\item Draw the $m$ child event times i.i.d. from $g$
\end{enumerate}

Note that Hawkes processes can lead to unstable systems; we check the stability of the system for each generated network using the maximum-eigenvalue criterion described in Appendix \ref{sec:stability}.

\section{Simulation Study}\label{sec:simulation_study_app}
This appendix provides supporting tables and figures for the simulation results in Section \ref{sec:simulationstudy}.
\subsection{Illustrative Example}\label{sec:illustrative_example_app}

The common parameters used in the illustrative example in Section \ref{sec:illustrativeexample} and the analysis of different scenarios and model settings in Section \ref{sec:scenarioanalysis} are summarised in Table \ref{tab:common_params}.

\iffalse
\begin{table}[htbp!]
\centering
\resizebox{0.43\textwidth}{!}{%
\begin{tabular}{|l|l|}
\hline
\textbf{Parameter} & \textbf{Value} \\ \hline
no of samples & 20500 \\ \hline
no of initial samples for RWMH & 500 \\ \hline
\multicolumn{2}{|c|}{\textbf{Network parameters}} \\ \hline
no of layers L & 2 \\ \hline
%allow self connections & TRUE \\ \hline
%integrate out zero edges & TRUE \\ \hline
$\alpha^0_{\rho}$ & 1 \\ \hline
$\beta^0_{\rho}$ & 1 \\ \hline
$\kappa^0$ & 1 \\ \hline
\multicolumn{2}{|c|}{\textbf{Prior for $\kappa$}} \\ \hline
distribution & Half-Cauchy \\ \hline
$a^0_{\kappa}$ & 0.001 \\ \hline
$b^0_{\kappa}$ & 10 \\ \hline
\multicolumn{2}{|c|}{\textbf{Prior for b}} \\ \hline
distribution & Half-Cauchy  \\ \hline
$a^0_{\kappa_b}$ & 0.001 \\ \hline
$b^0_{\kappa_b}$ & 10 \\ \hline
\multicolumn{2}{|c|}{\textbf{Background hyper parameters}} \\ \hline
$\lambda_{\alpha}$ & 1 \\ \hline
$\lambda_{\beta}$ & 1 \\ \hline
\multicolumn{2}{|c|}{\textbf{Impulse hyper parameters}} \\ \hline
$\mu^0_{\mu}$ & -1 \\ \hline
$\kappa^0_{\tau}$ & 10 \\ \hline
$\alpha^0_{\tau}$ & 10 \\ \hline
$\beta^0_{\tau}$ & 1 \\ \hline
\end{tabular}%
}
\caption{Model parameters and their values.}% \label{tab:common_params}
\end{table}
\fi

\begin{table}[htbp!]
\centering
\resizebox{0.43\textwidth}{!}{%
\begin{tabular}{|l|l|}
\hline
\textbf{Parameter} & \textbf{Value} \\ \hline
no of samples & 20500 \\ \hline
no of initial samples for RWMH & 500 \\ \hline
\multicolumn{2}{|c|}{\textbf{Network parameters}} \\ \hline
no of layers L & 2 \\ \hline
%allow self connections & TRUE \\ \hline
%integrate out zero edges & TRUE \\ \hline
$\alpha^0_{\rho}$ & 1 \\ \hline
$\beta^0_{\rho}$ & 1 \\ \hline
$\kappa^0$ & 1 \\ \hline
\multicolumn{2}{|c|}{\textbf{Prior for $\kappa$}} \\ \hline
distribution & Half-Cauchy \\ \hline
$a^0_{\kappa}$ & 0.001 \\ \hline
$b^0_{\kappa}$ & 10 \\ \hline
\end{tabular}%
}
\resizebox{0.33\textwidth}{!}{%
\begin{tabular}{|l|l|}
\hline
%\textbf{Parameter} & \textbf{Value} \\ \hline
\multicolumn{2}{|c|}{\textbf{Prior for b}} \\ \hline
distribution & Half-Cauchy  \\ \hline
$a^0_{\kappa_b}$ & 0.001 \\ \hline
$b^0_{\kappa_b}$ & 10 \\ \hline
\multicolumn{2}{|c|}{\textbf{Background hyper parameters}} \\ \hline
$\lambda_{\alpha}$ & 1 \\ \hline
$\lambda_{\beta}$ & 1 \\ \hline
\multicolumn{2}{|c|}{\textbf{Impulse hyper parameters}} \\ \hline
$\mu^0_{\mu}$ & -1 \\ \hline
$\kappa^0_{\tau}$ & 10 \\ \hline
$\alpha^0_{\tau}$ & 10 \\ \hline
$\beta^0_{\tau}$ & 1 \\ \hline
\end{tabular}%
}
\caption{Model parameters and their values.}
\label{tab:common_params}
\end{table}

\begin{table}[htbp!]
\centering
\resizebox{0.95\textwidth}{!}{%
\begin{tabular}{cccccccccccccc}
\hline
\textbf{Scenario}&\textbf{Setting}& \textbf{Est $b^0_{\kappa}$} & \textbf{Est $\kappa$} & \textbf{$\Delta t_{\text{max}}$} & \textbf{$a^0_{\kappa}$} & \textbf{$b^0_{\kappa}$} & \textbf{$a^0_{\kappa_b}$} & \textbf{$b^0_{\kappa_b}$} & \textbf{$\Sigma_0$} & \textbf{$L$} & \textbf{$\alpha^0_{\rho}$} & \textbf{$\beta^0_{\rho}$} & \textbf{Prior for $\kappa/\kappa_b$} \\ 
\hline
small & base & TRUE & TRUE & 0.038 & 0.001 & 10 & 0.001 & 10 & 10 & 2 & 1 & 1 & Halfcauchy \\
small & invGamma & TRUE & TRUE & 0.038 & 1 & 1 & 5 & 25 & 10 & 2 & 1 & 1 & InvGamma \\
small & 3layers & TRUE & TRUE & 0.038 & 0.001 & 10 & 0.001 & 10 & 10 & 3 & 1 & 1 & Halfcauchy \\
small & 0.1dt & TRUE & TRUE & 0.1 & 0.001 & 10 & 0.001 & 10 & 10 & 2 & 1 & 1 & Halfcauchy \\
small & 0.01dt & TRUE & TRUE & 0.01 & 0.001 & 10 & 0.001 & 10 & 10 & 2 & 1 & 1 & Halfcauchy \\
small-beta & base & TRUE & TRUE & 0.038 & 0.001 & 10 & 0.001 & 10 & 10 & 2 & 1 & 1 & Halfcauchy \\
small-beta & 3layers & TRUE & TRUE & 0.038 & 0.001 & 10 & 0.001 & 10 & 10 & 3 & 1 & 1 & Halfcauchy \\
small-beta & 0.1dt & TRUE & TRUE & 0.1 & 0.001 & 10 & 0.001 & 10 & 10 & 2 & 1 & 1 & Halfcauchy \\
small-beta & 0.01dt & TRUE & TRUE & 0.01 & 0.001 & 10 & 0.001 & 10 & 10 & 2 & 1 & 1 & Halfcauchy \\
small-beta & invGamma & TRUE & TRUE & 0.038 & 1 & 1 & 5 & 25 & 10 & 2 & 1 & 1 & InvGamma \\
medium & base & TRUE & TRUE & 10.5 & 0.001 & 10 & 0.001 & 10 & 10 & 2 & 1 & 1 & Halfcauchy \\
medium & invGamma & TRUE & TRUE & 10.5 & 1 & 1 & 5 & 25 & 10 & 2 & 1 & 1 & InvGamma \\
medium & sigma100 & TRUE & TRUE & 10.5 & 0.001 & 10 & 0.001 & 10 & 100 & 2 & 1 & 1 & Halfcauchy \\
medium & kappahypers100 & TRUE & TRUE & 10.5 & 0.001 & 100 & 0.001 & 100 & 10 & 2 & 1 & 1 & Halfcauchy \\
medium & estKappaHyper & FALSE & TRUE & 10.5 & 0.001 & 10 & 0.001 & 10 & 10 & 2 & 1 & 1 & Halfcauchy \\
medium & estKappa & FALSE & FALSE & 10.5 & 0.001 & 10 & 0.001 & 10 & 10 & 2 & 1 & 1 & Halfcauchy \\
medium & ab10 & TRUE & TRUE & 10.5 & 0.001 & 10 & 0.001 & 10 & 10 & 2 & 10 & 10 & Halfcauchy \\
medium & 3layers & TRUE & TRUE & 10.5 & 0.001 & 10 & 0.001 & 10 & 10 & 3 & 1 & 1 & Halfcauchy \\
medium & 5dt & TRUE & TRUE & 5.5 & 0.001 & 10 & 0.001 & 10 & 10 & 2 & 1 & 1 & Halfcauchy \\
medium & 20dt & TRUE & TRUE & 20.5 & 0.001 & 10 & 0.001 & 10 & 10 & 2 & 1 & 1 & Halfcauhy \\
medium-beta& base & TRUE & TRUE & 10.5 & 0.001 & 10 & 0.001 & 10 & 10 & 2 & 1 & 1 & Halfcauchy \\
medium-beta & invGamma & TRUE & TRUE & 10.5 & 1 & 1 & 5 & 25 & 10 & 2 & 1 & 1 & InvGamma \\
medium-beta & estKappaHyper & FALSE & TRUE & 10.5 & 0.001 & 10 & 0.001 & 10 & 10 & 2 & 1 & 1 & Halfcauchy \\
medium-beta & estKappa & FALSE & FALSE & 10.5 & 0.001 & 10 & 0.001 & 10 & 10 & 2 & 1 & 1 & Halfcauchy \\
large-beta & base & TRUE & TRUE & 10.5 & 0.001 & 10 & 0.001 & 10 & 10 & 2 & 1 & 1 & Halfcauchy \\
large-beta& invGamma & TRUE & TRUE & 10.5 & 1 & 1 & 5 & 25 & 10 & 2 & 1 & 1 & InvGamma \\
large-beta & sigma100 & TRUE & TRUE & 10.5 & 0.001 & 10 & 0.001 & 10 & 100 & 2 & 1 & 1 & Halfcauchy \\
large-beta & estKappaHyper & TRUE & FALSE & 10.5 & 0.001 & 10 & 0.001 & 10 & 10 & 2 & 1 & 1 & Halfcauchy \\
\hline
\end{tabular}}
\caption{Parameters used in scenario estimation.}
\label{tab:scenarios_estimation}
\end{table}

Table \ref{tab:w_lambda0} reports posterior estimates and HDIs for the nonzero weight elements and background intensities.

\begin{table}[htbp!]
\centering
\resizebox{0.36\textwidth}{!}{%
\begin{tabular}{ccccc}
\hline
\textbf{Parameter}        & \textbf{Generated} & \textbf{Median} & \textbf{95\% HDI}      \\ \hline
\(\rho\)                  & 0.600             & 0.067           & [0.03,0.12]                  \\
\(\lambda^{(0)}_0\)       & 0.2               & 0.25            & [0.16,0.36]                    \\
\(\lambda^{(0)}_1\)       & 0.2               & 0.2             & [0.11,0.28]                  \\
\(\lambda^{(0)}_2\)       & 0.2               & 0.21            & [0.12,0.3]                   \\
\(\lambda^{(0)}_3\)       & 0.2               & 0.13            & [0.06,0.2]                   \\
\(\lambda^{(0)}_4\)       & 0.2               & 0.18            & [0.1,0.26]                   \\
\(\lambda^{(0)}_5\)       & 0.2               & 0.24            & [0.15,0.34]                  \\
\(\lambda^{(0)}_6\)       & 0.2               & 0.26            & [0.17,0.37]                  \\
\(\lambda^{(0)}_7\)       & 0.2               & 0.22            & [0.14,0.32]                  \\
\(\lambda^{(0)}_8\)       & 0.2               & 0.31            & [0.21,0.43]                  \\
\(\lambda^{(0)}_9\)       & 0.2               & 0.25            & [0.16,0.36]                    \\ \hline
\end{tabular}%
}
%caption{Posterior median estimates and HDIs for  background intensities and sparsity parameter $\rho$.}
\resizebox{0.36\textwidth}{!}{%
\begin{tabular}{ccccc}
\hline
\textbf{Parameter} & \textbf{Generated} & \textbf{Median} & \textbf{95\% HDI}              \\ \hline
\(W_{3,5}\)        & 2.25              & 2.21            & [2.01, 2.45]                        \\
\(W_{4,3}\)        & 1.23              & 1.19            & [1.01, 1.41]                         \\
\(W_{6,3}\)        & 0.96              & 0.91            & [0.6, 1.36]                         \\
\(W_{6,4}\)        & 3.22              & 3.58            & [2.92, 4.37]                        \\
\(W_{7,0}\)        & 3.01              & 2.58            & [2.03, 3.39]                         \\
\(W_{7,2}\)        & 0.23              & 0.26            & [0.09, 0.54]                        \\ \hline
\end{tabular}%
}
\caption{Posterior median estimates and HDIs: for  background intensities and sparsity parameter $\rho$ (left), and for weights elements $W_{k'\rightarrow k}$ where median $A_{k'\rightarrow k}=1$ (right).}
\label{tab:w_lambda0}
\end{table}

In this table, $\lambda^{(0)}_k$ denotes the background intensity for process $k$; it is distinct from the stability quantity $\lambda_{\max}$, which is the spectral radius of the combined integrated excitation matrix defined in Appendix \ref{sec:stability}.

\begin{figure}[htbp!]
        \centering
        \includegraphics[width=\textwidth, angle=0, trim= 287 8 170 30,clip]{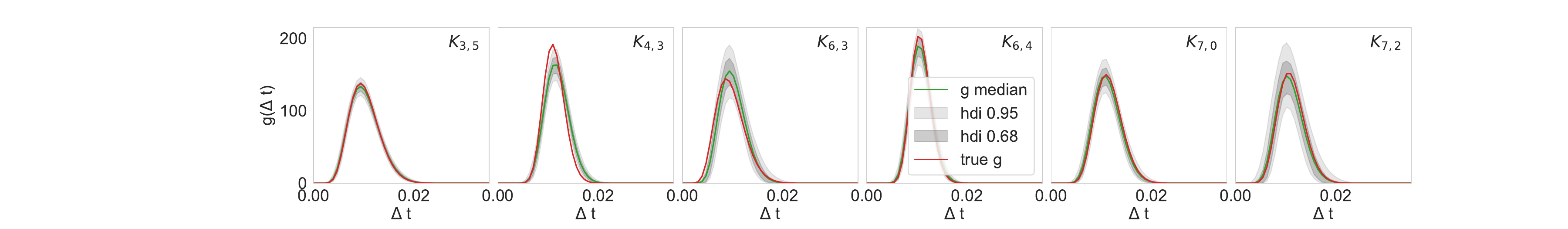}
    \caption{Generated versus estimated interaction kernel $g_{\theta_{k'\rightarrow k}}$ (posterior median) for elements where $A_{k'\rightarrow k}=1$.}
    \label{fig:interaction_kernel_app}
\end{figure}

The estimated interaction kernel as a function of different values of $\Delta t$ is presented in Figure \ref{fig:interaction_kernel_app}. The true function $g$ falls within the $95\%$ HDI across all edges for most $\Delta t$ values, except for edge $A_{4,3}$, which is underestimated for $\Delta t<0.012$ and overestimated for larger values.

\begin{figure}[htbp!]
\centering
\includegraphics[width=0.38\textwidth, angle=0, trim= 50 289 40 50,clip]{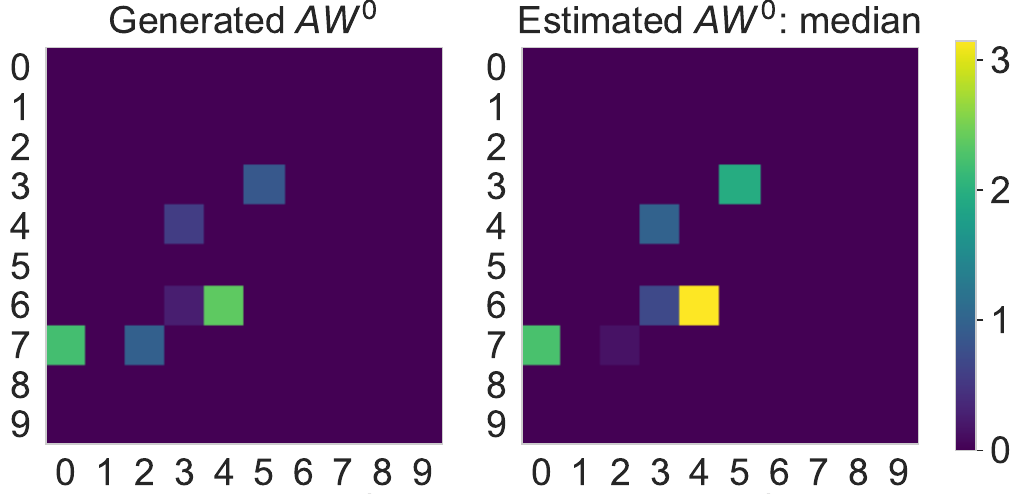}%
\includegraphics[width=0.38\textwidth, angle=0, trim= 50 55 40 289,clip]{img/simulation_study/2l_sigma10/estimated_network_layer_ecr_iterative2_p2.pdf}
\caption{Generated and estimated layer networks (posterior medians) after ECR relabelling.}
\label{fig:weights_layer_ecr}
\end{figure}

Figure \ref{fig:weights_layer_ecr} shows the generated and estimated layer networks after ECR relabelling for the illustrative example.

\subsection{Scenario Analysis}\label{sec:scenario_analysis_app}

Table \ref{tab:lambda0_2} reports the background-rate diagnostic moved out of the main simulation observations.

\begin{table}[htbp!]
\centering
\resizebox{0.62\textwidth}{!}{%
\begin{tabular}{lllllll}
\hline
\textbf{Scenario} & \textbf{$\lambda^0_{gen}$} &  \textbf{Avg $\lambda^0$} & \textbf{RMSE} & \textbf{Rel RMSE} & \textbf{Avg 98\% HDI /$\lambda^0_{gen}$}  \\ \hline
snall                 & 0.2                                                        & 0.225                                                      & 0.056          & 28\%              & 91\%                                                                                             \\
small-beta                & 0.2                                                        & 0.185                                                      & 0.036          & 18\%              & 82\%                                                                                             \\
medium                 & 0.005                                                      & 0.005                                                      & 0.001         & 13\%              & 41\%                                                                                             \\
medium-beta                & 0.01                                                       & 0.010                                                      & 0.001       & 15\%              & 42\%                                                                                             \\
large-beta                & 0.0100                                                     & 0.0100                                                     & 0.0015     & 15\%              & 25\%                                                                                             \\ \hline
\end{tabular}%
}
\caption{Averages of $\lambda^{0}$ posterior medians and relative HDIs across all node-level ($k$) processes and absolute and relative RMSEs across generated ``base'' scenarios.}
\label{tab:lambda0_2}
\end{table}

Tables \ref{tab:rho_invgamma_2} and \ref{tab:beta_invgamma_2} report prior-sensitivity checks for $\rho$, $\kappa$ and $\bm{\beta}$. They support the main-text conclusion that coefficient recovery is easier when layer-specific information is stronger and harder in sparse or weakly separated settings.

\begin{table}[htbp!]
\centering
\resizebox{0.48\textwidth}{!}{%
\begin{tabular}{ccccc}
\hline
\textbf{Scen} & \textbf{Name} & $\boldsymbol{\rho_{g}}$ & $\boldsymbol{\rho}$ & \textbf{95\% HDI}  \\
\hline
small & base & 0.06 & 0.067 & [0.03,0.12]  \\
small & invGamma & 0.06 & 0.069 & [0.02,0.12]  \\
small-beta & base & 0.06 & 0.066 & [0.02,0.12]  \\
small-beta & invGamma & 0.06 & 0.070 & [0.03,0.13]  \\
medium & base & 0.1 & 0.097 & [0.07,0.13]  \\
medium & invGamma & 0.1 & 0.099 & [0.07,0.13]  \\
medium-beta & base & 0.1 & 0.087 & [0.06,0.12]  \\
medium-beta & invGamma & 0.1 & 0.089 & [0.06,0.12]  \\
large-beta & base & 0.085 & 0.094 & [0.075,0.115]  \\
large-beta & invGamma & 0.085 & 0.095 & [0.077,0.117]  \\
\hline
\end{tabular}%
}
\centering
\resizebox{0.48\textwidth}{!}{%
\begin{tabular}{cccccc}
\hline
\textbf{Scen} & \textbf{Name} & \textbf{L} & \textbf{$\kappa_{g}$} & \textbf{$\kappa$} & \textbf{95\% HDI}  \\
\hline
small & base & 0 & 1 & 0.99 & [0,3.74]  \\
small & base & 1 & 1 & 0.98 & [0,4.23]  \\
small & invGamma & 0 & 1 & 1.97 & [0.43,5.54]  \\
small & invGamma & 1 & 1 & 2.04 & [0.38,6.14]  \\
small-beta & base & 0 & 1 & 0.64 & [0,3.88]  \\
small-beta & base & 1 & 1 & 1.11 & [0,6.05]  \\
small-beta & invGamma & 0 & 1 & 2.40 & [0.45,7.45]  \\
small-beta & invGamma & 1 & 1 & 2.71 & [0.44,9.31]  \\
medium & base & 0 & 1 & 0.60 & [0.01,1.28]  \\
medium & base & 1 & 1 & 0.70 & [0,1.28]  \\
medium & invGamma & 0 & 1 & 0.90 & [0.43,1.58]  \\
medium & invGamma & 1 & 1 & 0.92 & [0.42,1.56]  \\
medium-beta & base & 0 & 1 & 0.58 & [0.02,1.22]  \\
medium-beta & base & 1 & 1 & 0.95 & [0.19,1.86]  \\
medium-beta & invGamma & 0 & 1 & 0.84 & [0.36,1.58]  \\
medium-beta & invGamma & 1 & 1 & 1.15 & [0.44,1.97]  \\
large-beta & base & 0 & 1 & 0.039 & [0.002,0.286]  \\
large-beta & base & 1 & 1 & 0.834 & [0.558,1.229]  \\
large-beta & invGamma & 0 & 1 & 0.919 & [0.584,1.389]  \\
large-beta & invGamma & 1 & 1 & 0.556 & [0.262,1.054]  \\
\hline
\end{tabular}%
}
\caption{$\rho$ (left) and $\kappa$ (right) estimation accuracy under Half-Cauchy versus inverse Gamma prior specification. Posterior median estimates are presented as $\rho$ and $\kappa$.}
\label{tab:rho_invgamma_2}
\end{table}

\begin{table}[htbp!]
\centering
\resizebox{1\textwidth}{!}{%
\begin{tabular}{lllllllllllllll}
\hline
 \textbf{Scenario} & \textbf{Setting} & \textbf{L} & \textbf{${\beta_0}_{g}$} & \textbf{$\beta_0$} &\textbf{95\% HDI} & \textbf{68\% HDI} &\textbf{${\beta_1}_g$} & \textbf{$\beta_1$} & \textbf{95\% HDI} & \textbf{68\% HDI} &  \textbf{${\beta_2}_{g}$} & \textbf{$\beta_2$} & \textbf{95\% HDI} & \textbf{68\% HDI} \\
\hline
 small                                 & base                              & 0                              & 0                                                        & 0.07                                                 & \cellcolor[HTML]{C6EFCE}{\color[HTML]{006100} {[}-2.53,1.71{]}}  & \cellcolor[HTML]{C6EFCE}{\color[HTML]{006100} {[}-0.63,1.13{]}}  & 1                                                        & 0.23                                                 & {[}-5.82,6.36{]}                                                & {[}-2.95,3.35{]}                                                & 1                                                        & -0.09                                                & {[}-5.97,6.1{]}                                                  & {[}-3.09,2.98{]}                                                 \\
\hline
\end{tabular}%
}
\caption{$\bm{\beta}$ estimation accuracy under Half-Cauchy versus inverse Gamma prior specification. Posterior median estimates are presented as $\bm{\beta}$. Green cells indicate where HDI does not contain 0 if covariate is not zero, or contain zero for 0 covariates. Yellow cells indicate cells where HDI contains zero but is skewed with a minimum 75\% of HDI not containing 0.}
\label{tab:beta_invgamma_2}
\end{table}

\begin{table}[htbp!]
\centering
\resizebox{\textwidth}{!}{%
\begin{tabular}{cccccccccccccccccc}
\hline
\textbf{Setting} & \textbf{L} & \textbf{${\beta_0}_{g}$} & \textbf{$\beta_0$} & \textbf{95\% HDI} & \textbf{68\% HDI} & \textbf{${\beta_1}_{g}$} & \textbf{$\beta_1$} & \textbf{95\% HDI} & \textbf{68\% HDI} & \textbf{${\beta_2}_{g}$} & \textbf{$\beta_2$} & \textbf{95\% HDI} & \textbf{68\% HDI} & \textbf{$\kappa_{g}$} & \textbf{$\kappa$ } & \textbf{95\% HDI} & \textbf{68\% HDI} \\
\hline
base & 0 & -3.05 & -2.54 & [-3.44, -1.95] & [-3.04, -2.01] & 0.05 & -0.00 & [-0.07, 0.10] & [-0.05, 0.03] & 0 & -0.02 & [-0.11, 0.10] & [-0.08, 0.02] & 1 & 0.04 & [0.00, 0.29] & [0.01, 0.12] \\
base & 1 & -1.70 & -1.67 & [-2.16, -1.21] & [-1.93, -1.44] & 0 & 0.03 & [-0.04, 0.11] & [-0.01, 0.07] & -0.035 & -0.03 & [-0.11, 0.04] & [-0.07, 0.01] & 1 & 0.83 & [0.56, 1.23] & [0.67, 1.02] \\
sigma100 & 0 & -3.05 & -3.16 & [-3.91, -1.93] & [-3.67, -2.76] & 0.05 & 0.04 & [-0.05, 0.14] & [-0.00, 0.11] & 0 & -0.01 & [-0.10, 0.08] & [-0.04, 0.03] & 1 & 0.02 & [0.00, 0.30] & [0.00, 0.12] \\
sigma100 & 1 & -1.70 & -1.64 & [-2.12, -1.15] & [-1.88, -1.40] & 0 & 0.03 & [-0.04, 0.10] & [-0.00, 0.07] & -0.035 & -0.03 & [-0.10, 0.04] & [-0.07, 0.01] & 1 & 0.75 & [0.51, 1.11] & [0.62, 0.91] \\
invGamma & 0 & -3.05 & -1.75 & [-2.44, -0.96] & [-2.10, -1.37] & 0.05 & 0.00 & [-0.09, 0.10] & [-0.04, 0.05] & 0 & -0.02 & [-0.12, 0.07] & [-0.06, 0.03] & 1 & 0.92 & [0.58, 1.39] & [0.74, 1.13] \\
invGamma & 1 & -1.70 & -2.39 & [-3.75, -1.14] & [-3.04, -1.72] & 0 & 0.02 & [-0.14, 0.15] & [-0.06, 0.09] & -0.035 & -0.02 & [-0.18, 0.15] & [-0.10, 0.06] & 1 & 0.56 & [0.26, 1.05] & [0.38, 0.78] \\
estKappa & 0 & -3.05 & -1.87 & [-2.67, -1.10] & [-2.27, -1.49] & 0.05 & 0.01 & [-0.09, 0.12] & [-0.04, 0.06] & 0 & -0.02 & [-0.13, 0.09] & [-0.07, 0.03] & 1 & 1.00 & n/a & n/a \\
estKappa & 1 & -1.70 & -2.27 & [-3.50, -1.33] & [-2.75, -1.76] & 0 & 0.03 & [-0.11, 0.16] & [-0.03, 0.10] & -0.035 & -0.02 & [-0.15, 0.12] & [-0.09, 0.04] & 1 & 1.00 & n/a & n/a \\
\hline
\hline
\end{tabular}}
\caption{Scenario ``large-beta'': Estimation accuracy across specifications. Posterior median estimates are presented as $\bm{\beta}$ and $\kappa$. Subscript ``g'' denotes true generated values.}
\label{tab:beta_11}
\end{table}

\begin{table}[htbp!]
\centering
\resizebox{0.82\textwidth}{!}{%
\begin{tabular}{ccccccc}
\hline
\textbf{Scenario} & \textbf{Setting} & \textbf{$A=1$ gen} & \textbf{$A=1$} & \textbf{$AW_{gen}$} & \textbf{$AW$} & \textbf{95\% HDI}  \\
\hline
small & base & 6 & 6 & 0.109 & 0.107 & [0.087,0.135]  \\
small & invGamma & 6 & 6 & 0.109 & 0.108 & [0.087,0.136]  \\
small-beta & base & 6 & 6 & 0.318 & 0.304 & [0.264,0.364]  \\
small-beta & invGamma & 6 & 6 & 0.318 & 0.301 & [0.263,0.364]  \\
medium & base & 39 & 37 & 0.035 & 0.032 & [0.025,0.042]  \\
medium & invGamma & 39 & 37 & 0.035 & 0.032 & [0.025,0.042]  \\
medium-beta & base & 39 & 32 & 0.037 & 0.035 & [0.029,0.044]  \\
medium-beta & invGamma & 39 & 32 & 0.037 & 0.035 & [0.029,0.045]  \\
large-beta & base & 88 & 80 & 0.023 & 0.023 & [0.022,0.024]  \\
large-beta & invGamma & 88 & 79 & 0.023 & 0.023 & [0.022,0.024]  \\
\hline
\end{tabular}%
}
\caption{Average $\bm{AW}$ estimation accuracy under Half-Cauchy versus inverse Gamma prior specification. Posterior median estimates are presented as $A$ and $\bm{AW}$.}
\label{tab:AW_invgamma}
\end{table}

Further investigating model sensitivity (Observations 7 and 8), Tables \ref{tab:rho_accuracy_8} and \ref{tab:aw_accuracy_8} report the estimates for sparsity ($\rho$) and edge weights ($\bm{AW}$) under different prior specifications.

\begin{table}[htbp!]
\centering
\resizebox{0.45\textwidth}{!}{%
\begin{tabular}{cccc}
\hline
\textbf{Setting} & $\boldsymbol{\rho_{\text{gen}}}$ & $\boldsymbol{\rho}$ & \textbf{95\% HDI}  \\
\hline
base & 0.1 & 0.097 & [0.07,0.13]  \\
ab10 & 0.1 & 0.115 & [0.09,0.15]  \\
invGamma & 0.1 & 0.099 & [0.07,0.13]  \\
estKappa & 0.1 & 0.099 & [0.07,0.13]  \\
estKappaHyper & 0.1 & 0.097 & [0.07,0.13]  \\
kappahypers100 & 0.1 & 0.097 & [0.07,0.13]  \\
sigma100 & 0.1 & 0.096 & [0.07,0.13]  \\
3layers & 0.1 & 0.095 & [0.07,0.12]  \\
20dt & 0.1 & 0.096 & [0.07,0.13]  \\
5dt & 0.1 & 0.100 & [0.07,0.13]  \\
\hline
\end{tabular}}
\caption{Scenario ``medium''. $\rho$ estimation accuracy across different parameter specifications.  Posterior median estimates are presented as $\rho$.}
\label{tab:rho_accuracy_8}
\end{table}

\begin{table}[htbp!]
\centering
\resizebox{0.70\textwidth}{!}{%
\begin{tabular}{ccccccccc}
\hline
\textbf{Setting} & \textbf{$A=1$ gen} & \textbf{$A=1$} & \textbf{$AW_{gen}$} & \textbf{$AW$} & \textbf{95\% HDI} & \textbf{95\% HDI/$AW_{gen}$}  \\
\hline
base & 39 & 37 & 0.035 & 0.032 & [0.025,0.042] & 48\%  \\
ab10 & 39 & 37 & 0.035 & 0.032 & [0.025,0.042] & 48\%  \\
invGamma & 39 & 37 & 0.035 & 0.032 & [0.025,0.042] & 48\%  \\
estKappa & 39 & 38 & 0.035 & 0.032 & [0.025,0.042] & 48\%  \\
estKappaHyper & 39 & 37 & 0.035 & 0.032 & [0.025,0.042] & 47\%  \\
kappahypers100 & 39 & 37 & 0.035 & 0.032 & [0.025,0.042] & 47\%  \\
sigma100 & 39 & 37 & 0.035 & 0.032 & [0.025,0.042] & 47\%  \\
3layers & 39 & 37 & 0.035 & 0.030 & [0.025,0.041] & 46\%  \\
20dt & 39 & 37 & 0.035 & 0.033 & [0.025,0.043] & 52\%  \\
5dt & 39 & 38 & 0.035 & 0.031 & [0.024,0.041] & 49\%  \\
\hline
\end{tabular}}
\caption{Scenario ``medium''. $\bm{AW}$ estimation accuracy across different parameter specifications.  Posterior median estimates are presented as $\bm{AW}$.}
\label{tab:aw_accuracy_8}
\end{table}

\begin{figure}[H]
     \centering
     \begin{subfigure}{0.42\textwidth}
         \includegraphics[width=\textwidth, angle=0, trim= 36 15 55 34,clip]{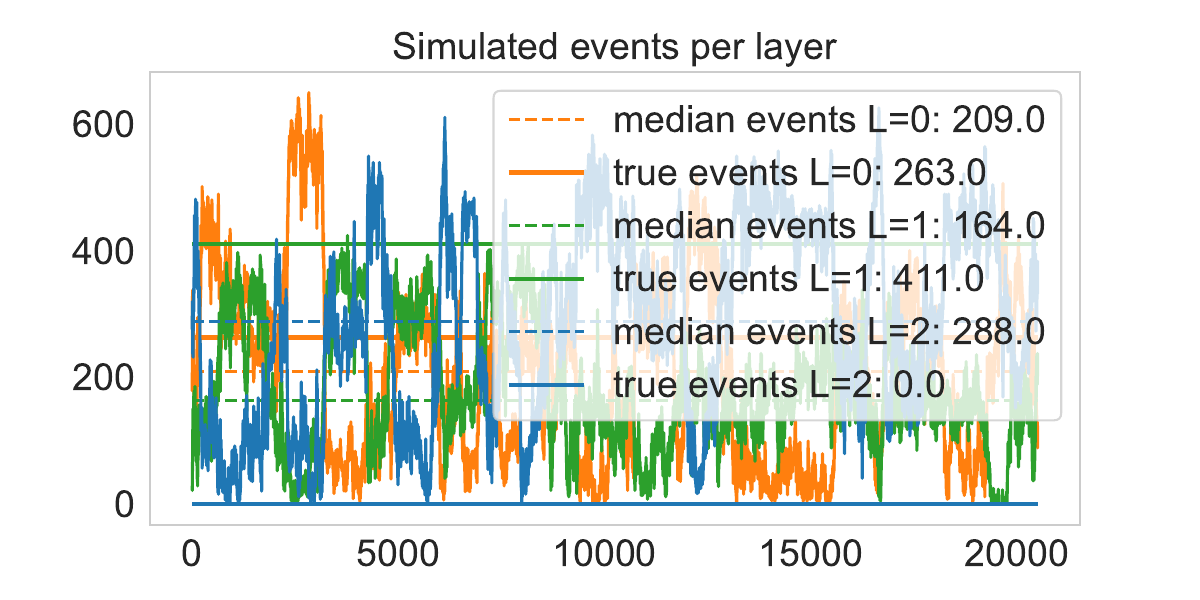}
          \caption{Without relabelling.}
     \end{subfigure} %
\begin{subfigure}{0.42\textwidth}
  \includegraphics[width=\textwidth, angle=0, trim= 70 15 55 34,clip]{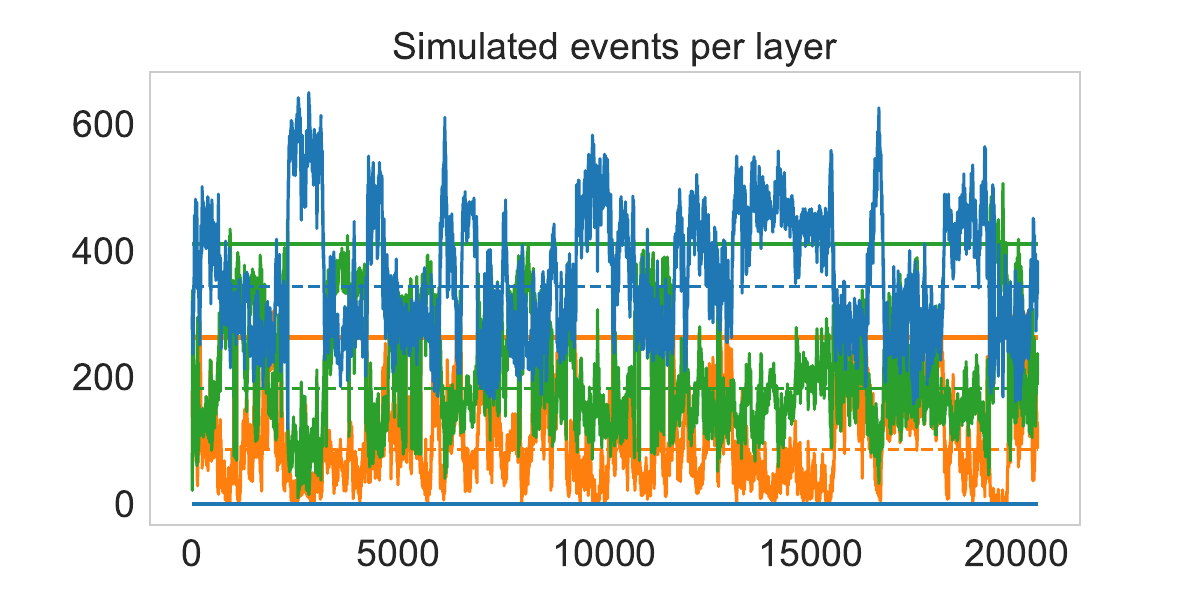}
    \caption{ECR iterative version 1.}
     \end{subfigure}
\caption{Scenario ``small-beta'' fitted with three layers: allocated events per layer without relabelling (left) and after the selected ECR iterative version 1 relabelling (right).}
\label{fig:ecr_samples_4_layer}
\end{figure}

Figure \ref{fig:ecr_samples_4_layer} shows the selected ECR comparison for the ``small-beta'' process fitted with an additional layer.

\FloatBarrier

\begin{figure}[H]
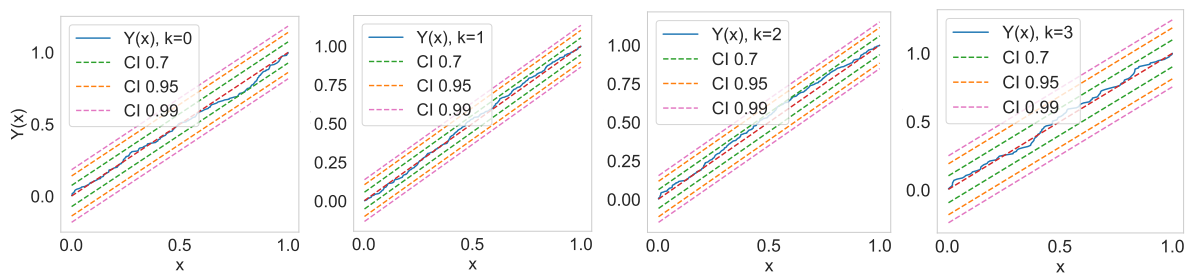

\centering
\includegraphics[width=0.25\textwidth, angle=0, trim= 0 6065 35 860, clip]{img/simulation_study/8l_base/step_multi.pdf}
\includegraphics[width=0.23\textwidth, angle=0, trim= 26 5790 35 1135, clip]{img/simulation_study/8l_base/step_multi.pdf}
\includegraphics[width=0.235\textwidth, angle=0, trim= 26 5518 35 1410, clip]{img/simulation_study/8l_base/step_multi.pdf}
\includegraphics[width=0.23\textwidth, angle=0, trim= 36 5245 35 1682, clip]{img/simulation_study/8l_base/step_multi.pdf}
\caption{Scenario ``medium''. Step functions for processes $0$--$3$.}
\label{fig:step_8}
\end{figure}

Graphical residual summaries were produced for the generated processes using the residual-process method described in Appendix \ref{sec:goodnessoffit}. For each fitted intensity, event times are transformed using its compensator. If the fitted intensity adequately describes the generated process, the resulting cumulative step function should remain close to the unit-rate Poisson reference line within the confidence bounds. Across the generated processes, these summaries indicate reasonable fit. Figure \ref{fig:step_8} presents selected processes in the ``medium'' scenario as an example; the displayed step functions do not indicate major divergences outside the 95\% confidence bounds.

\clearpage

\section{Empirical Results}\label{sec:empirical_results_app}
This appendix collects supporting tables and figures for the empirical model set-ups analysed in Section \ref{sec:empirical}.

\subsection{Model Set-Ups and Goodness of Fit}
The common/baseline parameters used in all model set-ups are summarised in Table \ref{tab:common_params_rd}.

\iffalse
\begin{table}[htbp!]
\centering
\resizebox{0.41\textwidth}{!}{%
\begin{tabular}{|l|l|}
\hline
\textbf{Parameter} & \textbf{Value} \\ \hline
no of samples & 20500 \\ \hline
no of initial samples for RWMH & 500 \\ \hline
time T & 4696 \\ \hline
no of events & 1905 \\ \hline
no of nodes K & 99 \\ \hline
\multicolumn{2}{|c|}{\textbf{Network parameters}} \\ \hline
allow self connections & TRUE \\ \hline
%integrate out zero edges & TRUE \\ \hline
$\alpha^0_{\rho}$ & 1 \\ \hline
$\beta^0_{\rho}$ & 1 \\ \hline
$\kappa^0$ & 1 \\ \hline
\multicolumn{2}{|c|}{\textbf{Prior for $\kappa$}} \\ \hline
distribution &  Inverse Gamma \\ \hline
$a^0_{\kappa}$ & 1 \\ \hline
$b^0_{\kappa}$ & n/a \\ \hline
\multicolumn{2}{|c|}{\textbf{Prior for b}} \\ \hline
distribution & Inverse Gamma  \\ \hline
$a^0_{\kappa_b}$ & 10 \\ \hline
$b^0_{\kappa_b}$ & 100 \\ \hline
\multicolumn{2}{|c|}{\textbf{Prior for $\bm{\beta}$}} \\ \hline
distribution & Normal  \\ \hline
$\mu$ & 0 \\ \hline
$\Sigma_0$ & 100 \\ \hline
\multicolumn{2}{|c|}{\textbf{Background hyper parameters}} \\ \hline
$\lambda_{\alpha}$ & 1 \\ \hline
$\lambda_{\beta}$ & 1 \\ \hline
\multicolumn{2}{|c|}{\textbf{Impulse hyper parameters}} \\ \hline
$\mu^0_{\mu}$ & -1 \\ \hline
$\kappa^0_{\tau}$ & 10 \\ \hline
$\alpha^0_{\tau}$ & 10 \\ \hline
$\beta^0_{\tau}$ & 1 \\ \hline
\end{tabular}%
}
\caption{Model parameters and their values.}% \label{tab:common_params_rd}
\end{table}
\fi

\begin{table}[htbp!]
\centering
\resizebox{0.43\textwidth}{!}{%
\begin{tabular}{|l|l|}
\hline
\textbf{Parameter} & \textbf{Value} \\ \hline
no of samples & 20500 \\ \hline
no of initial samples for RWMH & 500 \\ \hline
%time T & 4696 \\ \hline
%no of events & 1905 \\ \hline
no of nodes K & 99 \\ \hline
\multicolumn{2}{|c|}{\textbf{Network parameters}} \\ \hline
%allow self connections & TRUE \\ \hline
%integrate out zero edges & TRUE \\ \hline
$\alpha^0_{\rho}$ & 1 \\ \hline
$\beta^0_{\rho}$ & 1 \\ \hline
$\kappa^0$ & 1 \\ \hline
\multicolumn{2}{|c|}{\textbf{Impulse hyper parameters}} \\ \hline
$\mu^0_{\mu}$ & -1 \\ \hline
$\kappa^0_{\tau}$ & 10 \\ \hline
$\alpha^0_{\tau}$ & 10 \\ \hline
$\beta^0_{\tau}$ & 1 \\ \hline
\end{tabular}%
}
\resizebox{0.33\textwidth}{!}{%
\begin{tabular}{|l|l|}
\hline
%\textbf{Parameter} & \textbf{Value} \\ \hline
\multicolumn{2}{|c|}{\textbf{Background hyper parameters}} \\ \hline
$\lambda_{\alpha}$ & 1 \\ \hline
$\lambda_{\beta}$ & 1 \\ \hline
\multicolumn{2}{|c|}{\textbf{Prior for $\kappa$}} \\ \hline
distribution &  Inverse Gamma \\ \hline
$a^0_{\kappa}$ & 1 \\ \hline
$b^0_{\kappa}$ & n/a \\ \hline
\multicolumn{2}{|c|}{\textbf{Prior for b}} \\ \hline
distribution & Inverse Gamma  \\ \hline
$a^0_{\kappa_b}$ & 10 \\ \hline
$b^0_{\kappa_b}$ & 100 \\ \hline
\multicolumn{2}{|c|}{\textbf{Prior for $\bm{\beta}$}} \\ \hline
distribution & Normal  \\ \hline
$\mu$ & 0 \\ \hline
$\Sigma_0$ & 100 \\ \hline
\end{tabular}%
}
\caption{Model parameters and their values.}
\label{tab:common_params_rd}
\end{table}

Table \ref{tab:waic_all_scenarios_app} presents the WAIC point estimates alongside the log pointwise predictive density (lppd) and effective number of parameters ($p_{waic}$) for all empirically evaluated model set-ups. We compute WAIC using quarterly blocks of the observed extreme credit-spread jumps as the pointwise units. This block choice retains the temporal dependence among jumps occurring close together during periods of market stress, rather than treating each jump as an independent prediction unit. If $Q$ denotes the number of quarterly pointwise units and $\mathrm{WAIC}_q$ is the contribution of block $q$, the standard error is estimated as $\mathrm{SE}(\mathrm{WAIC})=\sqrt{Q\,\mathrm{Var}_q(\mathrm{WAIC}_q)}$. Consequently, close WAIC point estimates should be interpreted cautiously unless accompanied by the corresponding standard errors.

\begin{table}[htbp]
    \centering
    \resizebox{0.78\textwidth}{!}{%
    \begin{tabular}{lcccc}
    \hline
    \textbf{Model set-up} & \textbf{WAIC} & \textbf{lppd contribution} & \textbf{$p_{waic}$ contribution}  & \textbf{SE} \\\hline
    1l nocovs & 22313 & 19628 & 2685 & 1276.0 \\
    1l base & 22389 & 19588 & 2801 & 1274.1 \\
    1l prof & 22387 & 19582 & 2805 & 1273.4 \\
    1l solv & 22375 & 19570 & 2805 & 1271.9 \\
    1l asset & 22331 & 19597 & 2735 & 1266.5 \\
    1l asset invgamma10 & 22346 & 19616 & 2730 & 1267.8 \\
    1l asset invgamma100 & 22335 & 19601 & 2734 & 1267.1 \\
    1l asset kappa1 & 22383 & 19571 & 2812 & 1278.9 \\
    1l asset kappahyper10 & 22340 & 19577 & 2763 & 1270.0 \\
    2l nocovs & 22609 & 19524 & 3085 & 1295.3 \\
    2l base & 22677 & 19519 & 3158 & 1291.4 \\
    2l asset prof & 22661 & 19500 & 3161 & 1290.4 \\
    2l solv prof & 22676 & 19511 & 3165 & 1292.2 \\
    2l asset solv & 22651 & 19498 & 3153 & 1287.8 \\
    2l asset solv kappa1 & 22601 & 19517 & 3084 & 1291.9 \\
    3l nocovs & 22859 & 19644 & 3215 & 1316.0 \\
    3l base & 22892 & 19604 & 3288 & 1310.3 \\
    3l all & 22864 & 19582 & 3282 & 1305.2 \\
    3l all 20dt & 22758 & 20249 & 2510 & 1339.3 \\
    3l all 5dt & 22997 & 19393 & 3605 & 1279.2 \\
    3l all kappa1 & 22800 & 19602 & 3199 & 1310.3 \\\hline
    \end{tabular}%
    }
    \caption{Quarterly-block WAIC point estimates for all empirical model set-ups. We report the log pointwise predictive density (lppd) and the effective number of parameters ($p_{waic}$) as their respective contributions to the final WAIC based on the formula $\text{WAIC} = -2\,\text{lppd} + 2\,p_{waic}$, rather than their raw values.}
    \label{tab:waic_all_scenarios_app}
\end{table}

We estimate one-, two- and three-layer model set-ups with alternative covariate combinations, using WAIC primarily to choose among regression specifications with the same number of layers. Cross-layer WAIC differences are small relative to their standard errors, so we do not interpret them as evidence that one architecture is predictively dominant. Time-horizon layer-weight and covariate-coefficient estimates are reported in Tables \ref{tab:scenarios_weights_layer_rd_app} and \ref{tab:scenarios_covariates_rd_app}.

Table \ref{tab:scenarios_estimation_rd} details each model set-up used in the analysis.

\begin{table}[htbp!]
\centering
\resizebox{\textwidth}{!}{%
\begin{tabular}{lccccccccccc}
\hline
\textbf{Model set-up} & \textbf{Covariates} & \textbf{$\Delta t_{\text{max}}$} & \textbf{L} & \textbf{Gamma GLM} & \textbf{Prior for $\kappa$/$\kappa_b$} & \textbf{Est $\kappa$} & \textbf{$a^0_{\kappa}$} & \textbf{$b^0_{\kappa}$} & \textbf{Est $\beta_{\kappa}^0$} & \textbf{$a^0_{\kappa_b}$} & \textbf{$b^0_{\kappa_b}$} \\ \hline
1l nocovs & none & 10.5 & 1 & n/a & n/a & FALSE & n/a & n/a & FALSE & n/a & n/a \\ 
1l base & 0 & 10.5 & 1 & rwmh & InvGamma & TRUE & 1 & n/a  & TRUE & 10 & 100 \\ 
1l prof & prof & 10.5 & 1 & rwmh & InvGamma & TRUE & 1 & n/a  & TRUE & 10 & 100 \\
1l solv & solv & 10.5 & 1 & rwmh & InvGamma & TRUE & 1 & n/a  & TRUE & 10 & 100 \\
1l asset & asset & 10.5 & 1 & rwmh & InvGamma & TRUE & 1 & n/a  & TRUE & 10 & 100 \\
1l asset invgamma10 & asset & 10.5 & 1 & rwmh & InvGamma & TRUE & 10 & n/a  & TRUE & 10 & 10 \\
1l asset invgamma25 & asset & 10.5 & 1 & rwmh & InvGamma & TRUE & 1 & n/a  & TRUE & 5 & 25 \\
\hline
\end{tabular}%
}
\caption{Empirical model set-up parameters. Covariates: \textit{asset} (categorical), \textit{prof} (profitability), \textit{solv} (solvency), \textit{0} (base defaults) and \textit{none} (inverse Gamma prior weights without regression).}
\label{tab:scenarios_estimation_rd}
\end{table}

Additional estimates of model-layer hyperparameters are reported in Table \ref{tab:scenarios_kappa_rd}.

\begin{table}[htbp!]
\centering
\resizebox{0.68\textwidth}{!}{%
\begin{tabular}{lcccccc}
\hline
\textbf{Model set-up} & \textbf{Cov} & \textbf{L} & \(\kappa\) & \textbf{95\% HDI} & \(b_{\kappa}\) & \textbf{95\% HDI}  \\ \hline
%1l nocovs                                    & none                                   & 0                              &                                                 &                                    &                                                         &                                                    &                                    &                                                                                                                  \\
1l base                                      & 0                                      & 0                              & \cellcolor[HTML]{B1C7E4}0.15                       & {[}0.11,0.2{]}                        & \cellcolor[HTML]{E6ECF7}3.30                          & {[}2.33,4.29{]}  \\
1l prof                                      & prof                                   & 0                              & \cellcolor[HTML]{B2C8E5}0.15                       & {[}0.1,0.21{]}                        & \cellcolor[HTML]{E5ECF7}3.29                          & {[}2.39,4.39{]}  \\
1l solv                                      & solv                                   & 0                              & \cellcolor[HTML]{ADC4E3}0.15                       & {[}0.1,0.2{]}                         & \cellcolor[HTML]{E2E9F5}3.22                          & {[}2.33,4.29{]}  \\
1l asset                                     & asset                                  & 0                              & \cellcolor[HTML]{A6BFE0}0.14                       & {[}0.1,0.18{]}                        & \cellcolor[HTML]{DDE6F4}3.14                          & {[}2.29,4.11{]}  \\
\hline
\end{tabular}%
}
\caption{$\kappa$ and $b_{\kappa}$ medians and HDIs}
  \label{tab:scenarios_kappa_rd}
\end{table}

Table \ref{tab:scenarios_rho_rd_app} presents additional estimates of the network sparsity parameter.

\begin{table}[htbp!]
    \centering
    \resizebox{0.38\textwidth}{!}{%
    \begin{tabular}{lcc}
    \hline
   % \rowcolor{gray!20}
    \textbf{Model set-up} & \textbf{$\rho$} & \textbf{95\% HDI} \\ \hline
%1l nocovs                                    & \cellcolor[HTML]{F8696B}0.076                           & {[}0.065,0.088{]}                                         & 31\%                                                                                                 \\
1l base                                      & \cellcolor[HTML]{FBC6C8}0.051                           & {[}0.044,0.059{]}  \\
1l prof                                      & \cellcolor[HTML]{FBC6C9}0.051                           & {[}0.044,0.059{]}  \\
1l solv                                      & \cellcolor[HTML]{FBC6C9}0.051                           & {[}0.044,0.059{]}  \\
1l asset                                     & \cellcolor[HTML]{FBB9BC}0.055                           & {[}0.047,0.063{]}  \\
\hline
    \end{tabular}%
    }
    \caption{$\rho$ median and HDIs}
    \label{tab:scenarios_rho_rd_app}
\end{table}

\subsection{Contagion Channels and Covariates}
Tables \ref{tab:scenarios_lambda_rd_app}, \ref{tab:scenarios_mutau_rd_app}, \ref{tab:scenarios_covariates_rd_app} and \ref{tab:scenarios_weights_rd_app} present additional parameter and covariate estimates.

\begin{table}[htbp!]
\centering
\resizebox{0.78\textwidth}{!}{%
\begin{tabular}{lllllllll}
\hline
\multicolumn{1}{c}{\textbf{Model set-up}} & \multicolumn{1}{c}{\textbf{Avg $\lambda^{(0)}$}} & \multicolumn{1}{c}{\textbf{Sum $\lambda^{(0)}$}} & \multicolumn{1}{c}{\textbf{Min $\lambda^{(0)}$}} & \multicolumn{1}{c}{\textbf{Max $\lambda^{(0)}$}} & \multicolumn{1}{c}{\textbf{Avg 95\% HDI}} & \multicolumn{1}{c}{\textbf{Avg 95\% HDI/$\lambda^{(0)}$}} \\ \hline
%1l nocovs                                    & \cellcolor[HTML]{F8696B}0.0021                                                         & \cellcolor[HTML]{F8696B}0.2116                                                         & 0.0002                                                                                 & \cellcolor[HTML]{FAFBFD}0.0075                                                         & {[}0.0009,0.0037{]}                                           & 154\%                                                                                        & 77\%                                                                                         \\
1l base                                      & \cellcolor[HTML]{F88587}0.0022                                                         & \cellcolor[HTML]{F88587}0.2149                                                         & 0.0003                                                                                 & \cellcolor[HTML]{F0F8F5}0.0078                                                         & {[}0.0009,0.0037{]}                       & 153\%                                                                                        \\
\hline
\end{tabular}%
}
\caption{Averages, sums, minima and maxima of $\lambda^{(0)}$ medians and HDIs.}
\label{tab:scenarios_lambda_rd_app}
\end{table}

\begin{table}[htbp!]
\centering
\resizebox{0.65\textwidth}{!}{%
\begin{tabular}{lllllll}
\hline
\textbf{Model set-up} & \textbf{Avg $\mu$}             & \textbf{Avg $\tau$}         & \multicolumn{1}{c}{\textbf{Avg 95\% HDI $\mu$}} & \multicolumn{1}{c}{\textbf{Avg 95\% HDI $\tau$}} \\ \hline
%1l nocovs              & \cellcolor[HTML]{FBF1F4}(1.000) & \cellcolor[HTML]{B2C8E5}9.65 & {[}-1.21,-0.79{]}                            & {[}-1.1,-0.9{]}                                                          \\
1l base                & \cellcolor[HTML]{FBF6F9}(1.000) & \cellcolor[HTML]{C5D5EC}9.65 & {[}-1.21,-0.79{]}                            & {[}4.27,16.25{]}                              \\
1l prof                & \cellcolor[HTML]{FBF5F8}(1.000) & \cellcolor[HTML]{C3D4EB}9.65 & {[}-1.21,-0.79{]}                            & {[}4.28,16.25{]}                              \\
1l solv                & \cellcolor[HTML]{F7F8FD}(1.000) & \cellcolor[HTML]{BACEE8}9.65 & {[}-1.21,-0.79{]}                            & {[}4.27,16.24{]}                              \\
1l asset               & \cellcolor[HTML]{FBF6F9}(1.000) & \cellcolor[HTML]{E1E9F6}9.65 & {[}-1.21,-0.79{]}                            & {[}4.28,16.25{]}                              \\
\hline
\end{tabular}%
}
\caption{$\mu$ and $\tau$ medians and HDIs. Relative HDI values (HDI/$\mu$ and HDI/$\tau$) were identical across these model set-ups and are not displayed.}
\label{tab:scenarios_mutau_rd_app}
\end{table}

%\newpage
%\begin{landscape}
\begin{table}[htbp!]
\centering
\resizebox{0.95\textwidth}{!}{%
\begin{tabular}{lcccccccccccccccccccccc}
\hline
\textbf{Model set-up}& \textbf{Cov}&\textbf{L}&\textbf{$\beta_0$}&\textbf{95\% HDI}&\textbf{$\beta_1$}&\textbf{95\% HDI}&\textbf{$\beta_2$}&\textbf{95\%HDI}&\textbf{$\beta_3$}&\textbf{95\% HDI}\\
\hline
%1l nocovs                                    & none                                   & 0                              &                                                      &                                       &                                       &                                                      &                                                                  &                                                                  &                                                      &                                                                 &                                                                 &                                                      &                                                                 &                                                                 \\
1l base                                      & 0                                      & 0                              & \cellcolor[HTML]{FAAEB0}-2.19                        & {[}-2.29,-2.1{]}                      & \cellcolor[HTML]{F97F82}0.51                         & {[}-20.2,19.18{]}                                                &                                                      &                                                                 &                                                      &                                                                 \\
\hline
\end{tabular}%
}
\caption{Covariate coefficients. Green cells indicate the HDI excludes 0, and yellow indicates that the interval contains 0 but is skewed ($\geq$75\% on one side).}
\label{tab:scenarios_covariates_rd_app}
\end{table}
%\end{landscape}

%\newpage
%\begin{landscape}

\begin{table}[htbp!]
\centering
\resizebox{1\textwidth}{!}{%
\begin{tabular}{llllllllllll}
\hline
\textbf{Model set-up}         & \textbf{Sum $A$} & \textbf{Avg $AW$}                    & \textbf{Sum $AW$}                    & \textbf{Min $AW$}                  & \textbf{Max $AW$}                 & \textbf{95\%}      & \textbf{Len 95\%} & \textbf{95\%/$AW$} \\ \hline
%1l nocovs             & 143   & \cellcolor[HTML]{5A8AC6}0.081 & \cellcolor[HTML]{5A8AC6}11.55 & \cellcolor[HTML]{66BF7E}0.00 & \cellcolor[HTML]{F97C7E}0.41 & {[}0.006,0.24{]}  & {[}0.017,0.131{]} & 0.23             & 0.11             & 4.90                 & 2.15                 \\
1l base               & 140   & \cellcolor[HTML]{B9CDE7}0.095 & \cellcolor[HTML]{CBD9ED}13.31 & \cellcolor[HTML]{86CC99}0.04 & \cellcolor[HTML]{FBC0C2}0.28 & {[}0.008,0.19{]}  & 0.18             & 2.09                 \\
1l prof               & 145   & \cellcolor[HTML]{B1C7E4}0.094 & \cellcolor[HTML]{DDE6F4}13.59 & \cellcolor[HTML]{78C68D}0.02 & \cellcolor[HTML]{FBC0C2}0.28 & {[}0.008,0.193{]} & 0.18             & 2.20                 \\
1l solv               & 141   & \cellcolor[HTML]{C7D7EC}0.097 & \cellcolor[HTML]{E3EAF6}13.69 & \cellcolor[HTML]{83CB96}0.03 & \cellcolor[HTML]{FBC0C2}0.28 & {[}0.009,0.194{]} & 0.19             & 2.09                 \\
1l asset              & 145   & \cellcolor[HTML]{C7D6EC}0.097 & \cellcolor[HTML]{FCFCFF}14.07 & \cellcolor[HTML]{81CA95}0.03 & \cellcolor[HTML]{FAA1A3}0.34 & {[}0.01,0.193{]}  & 0.18             & 2.12                 \\
\hline
\end{tabular}%
}
\caption{Weight estimates for edges where $A_{i,j}=1$. Averages, sums, minima and maxima across all edges, as well as average HDI lengths, are shown for all empirical model set-ups. ``Average'' is abbreviated in headers for space.}
\label{tab:scenarios_weights_rd_app}
\end{table}

\newpage
\begin{table}[htbp!]
\centering
\resizebox{\textwidth}{!}{%
\begin{tabular}{llllllllllllllllllllllllllllllll}
\hline
\textbf{Model set-up} & \textbf{Avg$AW^{(0)}$} & \textbf{Avg$AW^{(1)}$} & \textbf{Avg$AW^{(2)}$} & \textbf{Sum$AW^{(0)}$} & \textbf{Sum$AW^{(1)}$} & \textbf{Sum$AW^{(2)}$} & \textbf{Min$AW^{(0)}$} & \textbf{Max$AW^{(0)}$} & \textbf{Min$AW^{(1)}$} & \textbf{Max$AW^{(1)}$} & \textbf{Min$AW^{(2)}$} & \textbf{Max$AW^{(2)}$} & \textbf{Len95\% $AW_0$} & \textbf{Len95\% $AW_1$} & \textbf{Len95\% $AW_2$} & \textbf{95\%/$AW^{(0)}$}  & \textbf{95\%/$AW^{(1)}$}  & \textbf{95\%/$AW^{(2)}$} \\ \hline
%1l nocovs & 143 & \cellcolor[HTML]{FCAB78}0.08 &  &  & \cellcolor[HTML]{FB9F76}11.55 &  &  & \cellcolor[HTML]{6CC183}0.00 & \cellcolor[HTML]{F9787A}0.41 &  &  &  &  & {[}0.01,0.24{]}  &                  &                  & {[}0.02,0.13{]}  &                  &                  & \cellcolor[HTML]{FB9975}0.23 &                              &                              & \cellcolor[HTML]{FBA076}0.11 &                              &                              & \cellcolor[HTML]{FFDD82}2.90 &                              &                              & \cellcolor[HTML]{FFDF82}1.42 &                              & \\
1l base & \cellcolor[HTML]{FA8471}0.10 &  &  & \cellcolor[HTML]{FA8471}13.31 &  &  & \cellcolor[HTML]{B5DFC2}0.04 & \cellcolor[HTML]{FAACAE}0.28 &  &  &  &  & \cellcolor[HTML]{FFDA81}0.18 &                              &                              & \cellcolor[HTML]{A2D07E}1.91 &                              & \\
1l prof & \cellcolor[HTML]{FA8871}0.09 &  &  & \cellcolor[HTML]{FA8070}13.59 &  &  & \cellcolor[HTML]{95D2A6}0.02 & \cellcolor[HTML]{FAACAE}0.28 &  &  &  &  & \cellcolor[HTML]{FED780}0.18 &                              &                              & \cellcolor[HTML]{AAD27F}1.97 &                              & \\
\hline
\end{tabular}%
}
\caption{Model-layer excitation-weight estimates for edges where $A_{i,j}=1$. Averages, sums, minima and maxima are across all edges.}
\label{tab:scenarios_weights_layer_rd_app}
\end{table}

%\end{landscape}

\subsection{Network Measures}\label{sec:empirical_app_network_measures}
Tables \ref{tab:degree_tests_app}, \ref{tab:scenarios_density_rd_app}, \ref{tab:scenarios_reciprocity_rd_app}, \ref{tab:scenarios_pearsoncorrelation_rd_app}, \ref{tab:scenarios_transitivity_rd_app}, \ref{tab:scenarios_averageclustering_rd_app}, \ref{tab:scenarios_acw_rd_app} and \ref{tab:scenarios_acw_layer_rd_app} present additional estimates of network measures for the fitted model set-ups. Table \ref{tab:topnodes_names} lists the names and frequencies, and Table \ref{tab:topnodes_list} details the nodes flagged in the top 10 list.

\begin{table}[htbp!]
\centering
\resizebox{0.495\textwidth}{!}{%
\begin{tabular}{lllllllll}
\hline
\multicolumn{1}{c}{\multirow{2}{*}{\textbf{Model set-up}}} & \multicolumn{1}{c}{\multirow{2}{*}{\textbf{Fit 1}}} & \multicolumn{1}{c}{\multirow{2}{*}{\textbf{Fit 2}}} & \multicolumn{3}{l}{\textbf{Degree}} & \multicolumn{3}{l}{\textbf{Weighted degree}} \\ \cline{4-9}
\multicolumn{1}{c}{}                                   & \multicolumn{1}{c}{}                                & \multicolumn{1}{c}{}                                & \textbf{LR}    & KS      & p-val    & LR            & KS           & p-val         \\ \hline
1l base                                                & KDE                                                 & Log normal                                          & (3)            & 0.47    & 0.00     & (2)           & 0.44         & 0.00          \\
1l base                                                & KDE                                                 & Power law                                           & 1,052          & 1.00    & 0.00     & (29)          & 0.49         & 0.00          \\
1l base                                                & KDE                                                 & Exponential                                         & 11             & 0.31    & 0.00     & 15            & 0.26         & 0.00          \\
1l base                                                & Log normal                                          & Power law                                           & 1,055          & 1.00    & 0.00     & (26)          & 0.46         & 0.00          \\
1l base                                                & Log normal                                          & Exponential                                         & 14             & 0.39    & 0.00     & 17            & 0.35         & 0.00          \\
\hline
\end{tabular}%
}
\resizebox{0.495\textwidth}{!}{%
\begin{tabular}{lllllllll}
\hline
\multicolumn{1}{c}{\multirow{2}{*}{\textbf{Model set-up}}} & \multicolumn{1}{c}{\multirow{2}{*}{\textbf{Fit 1}}} & \multicolumn{1}{c}{\multirow{2}{*}{\textbf{Fit 2}}} & \multicolumn{3}{l}{\textbf{Degree}} & \multicolumn{3}{l}{\textbf{Weighted degree}} \\ \cline{4-9}
\multicolumn{1}{c}{}                                   & \multicolumn{1}{c}{}                                & \multicolumn{1}{c}{}                                & \textbf{LR}    & KS      & p-val    & LR            & KS           & p-val         \\ \hline
2l asset solv                                          & KDE                                                 & Log normal                                          & (3)            & 0.41    & 0.00     & (3)           & 0.43         & 0.00          \\
2l asset solv                                          & KDE                                                 & Power law                                           & 737            & 0.99    & 0.00     & (65)          & 0.54         & 0.00          \\
2l asset solv                                          & KDE                                                 & Exponential                                         & 15             & 0.48    & 0.00     & 11            & 0.21         & 0.02          \\
2l asset solv                                          & Log normal                                          & Power law                                           & 740            & 0.97    & 0.00     & (62)          & 0.52         & 0.00          \\
2l asset solv                                          & Log normal                                          & Exponential                                         & 18             & 0.48    & 0.00     & 14            & 0.29         & 0.00          \\
\hline
\end{tabular}%
}
\caption{Degree distribution fit. LR is the log likelihood ratio between Fit 1 and Fit 2: positive values favour Fit 1, while parenthesised values denote negative values and favour Fit 2. KS and its p-value compare the two fitted distributions; they are not an absolute test of whether either candidate fits the observed degrees.}
\label{tab:degree_tests_app}
\end{table}

\begin{figure}[htbp]
\centering
%  -  -  -  -  - - Row 1  -  -  -  -  - -
\begin{subfigure}[t]{0.495\textwidth}
  \centering
  \resizebox{\textwidth}{!}{%
    \begin{tabular}{llllllll}
    \hline
    \textbf{Model set-up} & \textbf{$D$}                    & \textbf{95\% HDI} & \textbf{Len 95\%} & \textbf{95\%/$D$} \\ \hline
    1l base                  & \cellcolor[HTML]{F97F81}0.052 & {[}0.05,0.06{]}   & 0.012                 & 23\%                  \\
    1l prof                  & \cellcolor[HTML]{F98183}0.052 & {[}0.05,0.06{]}   & 0.012                 & 24\%                  \\
    1l solv                  & \cellcolor[HTML]{F8696B}0.052 & {[}0.05,0.06{]}   & 0.012                 & 24\%                  \\
    1l asset                 & \cellcolor[HTML]{F8696B}0.055 & {[}0.05,0.06{]}   & 0.013                 & 24\%                  \\
    2l base                  & \cellcolor[HTML]{F8F9FD}0.033 & {[}0.03,0.04{]}   & 0.008                 & 23\%                  \\
    2l asset prof            & \cellcolor[HTML]{FCFCFF}0.034 & {[}0.03,0.04{]}   & 0.008                 & 24\%                  \\
    2l solv prof             & \cellcolor[HTML]{EAEFF8}0.033 & {[}0.03,0.04{]}   & 0.007                 & 22\%                  \\
    2l asset solv            & \cellcolor[HTML]{FAFAFE}0.033 & {[}0.03,0.04{]}   & 0.008                 & 24\%                  \\
    3l base                  & \cellcolor[HTML]{5D8CC7}0.025 & {[}0.02,0.03{]}   & 0.006                 & 22\%                  \\
    3l all                   & \cellcolor[HTML]{5A8AC6}0.025 & {[}0.02,0.03{]}   & 0.006                 & 23\%                  \\ \hline
    \end{tabular}%
  }
  \caption{Medians of density metric (D).}
  \label{tab:scenarios_density_rd_app}
\end{subfigure}
\hfill
\begin{subfigure}[t]{0.495\textwidth}
  \centering
  \resizebox{\textwidth}{!}{%
    \begin{tabular}{llllllll}
    \hline
    \textbf{Model set-up}       & \textbf{$R$}           & \textbf{95\% HDI}                    & \textbf{Len 95\%} & \textbf{95\%/$R$} \\ \hline
    1l base                  & \cellcolor[HTML]{F97A7C}0.061 & {[}0.04,0.09{]}   & 0.05                  & 87\%                  \\
    1l prof                  & \cellcolor[HTML]{F97E80}0.060 & {[}0.03,0.09{]}   & 0.05                  & 89\%                  \\
    1l solv                  & \cellcolor[HTML]{F97A7C}0.061 & {[}0.04,0.09{]}   & 0.05                  & 87\%                  \\
    1l asset                 & \cellcolor[HTML]{F8696B}0.063 & {[}0.04,0.09{]}   & 0.05                  & 84\%                  \\
    2l base                  & \cellcolor[HTML]{F9FAFE}0.044 & {[}0.02,0.07{]}   & 0.05                  & 119\%                 \\
    2l asset prof            & \cellcolor[HTML]{F7F8FD}0.044 & {[}0.02,0.07{]}   & 0.05                  & 120\%                 \\
    2l solv prof             & \cellcolor[HTML]{F2F5FB}0.044 & {[}0.02,0.07{]}   & 0.05                  & 119\%                 \\
    2l asset solv            & \cellcolor[HTML]{FCFCFF}0.044 & {[}0.02,0.07{]}   & 0.05                  & 121\%                 \\
    3l base                  & \cellcolor[HTML]{5A8AC6}0.035 & {[}0.01,0.07{]}   & 0.05                  & 150\%                 \\
    3l all                   & \cellcolor[HTML]{6F99CD}0.036 & {[}0.01,0.07{]}   & 0.05                  & 143\%                 \\ \hline
    \end{tabular}%
  }
  \caption{Medians of reciprocity metric (R).}
  \label{tab:scenarios_reciprocity_rd_app}
\end{subfigure}

%\vspace{0.5cm}
%  -  -  -  -  - - Row 2  -  -  -  -  - -
\begin{subfigure}[t]{0.51\textwidth}
  \centering
  \resizebox{1.0\textwidth}{!}{%
    \begin{tabular}{llllllll}
    \hline
    \textbf{Model set-up}       & \textbf{$PC$}            & \textbf{95\% HDI} & \textbf{Len 95\%} & \textbf{95\%/$PC$} \\ \hline
    1l base                  & \cellcolor[HTML]{F8696B}(0.047) & {[}-0.12,0.03{]}  & 0.15                  & 315\%                 \\
    1l prof                  & \cellcolor[HTML]{F96B6D}(0.047) & {[}-0.12,0.03{]}  & 0.15                  & 312\%                 \\
    1l solv                  & \cellcolor[HTML]{FBD3D6}(0.056) & {[}-0.13,0.02{]}  & 0.15                  \\
    1l asset                 & \cellcolor[HTML]{FA9194}(0.051) & {[}-0.12,0.02{]}  & 0.14                  \\
    2l base                  & \cellcolor[HTML]{FCF8FB}(0.060) & {[}-0.15,0.03{]}  & 0.18                  \\
    2l asset prof            & \cellcolor[HTML]{EEF2FA}(0.060) & {[}-0.15,0.03{]}  & 0.18                  \\
    2l solv prof             & \cellcolor[HTML]{CCDAEE}(0.061) & {[}-0.15,0.03{]}  & 0.18                  \\
    2l asset solv            & \cellcolor[HTML]{5A8AC6}(0.064) & {[}-0.15,0.03{]}  & 0.18                  \\
    3l base                  & \cellcolor[HTML]{87AAD6}(0.063) & {[}-0.17,0.04{]}  & 0.21                  \\
    3l all                   & \cellcolor[HTML]{E8EEF8}(0.060) & {[}-0.16,0.05{]}  & 0.21                  \\ \hline
    \end{tabular}%
  }
  \caption{Medians of degree Pearson correlation (PC).}
  \label{tab:scenarios_pearsoncorrelation_rd_app}
\end{subfigure}
\hfill
\begin{subfigure}[t]{0.48\textwidth}
  \centering
  \resizebox{1\textwidth}{!}{%
    \begin{tabular}{llllllll}
    \hline
    \textbf{Model set-up}       & \textbf{$T$}           & \textbf{95\% HDI}                    & \textbf{Len 95\%} & \textbf{95\%/$T$} \\ \hline
    1l base                  & \cellcolor[HTML]{F97B7D}0.050 & {[}0.04,0.06{]}   & 0.02                  & 39\%                  \\
    1l prof                  & \cellcolor[HTML]{F97F81}0.049 & {[}0.04,0.06{]}   & 0.02                  & 40\%                  \\
    1l solv                  & \cellcolor[HTML]{F98082}0.049 & {[}0.04,0.06{]}   & 0.02                  & 40\%                  \\
    1l asset                 & \cellcolor[HTML]{F8696B}0.052 & {[}0.04,0.06{]}   & 0.02                  & 38\%                  \\
    2l base                  & \cellcolor[HTML]{F7F9FD}0.029 & {[}0.02,0.04{]}   & 0.02                  & 62\%                  \\
    2l asset prof            & \cellcolor[HTML]{FCFCFF}0.030 & {[}0.02,0.04{]}   & 0.02                  & 62\%                  \\
    2l solv prof             & \cellcolor[HTML]{E9EFF8}0.029 & {[}0.02,0.04{]}   & 0.02                  & 62\%                  \\
    2l asset solv            & \cellcolor[HTML]{F9FAFE}0.030 & {[}0.02,0.04{]}   & 0.02                  & 62\%                  \\
    3l base                  & \cellcolor[HTML]{5A8AC6}0.020 & {[}0.01,0.03{]}   & 0.02                  & 85\%                  \\
    3l all                   & \cellcolor[HTML]{5A8AC6}0.020 & {[}0.01,0.03{]}   & 0.02                  & 84\%                  \\ \hline
    \end{tabular}%
  }
  \caption{Medians of transitivity (T).}
  \label{tab:scenarios_transitivity_rd_app}
\end{subfigure}

%\vspace{0.5cm}
%  -  -  -  -  - - Row 3  -  -  -  -  - -
\begin{subfigure}[t]{0.48\textwidth}
  \centering
  \resizebox{1\textwidth}{!}{%
    \begin{tabular}{llllllll}
    \hline
    \textbf{Model set-up}       & \textbf{$AC$}          & \textbf{95\% HDI}                    & \textbf{Len 95\%} & \textbf{95\%/$AC$} \\ \hline
    1l base                  & \cellcolor[HTML]{F97A7C}0.062 & {[}0.05,0.07{]}   & 0.02                  & 39\%                  \\
    1l prof                  & \cellcolor[HTML]{F97C7E}0.062 & {[}0.05,0.07{]}   & 0.02                  & 39\%                  \\
    1l solv                  & \cellcolor[HTML]{F97C7E}0.062 & {[}0.05,0.07{]}   & 0.02                  & 39\%                  \\
    1l asset                 & \cellcolor[HTML]{F8696B}0.064 & {[}0.05,0.08{]}   & 0.02                  & 37\%                  \\
    2l base                  & \cellcolor[HTML]{F0F3FA}0.044 & {[}0.03,0.06{]}   & 0.03                  & 74\%                  \\
    2l asset prof            & \cellcolor[HTML]{FCFCFF}0.045 & {[}0.03,0.06{]}   & 0.03                  & 73\%                  \\
    2l solv prof             & \cellcolor[HTML]{F3F5FB}0.044 & {[}0.03,0.06{]}   & 0.03                  & 75\%                  \\
    2l asset solv            & \cellcolor[HTML]{FAFAFE}0.045 & {[}0.03,0.06{]}   & 0.03                  & 72\%                  \\
    3l base                  & \cellcolor[HTML]{5A8AC6}0.032 & {[}0.02,0.05{]}   & 0.03                  & 108\%                 \\
    3l all                   & \cellcolor[HTML]{6693CA}0.033 & {[}0.02,0.05{]}   & 0.04                  & 109\%                 \\ \hline
    \end{tabular}%
  }
  \caption{Medians of average clustering (AC).}
  \label{tab:scenarios_averageclustering_rd_app}
\end{subfigure}
\hfill
\begin{subfigure}[t]{0.51\textwidth}
  \centering
  \resizebox{1\textwidth}{!}{%
    \begin{tabular}{llllllll}
    \hline
    \textbf{Model set-up}       & \textbf{$ACW$}          & \textbf{95\% HDI} & \textbf{Len 95\%} & \textbf{95\%/$ACW$} \\ \hline
    1l base                  & \cellcolor[HTML]{F8696B}0.021 & {[}0.01,0.03{]}   & 0.01                  & 66\%                  \\
    1l prof                  & \cellcolor[HTML]{F97D7F}0.021 & {[}0.01,0.03{]}   & 0.01                  & 67\%                  \\
    1l solv                  & \cellcolor[HTML]{FA9698}0.020 & {[}0.01,0.03{]}   & 0.01                  & 66\%                  \\
    1l asset                 & \cellcolor[HTML]{E6ECF7}0.017 & {[}0.01,0.02{]}   & 0.01                  & 70\%                  \\
    2l base                  & \cellcolor[HTML]{FBC4C6}0.019 & {[}0.01,0.03{]}   & 0.02                  & 85\%                  \\
    2l asset prof            & \cellcolor[HTML]{FCFAFD}0.017 & {[}0.01,0.03{]}   & 0.02                  & 87\%                  \\
    2l solv prof             & \cellcolor[HTML]{F5F7FC}0.017 & {[}0.01,0.03{]}   & 0.02                  & 90\%                  \\
    2l asset solv            & \cellcolor[HTML]{F8F9FD}0.017 & {[}0.01,0.03{]}   & 0.01                  & 86\%                  \\
    3l base                  & \cellcolor[HTML]{94B2DA}0.015 & {[}0.01,0.03{]}   & 0.02                  & 116\%                 \\
    3l all                   & \cellcolor[HTML]{5A8AC6}0.014 & {[}0.01,0.02{]}   & 0.02                  & 118\%                 \\ \hline
    \end{tabular}%
  }
  \caption{Medians of weighted average clustering (ACW).}
  \label{tab:scenarios_acw_rd_app}
\end{subfigure}
\caption{Summary of network metrics for each empirical model set-up with corresponding HDIs and HDI lengths.}
\label{fig:all_metrics_app}
\end{figure}

\begin{figure}[t]
   \centering
  \resizebox{0.78\textwidth}{!}{%
    \begin{tabular}{lllllllll}
    \hline
    \multicolumn{1}{l|}{} & \multicolumn{4}{c|}{\textbf{$ACW$}} & \multicolumn{1}{c}{\textbf{full}} & \multicolumn{1}{c}{\textbf{L=0}} & \multicolumn{1}{c}{\textbf{L=1}} & \multicolumn{1}{c}{\textbf{L=2}} \\ \cline{2-9}
    \multicolumn{1}{l|}{\multirow{-2}{*}{\textbf{Model set-up}}} & \cellcolor[HTML]{FFFFFF}\textbf{full} & \textbf{L=0} & \textbf{L=1} & \textbf{L=2} & \textbf{95\% HDI/med} & \textbf{95\% HDI/med} & \textbf{95\% HDI/med} & \textbf{95\% HDI/med} \\ \hline
    1l base                         & \cellcolor[HTML]{F8696B}0.021 & \cellcolor[HTML]{F8696B}0.0213 & & & 66\% & & & \\
    1l prof                         & \cellcolor[HTML]{F9736D}0.021 & \cellcolor[HTML]{F9736D}0.0208 & & & 67\% & & & \\
    1l solv                         & \cellcolor[HTML]{FA8070}0.020 & \cellcolor[HTML]{FA8070}0.0201 & & & 66\% & & & \\
    1l asset                        & \cellcolor[HTML]{FDBC7B}0.017 & \cellcolor[HTML]{FDBC7B}0.0169 & & & 70\% & & & \\
    2l base                         & \cellcolor[HTML]{FB9774}0.019 & \cellcolor[HTML]{FDEA83}0.0143 & \cellcolor[HTML]{FFEB84}0.0144 & & 85\% & 95\% & 94\% & \\
    2l asset prof                   & \cellcolor[HTML]{FCB27A}0.017 & \cellcolor[HTML]{C8DB80}0.0124 & \cellcolor[HTML]{E7E482}0.0135 & & 87\% & 104\% & 101\% & \\
\hline          
    \end{tabular}%
  }
  \caption{Medians of model-layer-specific weighted average clustering coefficients (ACW). HDI lengths relative to median estimates are shown for all empirical model set-ups.}
  \label{tab:scenarios_acw_layer_rd_app}
\end{figure}

\begin{figure}[htbp!]
\centering
\includegraphics[width=0.76\linewidth, angle=0, trim= 20 20 30 31,clip]{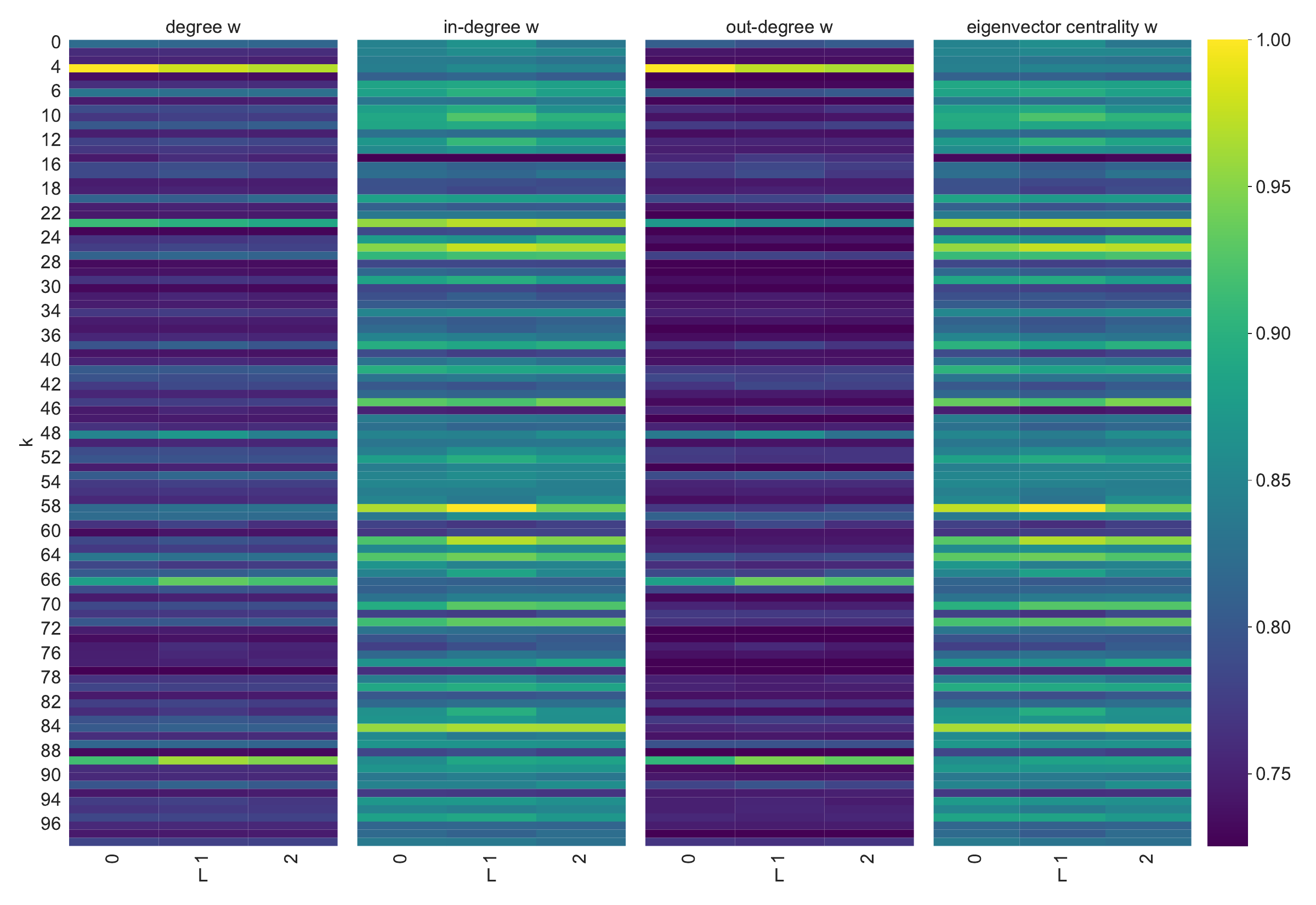}
\caption{Medians of weighted metrics for the ``3l all'' model set-up, normalized by maximum values across channels. From left: degree, in-degree, out-degree and eigenvector centrality.}
\label{fig:3l_all_l}
\end{figure}

\begin{table}[htbp!]
\centering
\resizebox{\textwidth}{!}{%
\begin{tabular}{rcrclllllrrr}
\hline
\textbf{Node} & \textbf{Freq. top 10} & \textbf{Freq. top 5} & \textbf{Ticker} & \textbf{Index} & \textbf{Name} & \textbf{Sector} & \textbf{Industry} & \textbf{Reg. of lg. rev.} & \textbf{\# Events} & \textbf{Solv} & \textbf{Prof} \\ \hline
22 & 19 & 18 & C   & SPX  & Citigroup Inc                    & Financials       & Banking                   & The Americas & 32 & 10 & 8 \\
63 & 15 & 1  & LNC & SPX  & Lincoln National Corp            & Financials       & Insurance                 & The Americas & 25 & 9  & 10 \\
57 & 14 & 7  & ISP & SX5E & Intesa Sanpaolo SpA              & Financials       & Banking                   & EMEA         & 37 & 1  & 9 \\
6  & 11 & 1  & ALV & SX5E & Allianz SE                       & Financials       & Insurance                 &              & 26 & 10 & 10 \\
3  & 11 & 10 & AIG & SPX  & American Intl. Group Inc         & Financials       & Insurance                 &              & 34 & 10 & 9 \\
66 & 11 & 11 & MAR & SPX  & Marriott Intl. Inc/MD            & Cons. Discr.     & Leisure Facilities \& Svcs & The Americas & 10 & 1  & 2 \\
48 & 11 & 10 & HEI & DAX  & HeidelbergCement AG              & Materials        & Construction Materials    & EMEA         & 20 & 9  & 4 \\
88 & 10 & 10 & TSN & SPX  & Tyson Foods Inc                  & Cons. Stpls.     & Food                      &              & 15 & 5  & 5 \\
58 & 10 & 0  & JPM & SPX  & JPMorgan Chase \& Co             & Financials       & Banking                   & The Americas & 26 & 9  & 8 \\
44 & 8  & 4  & GE  & SPX  & General Electric Co              & Industrials      & Diversified Industrials   & The Americas & 26 & 10 & 3 \\
84 & 8  & 8  & SPG & SPX  & Simon Property Group Inc         & Real Estate      & REIT                      &              & 22 & 5  & 5 \\
25 & 8  & 8  & CB  & SPX  & Chubb Ltd                        & Financials       & Insurance                 &              & 55 & 10 & 9 \\
69 & 7  & 0  & MET & SPX  & MetLife Inc                      & Financials       & Insurance                 & The Americas & 23 & 9  & 10 \\
61 & 7  & 5  & L   & SPX  & Loews Corp                       & Financials       & Insurance                 &              & 24 & 9  & 9 \\
71 & 7  & 0  & MUV & SX5E & Munich Re                        & Financials       & Insurance                 &              & 26 & 10 & 10 \\
0  & 7  & 0  & AAL & UKX  & Anglo American PLC               & Materials        & Metals \& Mining          & Asia Pacific & 20 & 7  & 9 \\
26 & 6  & 0  & CON & DAX  & Continental AG                   & Cons. Discr.     & Automotive                & EMEA         & 19 & 8  & 6 \\
86 & 4  & 0  & TGT & SPX  & Target Corp                      & Cons. Stpls.     & Retail - Cons. Stpls.     & The Americas & 24 & 7  & 6 \\
37 & 3  & 2  & DHR & SPX  & Danaher Corp                     & Health Care      & Medical Equip. \& Dev.    &              & 19 & 6  & 2 \\
53 & 3  & 0  & HPQ & SPX  & HP Inc                           & Technology       & Technology Hardw.         &              & 20 & 5  & 5 \\
19 & 2  & 0  & BP  & UKX  & BP PLC                           & Energy           & Oil \& Gas Producers      &              & 21 & 3  & 9 \\
65 & 1  & 0  & LUV & SPX  & Southwest Airlines Co            & Industrials      & Transportation \& Logistics & The Americas & 15 & 3  & 9 \\
9  & 1  & 0  & AZN & UKX  & AstraZeneca PLC                  & Health Care      & Biotech \& Pharma         & Asia Pacific & 39 & 4  & 8 \\
10 & 1  & 0  & BA  & SPX  & Boeing Co/The                    & Industrials      & Aerospace \& Defense      & The Americas & 21 & 4  & 7 \\
29 & 1  & 0  & CS  & SX5E & AXA SA                           & Financials       & Insurance                 & EMEA         & 21 & 6  & 10 \\
91 & 1  & 0  & VIA & SPX  & ViacomCBS Inc                    & Communications   & Entertainment Content     &              & 23 & 8  & 3 \\
24 & 1  & 0  & CAT & SPX  & Caterpillar Inc                  & Industrials      & Machinery                 & The Americas & 26 & 6  & 3 \\
8  & 1  & 0  & AV  & UKX  & Aviva PLC                        & Financials       & Insurance                 & EMEA         & 21 & 2  & 10 \\
40 & 1  & 0  & ENE & SX5E & Enel SpA                         & Utilities        & Electric Utilities        & EMEA         & 20 & 2  & 5 \\ \hline
\end{tabular}%
}
\caption{Details of the nodes flagged in at least one top-10 node-importance list.}
\label{tab:topnodes_list}
\end{table}

\subsection{Time Variation Supporting Tables}
Tables \ref{tab:timevariation_covariates}--\ref{tab:timevariation_layers} report the supporting covariate, network-measure and layer-level summaries for the regime-specific refits discussed in Section \ref{sec:empirical_time_variation_3all}. Table \ref{tab:timevariation_topnodes} reports the corresponding ticker-level top-node rankings.

\begin{table}[htbp!]
\centering
\resizebox{\textwidth}{!}{%
\begin{tabular}{lllcccccccccccc}
\arrayrulecolor{black}\hline
\textbf{Period} & \textbf{Cov.} & \textbf{L} & \textbf{$\beta_0$} & \textbf{95\% HDI} & \textbf{68\% HDI} & \textbf{$\beta_1$} & \textbf{95\% HDI} & \textbf{68\% HDI} & \textbf{$\beta_2$} & \textbf{95\% HDI} & \textbf{68\% HDI} & \textbf{$\beta_3$} & \textbf{95\% HDI} & \textbf{68\% HDI} \\
\arrayrulecolor{black}\hline
Full period & asset & 0 & -2.57 & \cellcolor[HTML]{C6EFCE}{\color[HTML]{006100} {[}-2.80,-2.32{]}} & \cellcolor[HTML]{C6EFCE}{\color[HTML]{006100} {[}-2.70,-2.44{]}} & -0.12 & {[}-0.86,0.59{]} & {[}-0.46,0.29{]} & 0.28 & {[}-0.61,1.17{]} & \cellcolor[HTML]{FFEB9C}{\color[HTML]{9C5700} {[}-0.06,0.77{]}} & 0.14 & {[}-0.23,0.59{]} & \cellcolor[HTML]{FFEB9C}{\color[HTML]{9C5700} {[}-0.03,0.36{]}} \\
Full period & solv & 1 & -2.23 & \cellcolor[HTML]{C6EFCE}{\color[HTML]{006100} {[}-2.86,-1.68{]}} & \cellcolor[HTML]{C6EFCE}{\color[HTML]{006100} {[}-2.56,-1.98{]}} & -0.07 & \cellcolor[HTML]{FFEB9C}{\color[HTML]{9C5700} {[}-0.13,0.00{]}} & \cellcolor[HTML]{C6EFCE}{\color[HTML]{006100} {[}-0.10,-0.03{]}} & 0.03 & {[}-0.03,0.09{]} & \cellcolor[HTML]{FFEB9C}{\color[HTML]{9C5700} {[}-0.00,0.06{]}} &  &  &  \\
Full period & prof & 2 & -2.28 & \cellcolor[HTML]{C6EFCE}{\color[HTML]{006100} {[}-2.89,-1.72{]}} & \cellcolor[HTML]{C6EFCE}{\color[HTML]{006100} {[}-2.59,-1.96{]}} & -0.05 & \cellcolor[HTML]{FFEB9C}{\color[HTML]{9C5700} {[}-0.11,0.02{]}} & \cellcolor[HTML]{C6EFCE}{\color[HTML]{006100} {[}-0.08,-0.01{]}} & 0.02 & {[}-0.05,0.08{]} & {[}-0.02,0.05{]} &  &  &  \\
Pre-GFC/GFC & asset & 0 & -1.94 & \cellcolor[HTML]{C6EFCE}{\color[HTML]{006100} {[}-2.34,-1.54{]}} & \cellcolor[HTML]{C6EFCE}{\color[HTML]{006100} {[}-2.13,-1.74{]}} & -0.33 & {[}-6.35,3.02{]} & {[}-1.30,0.73{]} & 0.16 & {[}-3.23,5.82{]} & {[}-1.01,1.19{]} & -0.09 & {[}-0.79,0.62{]} & {[}-0.46,0.22{]} \\
Pre-GFC/GFC & solv & 1 & -1.88 & \cellcolor[HTML]{C6EFCE}{\color[HTML]{006100} {[}-2.87,-0.96{]}} & \cellcolor[HTML]{C6EFCE}{\color[HTML]{006100} {[}-2.41,-1.45{]}} & -0.03 & {[}-0.15,0.08{]} & \cellcolor[HTML]{FFEB9C}{\color[HTML]{9C5700} {[}-0.09,0.02{]}} & 0.01 & {[}-0.08,0.12{]} & {[}-0.03,0.06{]} &  &  &  \\
Pre-GFC/GFC & prof & 2 & -1.98 & \cellcolor[HTML]{C6EFCE}{\color[HTML]{006100} {[}-2.98,-1.12{]}} & \cellcolor[HTML]{C6EFCE}{\color[HTML]{006100} {[}-2.44,-1.50{]}} & -0.03 & {[}-0.12,0.08{]} & \cellcolor[HTML]{FFEB9C}{\color[HTML]{9C5700} {[}-0.08,0.02{]}} & 0.02 & {[}-0.08,0.12{]} & \cellcolor[HTML]{FFEB9C}{\color[HTML]{9C5700} {[}-0.02,0.08{]}} &  &  &  \\
Euro crisis/calm & asset & 0 & 0.33 & {[}-19.50,19.04{]} & {[}-7.75,10.94{]} & 0.25 & {[}-18.27,19.34{]} & {[}-9.31,9.88{]} & -0.02 & {[}-20.35,18.52{]} & {[}-9.41,9.74{]} & 0.23 & {[}-18.42,21.29{]} & {[}-10.14,10.15{]} \\
Euro crisis/calm & solv & 1 & 0.08 & {[}-21.26,20.98{]} & {[}-10.70,10.62{]} & 0.46 & {[}-18.20,20.88{]} & {[}-9.92,9.75{]} & 0.19 & {[}-18.18,22.38{]} & {[}-9.10,10.00{]} &  &  &  \\
Euro crisis/calm & prof & 2 & -0.22 & {[}-18.98,20.49{]} & {[}-10.38,9.53{]} & 0.12 & {[}-18.95,17.75{]} & {[}-9.36,8.91{]} & 0.15 & {[}-19.89,19.00{]} & {[}-8.99,9.84{]} &  &  &  \\
COVID-19 & asset & 0 & -0.89 & \cellcolor[HTML]{C6EFCE}{\color[HTML]{006100} {[}-1.47,-0.38{]}} & \cellcolor[HTML]{C6EFCE}{\color[HTML]{006100} {[}-1.14,-0.65{]}} & -0.15 & {[}-2.03,1.35{]} & {[}-0.75,0.69{]} & 0.36 & {[}-10.50,14.88{]} & {[}-1.81,2.43{]} & 0.04 & {[}-0.86,0.98{]} & {[}-0.36,0.59{]} \\
COVID-19 & solv & 1 & -0.55 & {[}-1.92,0.83{]} & \cellcolor[HTML]{FFEB9C}{\color[HTML]{9C5700} {[}-1.15,0.11{]}} & -0.06 & {[}-0.21,0.10{]} & \cellcolor[HTML]{FFEB9C}{\color[HTML]{9C5700} {[}-0.14,0.01{]}} & 0.00 & {[}-0.15,0.18{]} & {[}-0.08,0.08{]} &  &  &  \\
COVID-19 & prof & 2 & -0.37 & {[}-1.67,1.00{]} & \cellcolor[HTML]{FFEB9C}{\color[HTML]{9C5700} {[}-1.04,0.32{]}} & -0.05 & {[}-0.20,0.09{]} & \cellcolor[HTML]{FFEB9C}{\color[HTML]{9C5700} {[}-0.12,0.03{]}} & -0.03 & {[}-0.18,0.11{]} & \cellcolor[HTML]{FFEB9C}{\color[HTML]{9C5700} {[}-0.11,0.03{]}} &  &  &  \\
\hline
\end{tabular}%
}
\caption{Channel-covariate coefficients for the time-variation refits. Posterior medians, 95\% HDIs and 68\% HDIs are reported in separate columns. Green cells indicate the HDI excludes 0, and yellow indicates that the interval contains 0 but is skewed ($>$75\% on one side).}
\label{tab:timevariation_covariates}
\end{table}

\begin{table}[htbp!]
\centering
\resizebox{0.90\textwidth}{!}{%
\begin{tabular}{lrrrrrrr}
\arrayrulecolor{black}\hline
\textbf{Period} & \textbf{$\rho$} & \textbf{$R$} & \textbf{$PC$} & \textbf{$T$} & \textbf{$AC$} & \textbf{$ACW$} & \textbf{Skew w-deg.} \\
\arrayrulecolor{black}\hline
Full period & \cellcolor[HTML]{F8696B}0.025 & \cellcolor[HTML]{F8696B}0.035 & \cellcolor[HTML]{FAA1A4}-0.055 & \cellcolor[HTML]{F8696B}0.028 & \cellcolor[HTML]{F8696B}0.018 & \cellcolor[HTML]{F8696B}0.011 & \cellcolor[HTML]{F99396}2.59 \\
Pre-GFC/GFC & \cellcolor[HTML]{FBE8EB}0.014 & \cellcolor[HTML]{5A8AC6}0.000 & \cellcolor[HTML]{F8696B}0.002 & \cellcolor[HTML]{5A8AC6}0.000 & \cellcolor[HTML]{5A8AC6}0.000 & \cellcolor[HTML]{5A8AC6}0.000 & \cellcolor[HTML]{F99193}2.61 \\
Euro crisis/calm & \cellcolor[HTML]{5A8AC6}0.000 & -- & -- & \cellcolor[HTML]{5A8AC6}0.000 & \cellcolor[HTML]{5A8AC6}0.000 & \cellcolor[HTML]{5A8AC6}0.000 & \cellcolor[HTML]{5A8AC6}0.00 \\
COVID-19 & \cellcolor[HTML]{D5E0F1}0.010 & \cellcolor[HTML]{5A8AC6}0.000 & \cellcolor[HTML]{5A8AC6}-0.296 & \cellcolor[HTML]{5A8AC6}0.000 & \cellcolor[HTML]{5A8AC6}0.000 & \cellcolor[HTML]{5A8AC6}0.000 & \cellcolor[HTML]{F8696B}3.03 \\
\hline
\end{tabular}%
}
\caption{Posterior-median graph summaries aligned with the network measures reported in Figure \ref{fig:all_metrics_app}: reciprocity ($R$), degree Pearson correlation ($PC$), transitivity ($T$), average clustering ($AC$), weighted average clustering ($ACW$), and weighted-degree concentration. Parentheses are avoided here; negative $PC$ values are shown with their sign.}
\label{tab:timevariation_network}
\end{table}

\begin{table}[htbp!]
\centering
\resizebox{\textwidth}{!}{%
\begin{tabular}{lrrrrrrrrrr}
\hline
\textbf{Period} & \textbf{Sum $A$} & \textbf{Avg $AW^{(0)}$} & \textbf{Avg $AW^{(1)}$} & \textbf{Avg $AW^{(2)}$} & \textbf{Sum $AW^{(0)}$} & \textbf{Sum $AW^{(1)}$} & \textbf{Sum $AW^{(2)}$} & \textbf{Max $AW^{(0)}$} & \textbf{Max $AW^{(1)}$} & \textbf{Max $AW^{(2)}$} \\
\hline
Full period & 113 & \cellcolor[HTML]{5A8AC6}0.07 & \cellcolor[HTML]{5E8DC8}0.08 & \cellcolor[HTML]{5B8AC6}0.07 & \cellcolor[HTML]{F87678}8.20 & \cellcolor[HTML]{F8696B}8.56 & \cellcolor[HTML]{F87476}8.24 & \cellcolor[HTML]{7CA2D2}0.15 & \cellcolor[HTML]{6592CA}0.13 & \cellcolor[HTML]{5A8AC6}0.12 \\
Pre-GFC/GFC & 57 & \cellcolor[HTML]{85A8D5}0.10 & \cellcolor[HTML]{84A8D5}0.10 & \cellcolor[HTML]{89ABD6}0.11 & \cellcolor[HTML]{FAC4C7}5.91 & \cellcolor[HTML]{FBC5C8}5.87 & \cellcolor[HTML]{FABFC2}6.05 & \cellcolor[HTML]{6A96CC}0.13 & \cellcolor[HTML]{608EC8}0.12 & \cellcolor[HTML]{6B96CC}0.13 \\
Euro crisis/calm & 0 & -- & -- & -- & \cellcolor[HTML]{5A8AC6}0.00 & \cellcolor[HTML]{5A8AC6}0.00 & \cellcolor[HTML]{5A8AC6}0.00 & -- & -- & -- \\
COVID-19 & 10 & \cellcolor[HTML]{F8696B}0.31 & \cellcolor[HTML]{F87779}0.29 & \cellcolor[HTML]{F87B7D}0.29 & \cellcolor[HTML]{CEDBEF}3.06 & \cellcolor[HTML]{C9D8ED}2.94 & \cellcolor[HTML]{C8D8ED}2.91 & \cellcolor[HTML]{F97E80}0.36 & \cellcolor[HTML]{F8797B}0.36 & \cellcolor[HTML]{F8696B}0.37 \\
\hline
\end{tabular}%
}
\caption{Layer-level excitation summaries using the same $AW$ quantities as Table \ref{tab:scenarios_weights_layer_rd}. Layer 0 is asset similarity, layer 1 solvency and layer 2 profitability.}
\label{tab:timevariation_layers}
\end{table}

\begin{figure}[htbp!]
\centering
\makebox[\textwidth][c]{\includegraphics[width=1.12\textwidth]{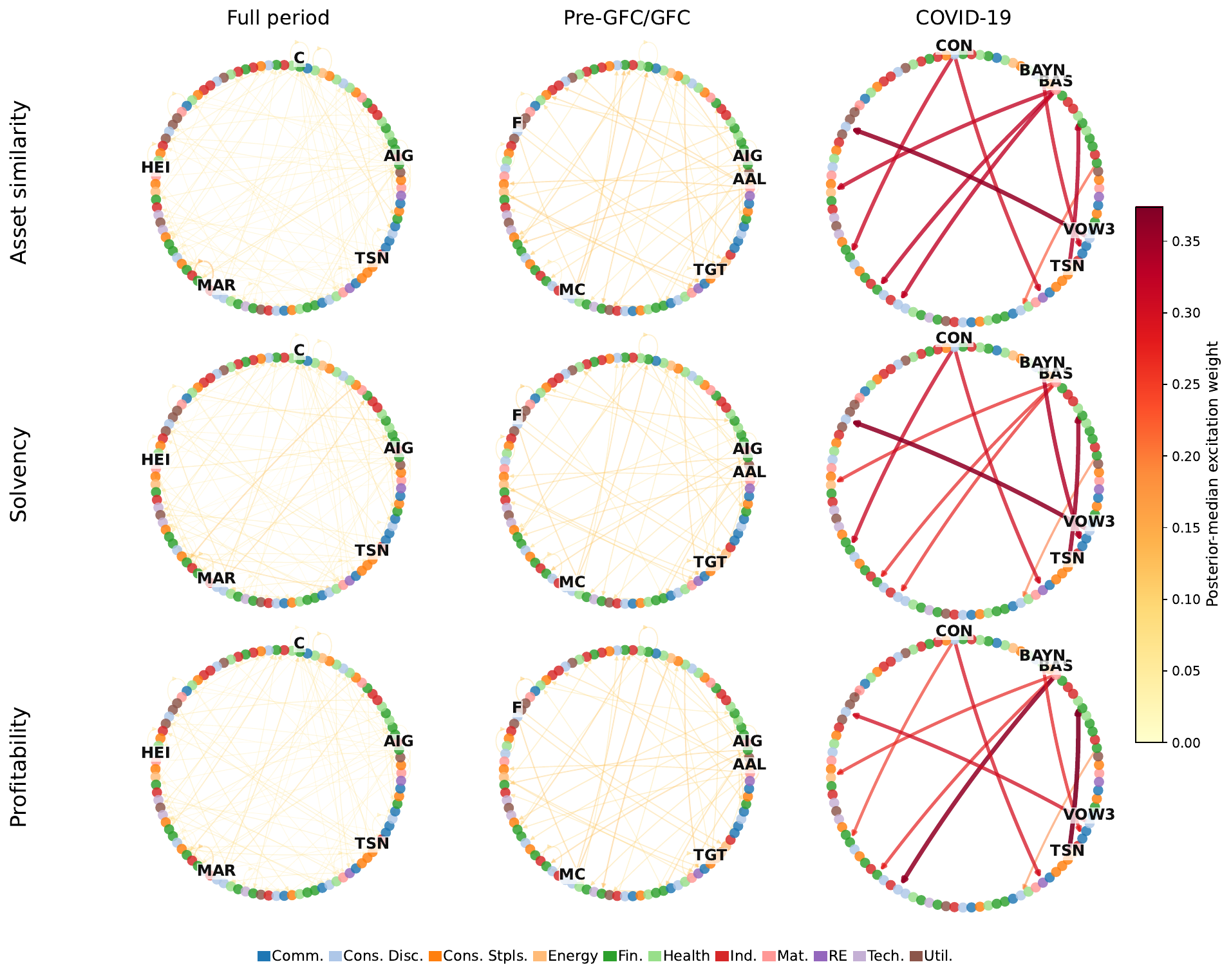}}
\caption{Layer-level posterior-median excitation networks for the selected full-period fit, the pre-GFC/GFC refit and the COVID-19 refit. The euro-crisis/calm refit is omitted because it has no positive posterior-median excitation edges. Period titles are shown above columns and channel names are shown on the left.}
\label{fig:timevariation_layer_networks}
\end{figure}

\begin{table}[htbp!]
\centering
\resizebox{0.8\textwidth}{!}{%
\begin{tabular}{rlllllllll}
\hline
\textbf{Rank} & \multicolumn{3}{c}{\textbf{Full period}} & \multicolumn{3}{c}{\textbf{Pre-GFC/GFC}} & \multicolumn{3}{c}{\textbf{COVID-19}} \\
\hline
 & \textbf{NDW} & \textbf{DW} & \textbf{ECW} & \textbf{NDW} & \textbf{DW} & \textbf{ECW} & \textbf{NDW} & \textbf{DW} & \textbf{ECW} \\
\hline
1 & \cellcolor[HTML]{2CA02C}{\color[HTML]{FFFFFF} AIG} & \cellcolor[HTML]{2CA02C}{\color[HTML]{FFFFFF} AIG} & \cellcolor[HTML]{2CA02C}{\color[HTML]{FFFFFF} C} & \cellcolor[HTML]{FF9896}{\color[HTML]{000000} AAL} & \cellcolor[HTML]{2CA02C}{\color[HTML]{FFFFFF} AIG} & \cellcolor[HTML]{2CA02C}{\color[HTML]{FFFFFF} AIG} & \cellcolor[HTML]{FF9896}{\color[HTML]{000000} BAS} & \cellcolor[HTML]{FF9896}{\color[HTML]{000000} BAS} & \cellcolor[HTML]{FF9896}{\color[HTML]{000000} BAS} \\
2 & \cellcolor[HTML]{AEC7E8}{\color[HTML]{000000} MAR} & \cellcolor[HTML]{2CA02C}{\color[HTML]{FFFFFF} C} & \cellcolor[HTML]{2CA02C}{\color[HTML]{FFFFFF} AIG} & \cellcolor[HTML]{2CA02C}{\color[HTML]{FFFFFF} AIG} & \cellcolor[HTML]{FF9896}{\color[HTML]{000000} AAL} & \cellcolor[HTML]{2CA02C}{\color[HTML]{FFFFFF} C} & \cellcolor[HTML]{AEC7E8}{\color[HTML]{000000} CON} & \cellcolor[HTML]{AEC7E8}{\color[HTML]{000000} CON} & \cellcolor[HTML]{AEC7E8}{\color[HTML]{000000} MAR} \\
3 & \cellcolor[HTML]{FF7F0E}{\color[HTML]{000000} TSN} & \cellcolor[HTML]{FF7F0E}{\color[HTML]{000000} TSN} & \cellcolor[HTML]{FF7F0E}{\color[HTML]{000000} TSN} & \cellcolor[HTML]{FF7F0E}{\color[HTML]{000000} TGT} & \cellcolor[HTML]{2CA02C}{\color[HTML]{FFFFFF} C} & \cellcolor[HTML]{AEC7E8}{\color[HTML]{000000} MC} & \cellcolor[HTML]{FF7F0E}{\color[HTML]{000000} TSN} & \cellcolor[HTML]{98DF8A}{\color[HTML]{000000} AZN} & \cellcolor[HTML]{2CA02C}{\color[HTML]{FFFFFF} LNC} \\
4 & \cellcolor[HTML]{FF9896}{\color[HTML]{000000} HEI} & \cellcolor[HTML]{AEC7E8}{\color[HTML]{000000} MAR} & \cellcolor[HTML]{AEC7E8}{\color[HTML]{000000} MAR} & \cellcolor[HTML]{AEC7E8}{\color[HTML]{000000} MC} & \cellcolor[HTML]{2CA02C}{\color[HTML]{FFFFFF} ISP} & \cellcolor[HTML]{2CA02C}{\color[HTML]{FFFFFF} LNC} & \cellcolor[HTML]{AEC7E8}{\color[HTML]{000000} VOW3} & \cellcolor[HTML]{FF7F0E}{\color[HTML]{000000} TSN} & \cellcolor[HTML]{FFBB78}{\color[HTML]{000000} HES} \\
5 & \cellcolor[HTML]{2CA02C}{\color[HTML]{FFFFFF} C} & \cellcolor[HTML]{FF9896}{\color[HTML]{000000} HEI} & \cellcolor[HTML]{2CA02C}{\color[HTML]{FFFFFF} WFC} & \cellcolor[HTML]{AEC7E8}{\color[HTML]{000000} F} & \cellcolor[HTML]{FF7F0E}{\color[HTML]{000000} TGT} & \cellcolor[HTML]{2CA02C}{\color[HTML]{FFFFFF} CB} & \cellcolor[HTML]{98DF8A}{\color[HTML]{000000} BAYN} & \cellcolor[HTML]{AEC7E8}{\color[HTML]{000000} F} & \cellcolor[HTML]{AEC7E8}{\color[HTML]{000000} CON} \\
\hline
\end{tabular}%
}
\caption{Top-five ticker rankings in the time-varying posterior-median networks. NDW ranks nodes by weighted out-degree minus weighted in-degree, DW is weighted degree and ECW is weighted eigenvector centrality computed on the symmetrised weighted network. Ticker cells are coloured by sector using the same palette as Figure \ref{fig:network_rd_3lall}. The euro-crisis/calm refit is omitted because it has no positive posterior-median excitation edges.}
\label{tab:timevariation_topnodes}
\end{table}

\subsection{Time Variation Degree-Distribution Checks}
Table \ref{tab:timevariation_degree_tests} repeats the degree-distribution comparison for the full-period and regime-specific networks, using the same KDE, log-normal, exponential and power-law comparisons as in Table \ref{tab:degree_tests_app}. As in the main empirical results, the evidence for scale-free behaviour is clearest for weighted degree rather than unweighted degree. The full-period and pre-GFC/GFC weighted-degree comparisons favour the power-law fit over the KDE, log-normal and exponential alternatives. The COVID-19 weighted-degree comparisons also point toward the power-law fit, but this result is based on a much smaller active posterior-median network and should therefore be interpreted cautiously. The euro-crisis/calm period has no positive posterior-median excitation network, so the comparison is not defined for that period.

\begin{table}[htbp!]
\centering
\resizebox{0.495\textwidth}{!}{%
\begin{tabular}{lllllllll}
\hline
\multicolumn{1}{c}{\multirow{2}{*}{\textbf{Period}}} & \multicolumn{1}{c}{\multirow{2}{*}{\textbf{Fit 1}}} & \multicolumn{1}{c}{\multirow{2}{*}{\textbf{Fit 2}}} & \multicolumn{3}{l}{\textbf{Degree}} & \multicolumn{3}{l}{\textbf{Weighted degree}} \\ \cline{4-9}
\multicolumn{1}{c}{} & \multicolumn{1}{c}{} & \multicolumn{1}{c}{} & \textbf{LR} & KS & p-val & LR & KS & p-val \\ \hline
Full period & KDE & Log normal & (489) & 0.66 & 0.00 & (9) & 0.51 & 0.00 \\
Full period & KDE & Power law & 31 & 0.50 & 0.00 & (230) & 0.88 & 0.00 \\
Full period & KDE & Exponential & (23) & 0.57 & 0.00 & (13) & 0.54 & 0.00 \\
Full period & Log normal & Power law & 520 & 0.29 & 0.00 & (221) & 0.79 & 0.00 \\
Full period & Log normal & Exponential & 466 & 0.66 & 0.00 & (4) & 0.11 & 0.71 \\
Full period & Power law & Exponential & (54) & 0.50 & 0.00 & 218 & 0.79 & 0.00 \\
Pre-GFC/GFC & KDE & Log normal & (569) & 0.58 & 0.00 & (30) & 0.58 & 0.00 \\
Pre-GFC/GFC & KDE & Power law & 7 & 0.58 & 0.00 & (164) & 0.85 & 0.00 \\
Pre-GFC/GFC & KDE & Exponential & (33) & 0.58 & 0.00 & (22) & 0.58 & 0.00 \\
Pre-GFC/GFC & Log normal & Power law & 576 & 0.58 & 0.00 & (134) & 0.65 & 0.00 \\
Pre-GFC/GFC & Log normal & Exponential & 536 & 0.58 & 0.00 & 8 & 0.50 & 0.00 \\
Pre-GFC/GFC & Power law & Exponential & (40) & 0.58 & 0.00 & 142 & 0.82 & 0.00 \\
\hline
\end{tabular}%
}
\resizebox{0.495\textwidth}{!}{%
\begin{tabular}{lllllllll}
\hline
\multicolumn{1}{c}{\multirow{2}{*}{\textbf{Period}}} & \multicolumn{1}{c}{\multirow{2}{*}{\textbf{Fit 1}}} & \multicolumn{1}{c}{\multirow{2}{*}{\textbf{Fit 2}}} & \multicolumn{3}{l}{\textbf{Degree}} & \multicolumn{3}{l}{\textbf{Weighted degree}} \\ \cline{4-9}
\multicolumn{1}{c}{} & \multicolumn{1}{c}{} & \multicolumn{1}{c}{} & \textbf{LR} & KS & p-val & LR & KS & p-val \\ \hline
Euro crisis/calm & KDE & Log normal & -- & -- & -- & -- & -- & -- \\
Euro crisis/calm & KDE & Power law & -- & -- & -- & -- & -- & -- \\
Euro crisis/calm & KDE & Exponential & -- & -- & -- & -- & -- & -- \\
Euro crisis/calm & Log normal & Power law & -- & -- & -- & -- & -- & -- \\
Euro crisis/calm & Log normal & Exponential & -- & -- & -- & -- & -- & -- \\
Euro crisis/calm & Power law & Exponential & -- & -- & -- & -- & -- & -- \\
COVID-19 & KDE & Log normal & (310) & 0.88 & 0.00 & (2) & 0.53 & 0.02 \\
COVID-19 & KDE & Power law & 8 & 0.88 & 0.00 & (6) & 0.65 & 0.00 \\
COVID-19 & KDE & Exponential & (15) & 0.88 & 0.00 & (2) & 0.65 & 0.00 \\
COVID-19 & Log normal & Power law & 318 & 0.88 & 0.00 & (4) & 0.29 & 0.47 \\
COVID-19 & Log normal & Exponential & 295 & 0.88 & 0.00 & (0) & 0.24 & 0.75 \\
COVID-19 & Power law & Exponential & (23) & 0.88 & 0.00 & 3 & 0.35 & 0.24 \\
\hline
\end{tabular}%
}
\caption{Degree distribution fit for the full-period and regime-specific networks, following the appendix implementation. LR is the log likelihood ratio between Fit 1 and Fit 2: positive values favour Fit 1, while parenthesised values denote negative values and favour Fit 2. The euro-crisis/calm period has no positive posterior-median edges, so the degree-fit comparisons are not defined.}
\label{tab:timevariation_degree_tests}
\end{table}

\subsection{Comparison to Previous Work}
Figure \ref{fig:topedges_combined} visualises the top edges connected to highly active nodes.

\begin{figure}[htbp!]
  \centering
  % Out-edges: k' is the transmitting node and k is the receiving node.
  \begin{minipage}[t]{1.0\textwidth}
    \centering
    \resizebox{\textwidth}{!}{%
      \begin{tabular}{lllllll}
      \hline
      \textbf{$k'$} & \textbf{$k$} & \textbf{$AW$} & \textbf{Ticker $k'$} & \textbf{Ticker $k$} & \textbf{Industry $k'$} & \textbf{Industry $k$} \\ \hline
      66 & 12 & 0.28 & MAR & BAC & Cons. Discr. Svcs. & Banking \\
      66 & 65 & 0.26 & MAR & LUV & Cons. Discr. Svcs. & Industrial Svcs. \\
      66 & 58 & 0.26 & MAR & JPM & Cons. Discr. Svcs. & Banking \\
      3  & 63 & 0.25 & AIG & LNC & Insurance          & Insurance \\
      88 & 22 & 0.24 & TSN & C   & Cons. Staple Prd.  & Banking \\
      3  & 51 & 0.23 & AIG & HIG & Insurance          & Insurance \\
      3  & 79 & 0.23 & AIG & PRU & Insurance          & Insurance \\
      22 & 22 & 0.22 & C   & C   & Banking            & Banking \\
      3  & 43 & 0.22 & AIG & FE  & Insurance          & Utilities \\
      66 & 74 & 0.22 & MAR & NWL & Cons. Discr. Svcs. & Cons. Discr. Svcs. \\
      3  & 3  & 0.21 & AIG & AIG & Insurance          & Insurance \\
      3  & 78 & 0.21 & AIG & PRU & Insurance          & Insurance \\
      88 & 84 & 0.21 & TSN & SPG & Cons. Staple Prd.  & Real Estate \\
      88 & 53 & 0.21 & TSN & HPQ & Cons. Staple Prd.  & Tech. Hardw. \& Semic. \\
      88 & 95 & 0.21 & TSN & WFC & Cons. Staple Prd.  & Banking \\
      22 & 95 & 0.21 & C   & WFC & Banking            & Banking \\
      22 & 57 & 0.20 & C   & ISP & Banking            & Banking \\
      88 & 90 & 0.20 & TSN & UNP & Cons. Staple Prd.  & Industrial Svcs. \\
      22 & 44 & 0.19 & C   & GE  & Banking            & Industrial Products \\
      \hline
      \end{tabular}%
      \quad
      \begin{tabular}{lllllll}
      \hline
      \textbf{$k'$} & \textbf{$k$} & \textbf{$AW$} & \textbf{Ticker $k'$} & \textbf{Ticker $k$} & \textbf{Industry $k'$} & \textbf{Industry $k$} \\ \hline
      3  & 69 & 0.18 & AIG & MET & Insurance          & Insurance \\
      88 & 60 & 0.18 & TSN & KR  & Cons. Staple Prd.  & Retail \& Whsl.-Stpls. \\
      88 & 1  & 0.17 & TSN & AD  & Cons. Staple Prd.  & Retail \& Whsl.-Stpls. \\
      3  & 22 & 0.16 & AIG & C   & Insurance          & Banking \\
      22 & 19 & 0.15 & C   & BP  & Banking            & Oil \& Gas \\
      3  & 36 & 0.15 & AIG & DGE & Insurance          & Cons. Staple Prd. \\
      3  & 46 & 0.15 & AIG & GSK & Insurance          & Health Care \\
      3  & 25 & 0.14 & AIG & CB  & Insurance          & Insurance \\
      22 & 89 & 0.13 & C   & TTE & Banking            & Oil \& Gas \\
      22 & 85 & 0.13 & C   & T   & Banking            & Telecommunications \\
      3  & 17 & 0.13 & AIG & BMY & Insurance          & Health Care \\
      3  & 55 & 0.10 & AIG & IBM & Insurance          & Software \& Tech. Svcs. \\
      3  & 92 & 0.10 & AIG & VOD & Insurance          & Telecommunications \\
      3  & 8  & 0.09 & AIG & AV  & Insurance          & Insurance \\
      66 & 22 & 0.09 & MAR & C   & Cons. Discr. Svcs. & Banking \\
      58 & 15 & 0.06 & JPM & BAYN & Banking           & Health Care \\
      88 & 44 & 0.06 & TSN & GE  & Cons. Staple Prd.  & Industrial Products \\
      58 & 4  & 0.06 & JPM & AIR & Banking            & Industrial Products \\
      3  & 57 & 0.05 & AIG & ISP & Insurance          & Banking \\
      \hline
      \end{tabular}%
    }
    \caption*{\textbf{(a)} Out-edges (sorted $AW$ descending)}
  \end{minipage}
  \vspace{2ex}
  % In-edges: k' is the transmitting node and k is the receiving node.
  \begin{minipage}[t]{1.0\textwidth}
    \centering
    \resizebox{0.6\textwidth}{!}{%
      \begin{tabular}{lllllll}
      \hline
      \textbf{$k'$} & \textbf{$k$} & \textbf{$AW$} & \textbf{Ticker $k'$} & \textbf{Ticker $k$} & \textbf{Industry $k'$} & \textbf{Industry $k$} \\ \hline
      66 & 58 & 0.26 & MAR & JPM & Cons. Discr. Svcs. & Banking \\
      88 & 22 & 0.24 & TSN & C   & Cons. Staple Prd. & Banking \\
      22 & 22 & 0.22 & C   & C   & Banking           & Banking \\
      42 & 88 & 0.21 & F   & TSN & Cons. Discr. Svcs. & Cons. Staple Prd. \\
      3  & 3  & 0.21 & AIG & AIG & Insurance         & Insurance \\
      3  & 22 & 0.16 & AIG & C   & Insurance         & Banking \\
      60 & 88 & 0.11 & KR  & TSN & Retail \& Whsl.-Stpls. & Cons. Staple Prd. \\
      66 & 22 & 0.09 & MAR & C   & Cons. Discr. Svcs. & Banking \\ \hline
      \end{tabular}%
    }
    \caption*{\textbf{(b)} In-edges (sorted $AW$ descending)}
  \end{minipage}

  \vspace{1ex}
  \caption{Median weights of the out- and in-edges of the top 5 nodes by weighted net-degree, shown vertically. Here $k'$ denotes the transmitting node and $k$ the receiving node.}
  \label{fig:topedges_combined}
\end{figure}

\subsection{Network Sparsity}\label{sec:networksparsity}
We report the posterior median of $\rho$ (lower values imply sparser inferred networks), with HDIs summarised in Table \ref{tab:scenarios_rho_rd}. The selected two- and three-channel model set-ups produce sparser inferred networks than the selected single-channel set-up, with the posterior median of $\rho$ decreasing from $0.055$ for ``1l asset'' to $0.033$ for ``2l asset solv'' and $0.025$ for ``3l all''. Because these model set-ups differ in both model layers and covariates, this contrast is descriptive rather than evidence that adding a model layer itself causes greater sparsity. Within the three-channel time-horizon sensitivity comparison, extending $\Delta t_{max}$ from $10$ to $20$ days reduces the posterior median from $0.025$ to $0.020$, while shortening it to $5$ days gives a denser estimated network. This comparison supports treating the horizon as a sensitivity choice.
\begin{table}[htbp!]
    \centering
    \resizebox{0.38\textwidth}{!}{%
    \begin{tabular}{lcc}
    \hline
   % \rowcolor{gray!20}
    \textbf{Model set-up} & \textbf{$\rho$} & \textbf{95\% HDI}  \\ \hline
%1l nocovs                                    & \cellcolor[HTML]{F8696B}0.076                           & {[}0.065,0.088{]}                     & 31\%                                                                                                              \\
1l base                                      & \cellcolor[HTML]{FBC6C8}0.051                           & {[}0.044,0.059{]}  \\
1l asset                                     & \cellcolor[HTML]{FBB9BC}0.055                           & {[}0.047,0.063{]}  \\
%1l asset invgamma10                          & \cellcolor[HTML]{FBBFC2}0.053                           & {[}0.046,0.061{]}                     & 29\%                                                                                                              \\
%1l asset    invgamma25                       & \cellcolor[HTML]{FBBDC0}0.054                           & {[}0.046,0.062{]}                     & 30\%                                                                                                              \\
%1l asset kappahyper10                        & \cellcolor[HTML]{FBB7BA}0.055                           & {[}0.047,0.063{]}                     & 29\%                                                                                                              \\
%1l asset kappa1                              & \cellcolor[HTML]{FA9699}0.064                           & {[}0.055,0.074{]}                     & 31\%                                                                                                              \\
%2l nocovs                                    & \cellcolor[HTML]{FCFBFE}0.037                           & {[}0.032,0.043{]}                     & 31\%                                                                                                              \\
2l base                                      & \cellcolor[HTML]{D6E1F1}0.033                           & {[}0.028,0.038{]}  \\
2l asset solv                                & \cellcolor[HTML]{D7E2F2}0.033                           & {[}0.028,0.038{]}  \\
%2l asset solv kappa1                         & \cellcolor[HTML]{FCFCFF}0.037                           & {[}0.031,0.043{]}                     & 32\%                                                                                                              \\
%3l nocovs                                    & \cellcolor[HTML]{83A6D4}0.024                           & {[}0.02,0.028{]}                      & 33\%                                                                                                              \\
3l base                                      & \cellcolor[HTML]{8BACD7}0.025                           & {[}0.021,0.029{]}  \\
3l all                                       & \cellcolor[HTML]{89ABD6}0.025                           & {[}0.021,0.029{]}  \\
3l all 20dt                                  & \cellcolor[HTML]{5A8AC6}0.020                           & {[}0.016,0.024{]}  \\
3l all 5dt                                   & \cellcolor[HTML]{A3BDDF}0.028                           & {[}0.023,0.032{]}  \\
%3l all kappa1                                & \cellcolor[HTML]{9DB9DD}0.027                           & {[}0.022,0.032{]}                     & 35\%                                                                                                             \\
    \hline
    \end{tabular}%
    }
    \caption{$\rho$ median and HDIs}
    \label{tab:scenarios_rho_rd}
\end{table}

\iffalse
\begin{figure}[htbp!]
\centering
\begin{subfigure}[b]{0.5\textwidth}
\includegraphics[width=1\textwidth, angle=0, trim= 825 350 750 750,clip]{img/empirical_application/scenario_1l_nocovs_10dt/adj.pdf}
\subcaption{1l nocovs}
\end{subfigure}%
\begin{subfigure}[b]{0.5\textwidth}
\includegraphics[width=1\textwidth, angle=0, trim=825 350 750 750,clip]{img/empirical_application/scenario1l_base_10dt_invgamma1000/adj.pdf}
\subcaption{1l base}
\end{subfigure}
\begin{subfigure}[b]{0.5\textwidth}
\includegraphics[width=1\textwidth, angle=0, trim= 825 350 750 750,clip]{img/empirical_application/scenario2l_nocovs_10dt/adj.pdf}
\subcaption{2l nocovs}
\end{subfigure}%
\begin{subfigure}[b]{0.5\textwidth}
\includegraphics[width=1\textwidth, angle=0, trim= 825 350 750 750,clip]{img/empirical_application/scenario2l_base_10dt_invgamma1000/adj.pdf}
\subcaption{2l base}
\end{subfigure}
\begin{subfigure}[b]{0.5\textwidth}
\includegraphics[width=1\textwidth, angle=0, trim= 825 350 750 750,clip]{img/empirical_application/scenario3l_nocovs_10dt/adj.pdf}
\subcaption{3l nocovs}
\end{subfigure}%
\begin{subfigure}[b]{0.5\textwidth}
\includegraphics[width=1\textwidth, angle=0, trim= 825 350 750 750,clip]{img/empirical_application/scenario3l_base_10dt_invgamma1000/adj.pdf}
\subcaption{3l base}
\end{subfigure}
\caption{Adjacency networks for model set-ups with no covariates.}% \label{fig:network_sparsity_adj_rd}
\end{figure}
\fi

\subsection{Goodness-of-fit}\label{sec:goodnessoffit_emp}

\begin{figure}[htbp!]
\centering
\begin{subfigure}[b]{0.243\textwidth}
\includegraphics[width=1\textwidth, angle=0, trim=  0 5 20 40,clip]{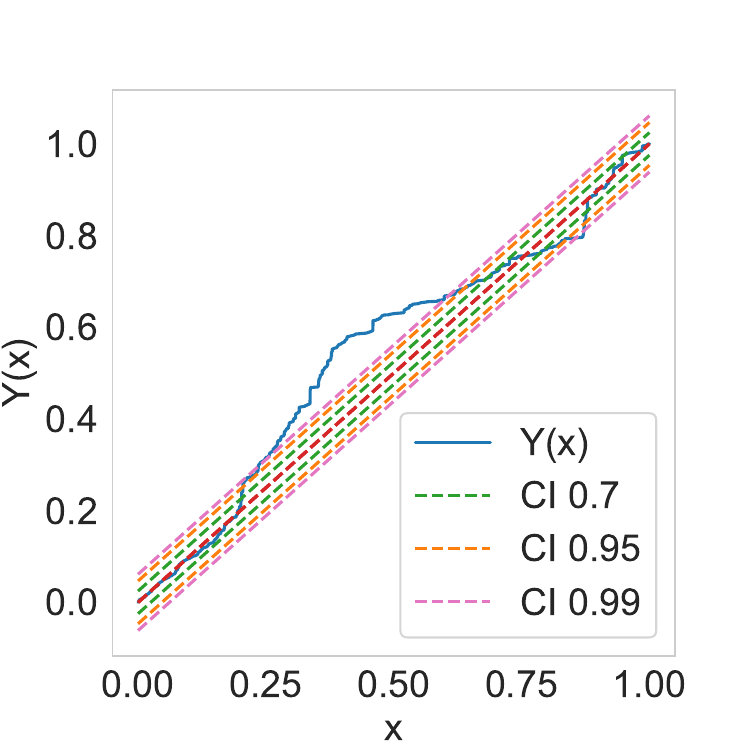}
\subcaption{3l all 20dt}
\end{subfigure}%
\begin{subfigure}[b]{0.25\textwidth}
\includegraphics[width=0.9\textwidth, angle=0, trim= 13 5 33 43,clip]
{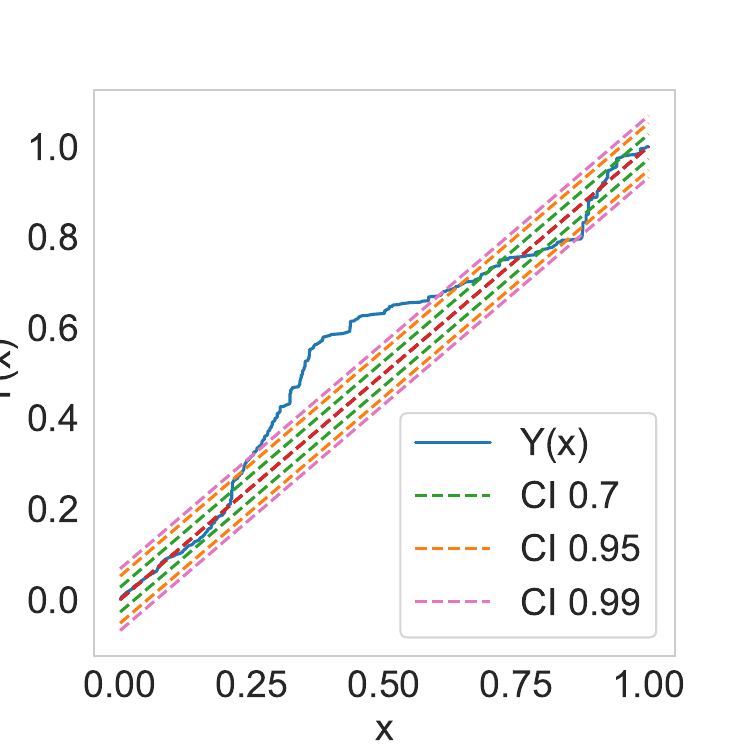}
\subcaption{1l asset}
\end{subfigure}%
\begin{subfigure}[b]{0.25\textwidth}
\includegraphics[width=0.9\textwidth, angle=0, trim=  13 5 33 43,clip]{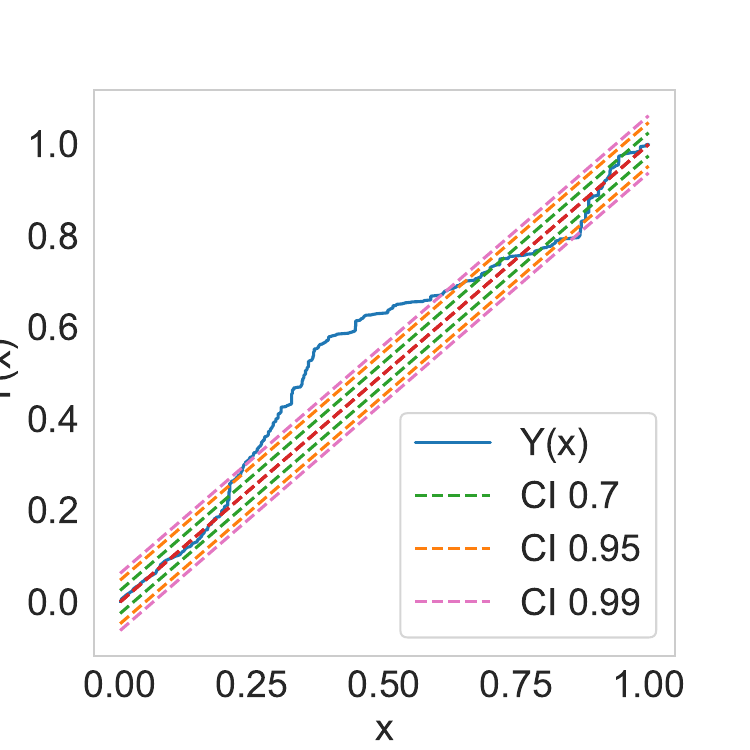}
\subcaption{3l all 10dt}
\end{subfigure}%
%\begin{subfigure}[b]{0.5\textwidth}
%\includegraphics[width=0.9\textwidth, angle=0, trim= 40 40 40 40,clip]%{img/empirical_application/scenario1l_cat_10dt_invgamma10/step.pdf}
%\subcaption{1l asset invgamma10}
%\end{subfigure}
\begin{subfigure}[b]{0.25\textwidth}
\includegraphics[width=0.9\textwidth, angle=0, trim=  13 5 33 43,clip]{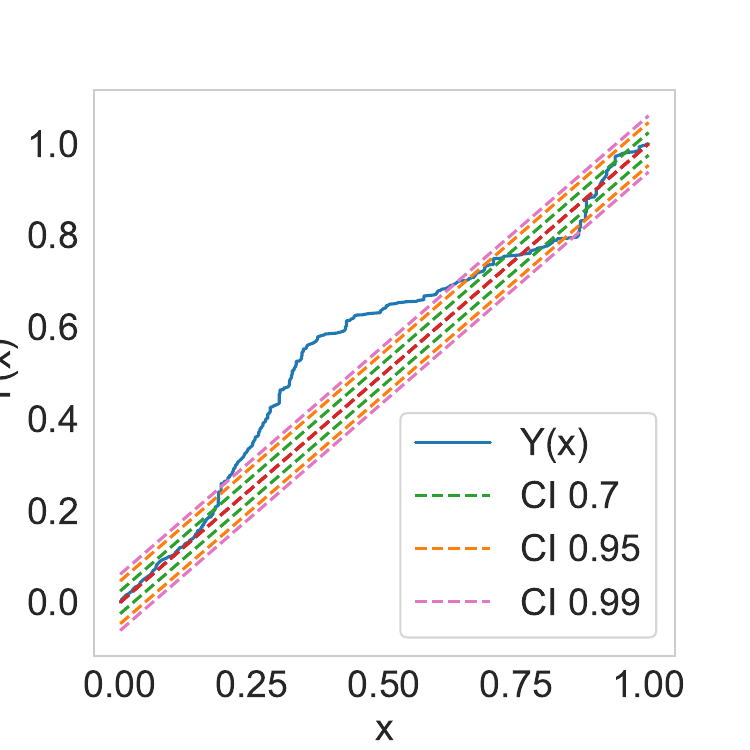}
\subcaption{3l all 5dt}
\end{subfigure}%
%\begin{subfigure}[b]{0.5\textwidth}
%\includegraphics[width=0.9\textwidth, angle=0, trim=  40 40 40 40,clip]{img/empirical_application/scenario1l_cat_10dt_kappa1/step.pdf}
%\subcaption{1l asset kappa1}
%\end{subfigure}
  \caption{Cumulative step function and confidence intervals for the total process, comparing three-layer model set-ups with $\Delta t_{max}$ equal to 5, 10 and 20 days and a selected single-layer set-up.}
\label{fig:step_rd2}
\end{figure}

To evaluate goodness-of-fit via residual process analysis (see Appendix \ref{sec:goodnessoffit}), we examine the cumulative step-function plots in Figure \ref{fig:step_rd2}. A hump outside the 99\% confidence interval followed by a subsequent drop indicates that a constant background rate does not fully capture broad time variation in extreme credit-spread jump activity. This is consistent with periods of increased jump activity during the Global Financial Crisis and subsequent market-stress episodes (Figure \ref{fig:process_rd2}).

The same broad departure occurs across the displayed model set-ups; changing the model-layer structure or covariates does not remove it. Changing the influence horizon $\Delta t_{max}$ modifies the residual pattern, with the 20-day fit spending less time outside the confidence bounds than the shorter-horizon fits. We therefore use these plots as a sensitivity diagnostic rather than as evidence that a particular model-layer structure resolves temporal misspecification. A time-varying background intensity could address this limitation, but is outside the scope of the present network-focused analysis.

\iffalse
\begin{figure}[htbp!]
\centering
\begin{subfigure}[b]{0.5\textwidth}
\includegraphics[width=0.9\textwidth, angle=0, trim= 40 40 40 40,clip]{img/empirical_application/scenario_1l_nocovs_10dt/step.pdf}
\subcaption{1l nocovs}
\end{subfigure}%
\begin{subfigure}[b]{0.5\textwidth}
\includegraphics[width=0.9\textwidth, angle=0, trim= 40 40 40 40,clip]{img/empirical_application/scenario1l_base_10dt_invgamma1000/step.pdf}
\subcaption{1l base}
\end{subfigure}
\begin{subfigure}[b]{0.5\textwidth}
\includegraphics[width=0.9\textwidth, angle=0, trim=  40 40 40 40,clip]{img/empirical_application/scenario2l_nocovs_10dt/step.pdf}
\subcaption{2l nocovs}
\end{subfigure}%
\begin{subfigure}[b]{0.5\textwidth}
\includegraphics[width=0.9\textwidth, angle=0, trim=  40 40 40 40,clip]{img/empirical_application/scenario2l_base_10dt_invgamma1000/step.pdf}
\subcaption{2l base}
\end{subfigure}
\begin{subfigure}[b]{0.5\textwidth}
\includegraphics[width=0.9\textwidth, angle=0, trim=  40 40 40 40,clip]{img/empirical_application/scenario3l_nocovs_10dt/step.pdf}
\subcaption{3l nocovs}
\end{subfigure}%
\begin{subfigure}[b]{0.5\textwidth}
\includegraphics[width=0.9\textwidth, angle=0, trim=  40 40 40 40,clip]{img/empirical_application/scenario3l_base_10dt_invgamma1000/step.pdf}
\subcaption{3l base}
\end{subfigure}
  \caption{Cumulative step function and the corresponding confidence intervals for a total process, comparing ``nocovs'' and ``base'' model set-ups across one-, two- and three-layer specifications.}
\label{fig:step_rd1}
\end{figure}
\fi

\fi

\end{document}